\documentclass[12pt,tightenlines,eqsecnum,floats,showpacs,nofootinbib,amsmath,amssymb,aps,preprint,superscriptaddress]{revtex4-1}

\usepackage{graphicx}
\usepackage{amsmath,amssymb}
\usepackage{hyperref}
\usepackage{mathrsfs}

\begin{document}
	
\title{Quantum linear scalar fields with time dependent potentials: Overview
and applications to cosmology}
	
\author{Jer\'onimo Cortez}
\email{jacq@ciencias.unam.mx}
\affiliation{Departamento de F\'isica, Facultad de Ciencias, Universidad Nacional Aut\'onoma de M\'exico, Ciudad de M\'exico 04510, M\'exico}
\author{Guillermo A. Mena Marug\'an}
\affiliation{Instituto de Estructura de la Materia, IEM-CSIC,
Serrano 121, 28006 Madrid, Spain.}
\email{mena@iem.cfmac.csic.es}
\author{Jos\'e  Velhinho}
\affiliation{Faculdade de Ci\^encias, Universidade
da Beira Interior, R. Marqu\^es D'\'Avila e Bolama,
6201-001 Covilh\~a, Portugal.}
\email{jvelhi@ubi.pt}

\begin{abstract}
In this work, {we present an overview of uniqueness results derived in recent years} for the quantization of Gowdy cosmological models {and for}  (test) Klein-Gordon fields minimally coupled to {Friedmann-Lema\^{\i}tre-Robertson-Walker}, de Sitter, and Bianchi I spacetimes. These {results are attained by imposing the criteria} of symmetry invariance and {of unitary implementability} of the dynamics. {This powerful combination of criteria allows not only to address the ambiguity in the representation of the} {canonical commutation relations,} {but also to single} out a preferred set of fundamental variables. For the sake of clarity and completeness in the presentation (essentially as a background {and} complementary material), {we}  {first review} the classical and quantum theories of a scalar field in globally hyperbolic spacetimes. Special emphasis is made on complex structures and the unitary implementability of symplectic transformations.

\end{abstract}


\maketitle


\tableofcontents

\section{Introduction}
\label{sec:Intro}
As it is well known, the quantization of systems with field-like degrees of freedom involves choices that generically lead to inequivalent theories within the standard Hilbert space approach \cite{Wald-book}. In contrast with {the situation found for mechanical systems with a finite number of degrees of freedom}, where the Stone-von Neumann theorem guarantees the unitary equivalence between strongly continuous, irreducible, and unitary representations of the Weyl relations \cite{Simon}, in quantum field theory no general uniqueness theorem exists and ``physical results'' depend on the representation adopted, a fact that brings into question their significance. So, in order to regain robustness in the quantum predictions, one has to look for physically plausible additional criteria, usually based on the classical symmetries of the system, to warrant uniqueness. For instance, in background independent quantum gravity \cite{lqg,lqg1,lqg2,lqg3,lqg4} the requirement of spatial diffeomorphism invariance provides a unique representation of the kinematical holonomy-flux algebra \cite{Lewandowski:2005jk}. For field theories in Minkowski spacetime, the criterion of Poincar\'{e} invariance is employed to arrive at a unique representation. {For example}, if the field theory corresponds to a Klein-Gordon (KG) field, Poincar\'{e} invariance, adapted to the dynamics of the considered theory, selects a complex structure, which is the mathematical object encoding the ambiguity in the representation of the canonical commutation relations (CCRs), and determines the vacuum state of the Fock representation. In  {more} general but still stationary spacetimes, the time translation symmetry is exploited to formulate the so-called energy criterion and then single out a preferred complex structure \cite{Ashtekar:1975zn,Ashtekar:1980}, determining a unique (up to unitary equivalence) Fock representation. However, when the symmetries are severely restricted, as it is the case for generic spacetimes or manifestly non-stationary systems, new requirements must be imposed to complete the quantization process. 

{In} addition, let us remark that the issue of uniqueness not only concerns {the} choice of a privileged representation of the CCRs, but {also} the choice of a preferred set of classical fundamental variables. Indeed, since classical canonical transformations will not all become into unitary transformations, there are different fundamental variables giving rise to {inequivalent} quantum theories \cite{Haag}. 

An interesting system, with applications in cosmology, is that of a scalar field with time dependent mass (or, equivalently, subject to a time dependent quadratic potential) propagating in a spatially compact, static spacetime. More precisely, {let us consider} a scalar field $\psi$ obeying an equation of the form
\begin{equation}
\label{sf-with-tdp}
\partial_t^2{\psi}-\Delta \psi +s(t)\psi=0,
\end{equation}
{in} a static spacetime {manifold} with topology $\mathbb{I}\times \Sigma$, where $\mathbb{I}$ is an interval of the real line, $\Sigma$ is a compact Riemannian surface of dimension $d\leq 3$, $\Delta$ is the Laplace-Beltrami (LB) operator on $\Sigma$, and $s(t)$ is a sufficiently regular function of {the time $t$}. Since this system is manifestly non-stationary, {given the time dependence of $s(t)$,} neither the Poincar\'{e} nor the energy criteria can be used to specify a unique favored quantization, and it is then necessary to seek for extra requirements in order to complete the quantization. Remarkably, as it has been shown in {Refs.}  \cite{CMSV,CMV-PRD81,CMOV-JCAP10,CMOV-cqg,CMOV-PRD83,CMOV-PRD86}, {one can demand that}
\begin{enumerate}
\item{the vacuum state be invariant under the isometries of the spatial manifold $\Sigma$,}
\item{{the} dynamics dictated by the field equation (\ref{sf-with-tdp}) be unitarily {implementable},}
\end{enumerate}
{in order to single out} a unique preferred Fock representation {for} the system. Notably, {these combined criteria} of symmetry invariance and {of} unitary dynamics {select} a unique preferred field description as well, {specifying (in a certain context) a canonical pair of field variables, so that they actually remove the two kinds} of ambiguities present {in} the quantization of the field system. To attain this uniqueness result, it suffices that $s(t)$ be twice differentiable and with a second derivative that is integrable {over} each compact subinterval of the time domain $\mathbb{I}$.

It is worth remarking that there is a variety of interesting situations where the study of scalar fields with time dependent mass {finds} application in cosmology. For instance, in the framework of symmetry reduced models in gravity, one can see that in (linearly polarized) Gowdy cosmological spacetimes \cite{Gowdy}, which are the simplest inhomogeneous, empty, spatially closed cosmological models, the local degrees of freedom characterizing the inhomogeneities can be described in terms of scalar fields obeying  equations of the form (\ref{sf-with-tdp}) \cite{CMSV}. Specifically, for Gowdy cosmologies with the topology of a three-torus, the wave equation corresponds to a scalar field with time dependent mass propagating in a static $(1+1)$-dimensional fictitious spacetime, {for which} the spatial manifold $\Sigma$ is a circle  \cite{CMSV,CCM-PRD73-i,CCM-PRD73-ii,CCMV-gt3-uniq,CMV-PRD75,CCMV-GT3-SR}. For the three-sphere and the three-handle, which are the remaining two possible spatial topologies in the Gowdy models, the local gravitational degrees of freedom are described by an axisymmetric KG field with time dependent mass in a static $(2+1)$-dimensional auxiliary spacetime, {such that the} spatial slices are two-spheres \cite{BVV0,BVV,CMV-GowS1S2S3}. Let us recall, in addition, that in non-stationary scenarios like those encountered in cosmology, it is customary to scale the field configurations by time varying functions when one allows that part of its evolution be assigned to the time dependent spacetime in which the propagation takes place. This is the situation found for free fields in (e.g.) {Friedmann-Lema\^{\i}tre-Robertson-Walker (FLRW)} universes, and in de Sitter spacetime, where the use of conformal time combined with a scaling (by the time dependent conformal factor) of the original field variable transforms the original free field equation into an equation of the form (\ref{sf-with-tdp}). This type of scaling and field equation are found in the treatment of quantum perturbations in cosmology, with the homogeneous background kept as a classical scenario (see, for instance, {Refs.} \cite{Mukhanov,Mukhanov-etal,Bardeen-PRD22,Halli-Haw-PRD31,FMMMOV}), as well as in the full quantization of inhomogeneous models via the hybrid approach \cite{hybrid-app}, where a loop representation is adopted for the homogeneous gravitational sector and a Fock quantization is used for the inhomogeneities. Thus, apart from the {general interest that the quantization of KG systems (\ref{sf-with-tdp})} {may} {have within the formal and mathematical physics apparatus of {quantum field theory} in curved spacetime, the exploration of this mathematically rigorous approach and the possible quantum outcomes have an important impact in the arena of modern} cosmology.

Let us emphasize that the specification of a complex structure (compatible with the symplectic form) is a key ingredient in order to attain a Fock quantization; in fact, it is from a complex structure that a Fock space representation of the CCRs is constructed \cite{Wald-book,Kay,ccq-annphys}, since the relevant information on the choice of annihilation and creation-like variables is encoded in the complex structure. In turn, a choice of annihilation and creation operators selects a specific vacuum, which is typically the particle physics perspective on how to select a specific Fock representation of the CCRs (or, strictly speaking, of the corresponding Weyl relations). In general, distinct complex structures will define different (i.e. {not} unitarily equivalent) Fock representations. {Recall} that the choice of a complex structure with physical content is by no means a straightforward process, and for a general spacetime, or a manifestly non-stationary system, there is a priori {no} criteria to select one. In the absence of stationarity, {and specifically in the case of a scalar} field with generic time varying mass (\ref{sf-with-tdp}), a natural strategy is to look for compatible complex structures allowing for a unitary implementation of the spatial symmetries. The simplest choice is the compatible complex structure $j_0$ associated with the free massless field representation, which in the space of Cauchy data {for the field and its momentum, $(\varphi,\pi)$,} reads explicitly as $j_{0}(\varphi,\pi)=(-[-h\Delta]^{-1/2}\pi,[-h\Delta]^{1/2}\varphi)$, {where $h$ is the determinant of the induced metric $h_{ab}$ {($a,b$=1,2,3)} on the (Cauchy) spatial section $\Sigma$.} This compatible complex structure, while constructed from the LB operator ({and $h$}), commutes with the isometries of the spatial manifold $\Sigma$ and, consequently, {defines a Fock representation that is invariant}{\footnote{Invariance of the Fock representation refers to the invariance of the vacuum state.}} {under these symmetries}. Though the invariance requirement {reduces} the set of admissible Fock quantizations, it should be stressed that (in general) there is still a plethora of invariant Fock representations {that are not unitarily equivalent}, and additional criteria must then be imposed in order to {specify} the quantization. A natural extra requirement is to demand that the {symplectomorphisms that dictate time evolution be mapped into unitary operators} in the quantum theory. Indeed, although time translation symmetry is broken, it seems quite reasonable to retain {unitary time evolution, by minimally relaxing the requirement of invariance under the evolution and replace it with the weaker condition that the dynamics be unitarily} implementable. The aim of this requirement is to {grant a privileged role to} those representations (if there exist) preserving the standard probabilistic interpretation of the quantum theory, including the evolution of the observables, and thus ensuring the availability of the Schr\"{o}dinger picture. Notably, as mentioned earlier, the unitarity requirement suffices to single out a unique (up to unitary transformations) preferred invariant Fock representation for the scalar field (\ref{sf-with-tdp}), namely the $j_0$-Fock representation. Moreover, the criteria of symmetry invariance and of unitary dynamics pick up a unique (modulo irrelevant constant scalings) preferred field description {for the system}. In total, the uniqueness result reads as follows. Up to irrelevant constant scalings and unitary transformations, there is a unique field description, the $\psi$-description, {that admits} a family of invariant Fock representations with unitary dynamics. This family {is formed by representations that are all equivalent among them}, and the $j_0$-Fock quantization is a {member of this family}. 

The {combined criteria} of invariance under spatial symmetries and of unitary dynamics were introduced for the first time in the quantization of Gowdy cosmological models \cite{CCM-PRD73-i,CCM-PRD73-ii,CMV-PRD75,CCMV-GT3-SR,CMV-GowS1S2S3}, just after the {reasons for} the failure of unitarity in the Gowdy $T^{3}$ model reported in Refs. \cite{CCH-GT3,Torre-GT3} were {understood} \cite{CM}. The {criteria were then successively applied to KG} fields with time dependent mass defined on the circle \cite{CMSV}, on the three-sphere \cite{CMV-PRD81,CMOV-JCAP10,CMOV-PRD83}, and, {after a notable generalization}, on spatial manifolds of arbitrary compact topology in three or less dimensions \cite{CMOV-cqg,CMOV-PRD86}. This comprises the {relevant} cosmological case (inasmuch as spatial flatness is favored by current {observations} \cite{Univ-obs,Univ-obs1,Univ-obs2}) of compact sections {with} three-torus topology \cite{CCMMV-T3,MenaMarugan:2013tba,CCMMV-T3-ii}. Apart from addressing the uniqueness {of} the quantization of free real scalar fields in {FLRW} spacetimes \cite{CMV-PRD81,CMOV-JCAP10,CMOV-PRD83,CCMMV-T3,MenaMarugan:2013tba,CCMMV-T3-ii}, the {criteria} were successfully employed to remove the ambiguities in the quantization of free (test) KG fields minimally coupled to a de Sitter background \cite{CdBMV-deSitter}, as well as in anisotropic Bianchi I spacetimes \cite{Unique-Bianchi-I}. It is worth remarking that the {criteria} of invariance and of unitarity have also been fruitfully exploited to single out a unique preferred quantum description for fermion fields in cosmological scenarios \cite{Uniq-for-DF,Uniq-for-DF1,Uniq-for-DF3,Uniq-for-DF4,Uniq-for-DF2,EMP,EMP2,EMP3}.

In this work, we overview the uniqueness {results obtained by us and our collaborators} for the quantization of Gowdy cosmological models and free scalar fields minimally coupled to cosmological backgrounds, {more concretely for KG} fields propagating in FLRW, de Sitter, and Bianchi I spacetimes. The presentation follows {the historical timeline in which these results were deduced,} {so that \em{a posteriori} generalizations} will not be {included nor discussed}. For the sake of clarity and completeness, {we first provide a brief introduction to} the classical and quantum theories of a scalar field in a globally hyperbolic spacetime. More specifically, the paper is divided in two parts. The first one, which comprises Sec. \ref{sec:preliminaries} and Sec. \ref{sec:q-theory},  {contains} a general discussion of the classical theory of a real scalar field (Sec. \ref{sec:preliminaries}) and its quantization (Sec. \ref{sec:q-theory}). These sections pay a special attention {to the definition of a complex structure on} phase space, the specification of basic observables, the analysis of classical and quantum time evolution, and the role that complex structures have in the construction of a quantum theory {on a} Hilbert space ({both from the covariant and the canonical perspectives}). The $j_0$-Fock representation for a scalar field with time dependent mass is summarized at the end of the first part. The second part of the work is entirely dedicated to an overview of the uniqueness results obtained by imposing the {criteria} of invariance under spatial symmetries and {of unitary implementability} of the dynamics, {applied to} the quantization of Gowdy models and (test) KG fields in cosmological spacetimes. The quantization of Gowdy models is discussed in 
Sec. \ref{sec:Gowdy}. The quantization of (test) scalar fields propagating in FLRW, de Sitter, and Bianchi I spacetimes is presented, respectively, in Sec. \ref{sec:FLRW}, Sec. \ref{sec:deSitter}, and Sec. \ref{sec:BI-univ}. We will end with a summary of our results in 
Sec. \ref{sec:diss}. In the following, we set $c=\hbar=1$.

\section{The classical setting}
\label{sec:preliminaries}

This section {contains some} background material. The section has three parts. In the first one, we review some key aspects about complex structures. {Complex} structures play a central role not only in the quantization of scalar field theories, but {also in the quantization of other linear field theories (like e.g. the Maxwell and Dirac fields \cite{J-MaxF,Uniq-for-DF,Uniq-for-DF1,Uniq-for-DF3,Uniq-for-DF4,Uniq-for-DF2}), {as well as} finite dimensional mechanical systems. In fact, complex structures have been employed in mechanical systems to specify (in an appropriate limit) a polymer quantum mechanics \cite{CVZ-PRD}, which is a non-standard representation of the finite dimensional CCRs} where the Stone-von Neumman uniqueness theorem is simply not valid because of the lack of strong continuity. In the second part of this section, we outline the classical theory of a KG field propagating in a globally hyperbolic spacetime. In particular, we introduce the canonical and covariant phase space {descriptions}, the basic observables, and the symplectomorphisms {that dictate time evolution. We discuss} the role played by compatible complex structures on phase space, their evolution in time, and how complex structures on the covariant and canonical phase spaces are related. In the third {and} final part of this section, {we consider} the case of a (test) massive scalar field in FLRW spacetimes and the scaling of the field with {respect to the scale} factor {of the metric}. Complex structures for the original and the scaled field are {provided explicitly}.

\subsection{Complex structure: Definition and some key results}
\label{subsec:cs-definitions}

Let us start by defining {the notion of} complex structure \cite{McDuff-Salamon,dGosson}. Let $V$ be a real vector space. A complex structure on $V$ is a linear isomorphism $J:V\to V$ such that $J^{2}=-I$, with $I$ being the identity map on $V$. 

Consider first a real, finite dimensional vector space $V$. Since ${\rm{det}}\left(J^{2}\right)=(-1)^{{\rm{dim}}V}>0$, we conclude that a complex structure can be specified on $V$ only if it is even dimensional. In addition, we notice that a complex structure on a vector space $V$ (not necessarily finite dimensional) splits it into two complementary vector subspaces. For {definiteness, let us take} ${\rm{dim}} V=2m$. Then, it can be shown that there exists $m$ linearly independent vectors $e_{i}$ ($i=1,\dots,m$) such that $e_{i}$ and $e_{m+i}=Je_{i}$ form a basis set for $V$; that is, there exist $m$-dimensional vector subspaces $V_{1}$ and $V_{2}$ of $V$ spanned by $e_i$ and by $e_{m+i}=Je_{i}$, respectively, and such that $V=V_{1}\oplus V_{2}$ with $v_{2}=Jv_{1}$ for all $v_{1}\in V_{1}$ and $v_{2}\in V_{2}$. The set of the $m$ linearly independent vectors $e_{i}$ used to {construct a basis of} $V$ with $J$ is by no means unique, so the splitting is not canonical. For instance, the $m$-dimensional vector subspaces $V'_{1}$ spanned by $e'_{i}=(e_{1},e_{2},...,Je_{m})$ and $V'_{2}$ spanned by $e'_{m+i}=Je'_{i}$ provide a distinct splitting $V=V'_{1}\oplus V'_{2}$.

A key feature of complex structures is that they allow to specify ``multiplication by $i$'' on $V$, {and hence endow} the space with a structure of complex vector space. Explicitly, the real linear operation $(a+ib)v=av+bJv$ defines multiplication by complex numbers $(a+ib)$ on $V$. It is a simple matter to check that this multiplication rule transforms $V$ into a complex vector space. In addition, it is worth {pointing out} the close relationship between the introduction of the mapping $J$ and the complexification of $V$, $V_{\mathbb{C}}=V\oplus iV$. From $V$ and $J$, we construct the complex linear spaces
\begin{equation}
\label{pm-vect-sp}
V^{\pm}_{J}=\big\{v^{\pm}\, \big\vert \, v^{\pm}=\tfrac{1}{2}(v\mp iJv)\, , \: v\in V \big\}.
\end{equation}
By taking the direct sum of these two spaces, we get the complex vector space $V^{+}_{J}\oplus V^{-}_{J}$, {the elements of which} can be written as $(x^{+}+y^{-})$, with $x^{+}$ in $V^{+}_{J}$ and $y^{-}$ in $V^{-}_{J}$. A direct inspection shows that $V^{+}_{J}\oplus V^{-}_{J}$ turns out to be {the same as} 
$V_{\mathbb{C}}$, so that the complex vector spaces {defined in Eq.} (\ref{pm-vect-sp}) provide a splitting for the complexification of $V$. Besides, notice that every $v$ in $V$ can be decomposed as $v=v^{+}+v^{-}$, with {$v^{\pm}\in V^{\pm}_{J}$}. Clearly, different complex structures will {lead to} distinct splittings for $V_{\mathbb{C}}$ and, consequently, to different decompositions for $v\in V$. By extending the action of $J$ from $V$ to $V_{\mathbb{C}}$ by complex linearity, we obtain that $v^{+}$ and $v^{-}$ are eigenvectors of $J$ with eigenvalues $i$ and $-i$,
\begin{equation}
\label{pm-eigenvect}
Jv^{+}=iv^{+},\qquad Jv^{-}=-iv^{-} .
\end{equation}
Given another complex structure, say $\tilde{J}$, its eigenvectors will satisfy relationships (\ref{pm-eigenvect}) with $J$ replaced with $\tilde{J}$. The eigenvectors of $\tilde{J}$ and $J$ are related by $\tilde{v}^{\pm}=v^{\pm}\pm \tfrac{i}{2}(J-\tilde{J})v$.

Let us equip now the vector space $V$ with a symplectic form. That is, a two-form $\mathbf{\Omega}$ which is (i) closed (i.e., $d\mathbf{\Omega}=0$) and (ii) non-degenerate [i.e., if $\mathbf{\Omega}(v,w)=0$ for all vectors $v\in V$, then $w=0$]. The space $V$ equipped with a symplectic form $\mathbf{\Omega}$ is called a symplectic vector space $(V,\mathbf{\Omega})$. {Suppose that $V$ can be identified with $
{\mathbb{R}}^{2m}$, with coordinates $\{(x_{1},\dots,x_{m},y_{1},\dots,y_{m})\}$. The standard (also called canonical) symplectic form is then given by $\mathbf{\Omega}=\sum_{i=1}^{m}dx_{i}\wedge dy_{i}$ \cite{McDuff-Salamon}. Equivalently, the standard symplectic form defines a skew-symmetric bilinear function on $V$  \cite{Wald-book}, 
\begin{equation}
\Omega:V\times V\to \mathbb{R},\quad  (v_{1},v_{2})\mapsto \Omega(v_{1},v_{2})=
\sum_{i=1}^{m}(y_{1i}x_{2i}-y_{2i}x_{1i}),
\end{equation}
where $(x_{K1},\dots,x_{Km},y_{K1},\dots,y_{Km})$} {are the coordinates of $v_{K}\in V$, and $K=1,2$.} 

{For a typical mechanical system, the space of classical states corresponds to the cotangent bundle $\Gamma_{c}=T^{*}C$, with the configuration space $C$ being described, say, by variables $\{(q^{1},\dots,q^{m})\}$, whereas the fibers $T^{*}_{q}$ { at $q\in C$} are coordinatized by the momentum variables  $\{(p_{1},\dots,p_{m})\}$. The phase space of the theory corresponds to the symplectic  space $(\Gamma_{c},\mathbf{\Omega})$, where $\mathbf{\Omega}$ is the canonical symplectic form on $\Gamma_{c}$, $\mathbf{\Omega}=\sum_{i=1}^{m}dq^{i}\wedge dp_{i}$. In the special case that $C$ be itself a vector space, it follows that the phase space  $\Gamma_{c}$ is a linear space that can be isomorphically identified with  $\Gamma=\mathbb{R}^{2m}$, as {we did} above.} 

Let us now abandon {the restriction of finite dimensionality, and consider infinite dimensional symplectic linear spaces $(V,\Omega)$} as well. For instance, the phase space of a free KG field in a globally hyperbolic spacetime can be described (in the canonical approach) by the symplectic vector space $(\Gamma,\Omega)$, where $\Gamma$ is the (infinite-dimensional) linear space coordinatized by the configurations and momenta of the field, $\{(\varphi(y),\pi(y))\}$ {where $y\in \Sigma$,} and $\Omega$ is the canonical symplectic structure thereon,\footnote{{See Sec. \ref{subsec:classical-sft} for a comprehensive disscussion.}}
\begin{equation}
\label{sympl-struc-KG-field}
\Omega(\, (\varphi,\pi)_{1},(\varphi,\pi)_{2}\,)=\int_{\Sigma}(\pi_{1}\varphi_{2}-\pi_{2}\varphi_{1})\, d^{3}y.
\end{equation}

A particularly important class of complex structures, {establishing a relation between} complex manifolds and symplectic geometry, is the class of the so-called $\Omega$-compatible complex structures. Let $(V,\Omega)$ be a real symplectic vector space. A complex structure $J$ on $(V,\Omega)$ is said to be compatible with $\Omega$ if it is a symplectic map (i.e., $J^{*}\Omega=\Omega$) and $\Omega(Jv,v)>0$ for all {non-zero} $v\in V$ (see, for instance, Ref. \cite{McDuff-Salamon}). We denote the set of complex structures on $V$ {compatible with $\Omega$} by ${\cal{J}}(\Omega,V)$.

Let $J$ be a complex structure on the real symplectic vector space $(V,\Omega)$, and suppose that $J$ is a symplectic map thereon. Let us consider the complexification of $V$. By extending the actions of $J$ and $\Omega$ from $V$ to $V_{\mathbb{C}}$ by complex linearity, it is not difficult to verify that
\begin{equation}
\label{pm--comp-sympl}
\Omega(v^{+},w^{+})=0,\quad \Omega(v^{-},w^{-})=0,\quad \Omega(v^{-},w^{+})=\tfrac{i}{2}\left[\Omega(Jv,w)-i\Omega(v,w)\right]
\end{equation}
for all $v^{\pm},w^{\pm}\in V^{\pm}_J$. Obviously, relationships (\ref{pm--comp-sympl}) hold for every $J\in {\cal{J}}(\Omega,V)$.

{For each $J\in{\cal{J}}(\Omega,V)$, it} can be easily seen that the real-valued, symmetric bilinear mapping
\begin{equation}
\label{inner-prod-svp}
\mu_{J}:V\times V\to \mathbb{R},\quad (v,w)\mapsto \mu_{J}(v,w)=\Omega(Jv,w),
\end{equation}
defines an inner product on the symplectic space. Since $J$ is, in particular, a symplectic map, we have that $\mu_{J}(v,w)=\mu_{J}(Jv,Jw)$ (i.e. $\mu_{J}$ is a $J$-invariant mapping) and that $\mu_{J}(Jv,w)=\mu_{J}(v,-Jw)$ (i.e. $J$ is skew-adjoint with respect to the inner product $\mu_{J}$). We define now the complex-valued mapping
\begin{equation}
\label{herm-inner-prod-svp}
\langle \,\cdot\, ,\, \cdot\, \rangle_{J}:V\times V\to \mathbb{C},\quad (v,w)\mapsto 
\langle v , w \rangle_{J}=\tfrac{1}{2}\mu_{J}(v,w)-\tfrac{i}{2}\Omega(v,w).
\end{equation}
By using the antisymmetry of $\Omega$, the properties defining $\Omega$-compatible complex structures, as well as the multiplication by complex numbers defined by $J$ on $V$ ($V$ is understood here as a complex vector space, {with the} structure provided precisely by $J$), it is not difficult to check that the mapping (\ref{herm-inner-prod-svp}) is a Hermitian inner product on $(V,\Omega)$. From Eqs. (\ref{pm--comp-sympl}) and (\ref{herm-inner-prod-svp}), it follows that $\langle v , w \rangle_{J}=\Omega(\overline{Jv^{+}},w^{+})$, where the bar denotes complex conjugation on $\mathbb{C}$, and where we have used that $Jv^{-}=\overline{Jv^{+}}$. A straightforward inspection shows that $\langle v , w \rangle_{J}$ defines a Hermitian inner product on $V^{+}_J$; that is,
\begin{equation}
\label{herm-inner-prod-pfs}
\langle \,\cdot\, ,\, \cdot\, \rangle_{V^{+}_{J}}:{V^{+}_J\times V^{+}_J}\to \mathbb{C},\quad (v^{+},w^{+})\mapsto 
\langle v^{+} , w^{+} \rangle_{V^{+}_{J}}=\Omega(\overline{Jv^{+}},w^{+})
\end{equation}
is a Hermitian inner product. This, together with the fact that any element of $V^{+}_J$ is uniquely represented by an element of $V$ (and vice versa), implies that the complex vector space $V$, with Hermitian inner product (\ref{herm-inner-prod-svp}), and the complex vector space ${V^{+}_J}$, with Hermitian inner product (\ref{herm-inner-prod-pfs}), are (essentially) the same inner product spaces.

\subsection{The scalar field: Classical theory}
\label{subsec:classical-sft}

Let us consider a free, massive real scalar field $\phi$ propagating in a four-dimensional globally hyperbolic spacetime $(M,g_{\alpha\beta})$ ($\alpha,\beta$=0,1,2,3). Here, $M$ has topology $\mathbb{I}\times \Sigma$ for some $\mathbb{I}\subset \mathbb{R}$, and can be foliated by a one-parameter family of Cauchy surfaces that are diffeomorphic to $\Sigma$. The phase space of the system is the symplectic linear space $(\Gamma,\Omega_{ij})$, where $\Gamma$ is the real vector space $\Gamma=\{(\varphi,\pi)\, \vert \, \varphi,\pi\in C_{0}^{\infty}(\Sigma)\}$ [$C_{0}^{\infty}(\Sigma)$ denotes the space of smooth real functions with compact support on $\Sigma$] and $\Omega_{ij}=\int_{\Sigma}(d\pi)_{i}\wedge (d\varphi)_{j}$ is the canonical symplectic form. This form defines a unique symplectic structure \cite{Wald-book}, given by Eq. 
(\ref{sympl-struc-KG-field}), which is known as the canonical (or standard) symplectic structure. The phase space of the theory can be alternatively described as the symplectic vector space $(\Gamma,\Omega)$. 

The symplectic structure (\ref{sympl-struc-KG-field}) defines natural coordinate functionals of configuration and momentum type, namely $\varphi[f]=\Omega((0,f),\,\cdot\, )$ and  $\pi[g]=\Omega((-g,0),\,\cdot\, )$  with $(-g,f)\in \Gamma$. On the other hand, the symplectic form defines the Poisson brackets (PB) on the real vector space of observables $\cal{O}$ (i.e. the linear space of smooth, real-valued functionals on phase space) $\{F,G\}=\Omega^{ij}(dF)_{i}(dG)_{j}$, where $\Omega^{ij}$ is the inverse of the symplectic form $\Omega_{ij}$. Thus, a direct calculation shows that the PB between the configuration and momentum observables are given by $\{\varphi[f],\pi[g]\}=-\Omega((0,f),(-g,0))$. Explicitly, $\{\int f\varphi,\int g\pi\}=\int fg$, that is the smeared version of the well-known but mathematically ill-defined expression $\{\varphi(x),\pi(x')\}=\delta(x-x')$, where $\delta(x)$ is the Dirac delta on {$\Sigma$}. By linearity, we get that generic linear observables are given by  $\Omega((g,f),\,\cdot\,)=\varphi[f] - \pi[g]$, for all $(g,f)\in \Gamma$. From the PB between the basic configuration and momentum observables, the linearity of $\Omega$, as well as the bilinear and skew-symmetric properties of $\{\,\cdot\, , \, \cdot\,\}$, it immediately follows that 
\begin{equation}
\label{PB-cano}
\{\,\Omega((g,f)_{1},\,\cdot\,)\, ,\, \Omega((g,f)_{2},\,\cdot\,)\, \}=-\Omega((g,f)_{1},(g,f)_{2}).
\end{equation}

A foliation of spacetime $(M,g_{{\alpha\beta}})$ by Cauchy surfaces $\Sigma_{t}$ parametrized by a global time function $t$ defines a one-parameter family of embeddings $E_{t}$ of $\Sigma$ as Cauchy surfaces in $M$, $\Sigma \mapsto E_{t}(\Sigma)=\Sigma_{t}$. Let $t=t_0$ be a fixed (but arbitrary) initial reference time. Let $S$ be the linear space of smooth solutions to the KG equation ${(g^{\alpha\beta}\nabla_{\alpha}\nabla_{\beta}-m^{2})\phi} =0$ which arises from initial data $(\varphi,\pi)_{t_0}$ in $\Gamma$, $\varphi=E^{*}_{t_0}\phi$ and $\pi=E^{*}_{t_0}(\sqrt{h}L_{n}\phi)$. Here, $L_{n}$ stands for the Lie derivative along the normal to the initial Cauchy surface $\Sigma_{t_0}$, whereas {we recall that} $h$ is the determinant of the induced metric $h_{ab}$ on such a surface. Every set of Cauchy data gives rise to a solution, and different initial Cauchy data indeed give rise to distinct solutions. Thus, by construction, solutions in $S$ are in a one-to-one and onto correspondence with initial data in $\Gamma$; i.e. $I_{t_0}:S\to\Gamma$, $I_{t_0}(\phi)=(E^{*}_{t_0}\phi,E^{*}_{t_0}[\sqrt{h}L_{n}\phi])$ is a bijection. In fact, associated with every embedding $E_t$, there is a bijection $I_{t}:S\to \Gamma$ relating solutions with their corresponding Cauchy data at time $t$, 
$\varphi=E^{*}_{t}\phi$ and $\pi=E^{*}_{t}(\sqrt{h}L_{n}\phi)$. Given a solution $\phi\in S$, we {see that the associated} dynamical trajectory in $\Gamma$ {is formed by the family of data} $(\varphi,\pi)_{t}=I_{t}(\phi)$ with $t\in {\mathbb{I}\subset \mathbb{R}}$. Since $t=t_0$ is the initial reference time, we have that $(\varphi,\pi)_{t}=\tau_{(t,t_{0})}(\varphi,\pi)_{t_0}$, where $\tau_{(t,t_{0})}=I_{t}\circ I_{t_{0}}^{-1}$ is a two-parameter family of linear symplectomorphisms,{\footnote{{The family of mappings $\tau_{(t,t_{0})}$ will form a one-parameter group of symplectomorphisms whenever the Hamiltonian does not depend explicitly on time, so that the system is invariant under time reparametrizations, and hence $\tau_{(t,t_{0})}=\tau_{t-t_{0}}$. Otherwise, the family $\tau_{(t,t_{0})}$ is a two-parameter family of symplectomorphisms \cite{dGosson}.}}} with $\tau_{(t_{0},t_{0})}$ the identity map on $\Gamma$. In view of the isomorphic relation between $S$ and $\Gamma$, the canonical symplectic structure (\ref{sympl-struc-KG-field}) induces a symplectic structure $\Omega'$ on $S$, namely $\Omega'=I^{*}_{t_0}\Omega$. Since $\tau_{(t,t_{0})}$ is a symplectomorphism, it follows that $I^{*}_{t_0}\Omega=I^{*}_{t}\Omega$ (i.e. $\Omega'$ is {time} independent). Alternatively to $(\Gamma,\Omega)$, we can consider the symplectic vector space $(S,\Omega')$ as the phase space of the theory. The time evolution in the phase space $(S,\Omega')$ is given by the two-parameter family of linear symplectomorphisms $T_{(t,t_{0})}=I^{-1}_{t_{0}}\circ\tau_{(t,t_{0})}\circ I_{t_0}$, with $t\in{\mathbb{I}\subset \mathbb{R}}$, which can be rewritten simply as $T_{(t,t_{0})}=I^{-1}_{t_{0}}\circ I_{t}$. In order to simplify the notation, we will denote $\Omega'$ also by $\Omega$ from now on. The symplectic vector spaces $(S,\Omega)$ and $(\Gamma,\Omega)$ will be referred to as the covariant and the canonical phase space, respectively. The symplectic structure on $S$ is explicitly given by
\begin{equation}
\label{sympl-struc-cov}
\Omega(\phi_{1},\phi_{2}\,)=\int_{\Sigma_{t_0}}(\phi_{2}L_{n}\phi_{1}-\phi_{1}L_{n}\phi_{2})\, \sqrt{h}\,d^{3}x.
\end{equation}
The time independence of $\Omega$ guarantees that the integration in Eq. (\ref{sympl-struc-cov}) is independent of the choice of Cauchy surface to perform it.

The covariant counterpart of the natural observables in $\Gamma$ are the real-valued linear functionals $\Omega(\phi,\,\cdot\,):S\to\mathbb{R}$, $\forall\phi\in S$. Given a bijection, say $I_{t_0}$, there is a one-to-one, onto correspondence between linear observables in the canonical and the covariant approaches: the observable  $\Omega(\phi,\,\cdot\,)$ on $S$ corresponds to (and it is the corresponding observable of) the observable $\Omega((g,f),\,\cdot\,)$ on $\Gamma$ for $(g,f)=I_{t_0}\phi$. 

The PB between {any pair of} observables $F$ and $G$ on {the} phase space $(S,\Omega)$ are given by $\{F,G\}=\Omega^{ij}(dF)_{i}(dG)_{j}$, where $\Omega^{ij}$ is the inverse of the symplectic form on $S$ induced by the canonical symplectic form on $\Gamma$. The PB between natural observables $\Omega(\phi,\,\cdot\,)$ --i.e. the analogue of Eq. (\ref{PB-cano})-- are given by 
\begin{equation}
\label{PB-cov}
\{\,\Omega(\phi,\,\cdot\,)\, ,\, \Omega(\tilde{\phi},\,\cdot\,)\, \}=-\Omega(\phi,\tilde{\phi}).
\end{equation}

{Let us introduce a compatible complex structure $j$ on $(\Gamma,\Omega)$ [i.e. $j\in {\cal{J}}(\Omega,\Gamma)$]. According to our previous discussion (Sec. \ref{subsec:cs-definitions}), we know} that $j$ will equip the linear symplectic space  $(\Gamma,\Omega)$ with a real inner product $\mu_{j}(\, \cdot\, , \, \cdot\, )=\Omega(j\, \cdot\, , \, \cdot\, )$ [see Eq. (\ref{inner-prod-svp})]. Let $R$ be a linear symplectomorphism on $\Gamma$. From $j$ and $R$ we then construct the compatible complex structure $j_{R}=R\,j\,R^{-1}$, and hence the real inner product $\mu_{j_{R}}(\, \cdot\, , \, \cdot\, )=\Omega(j_{R}\, \cdot\, , \, \cdot\, )$. The inner products $\mu_{j_{R}}$ and $\mu_{j}$ (which are in general distinct) are related by $\mu_{j_R}(Ru,R\tilde{u})=\mu_{j}(u,\tilde{u})$, for all $u,\tilde{u}\in\Gamma$. {In particular, given a reference time $t_0$}, the evolution map $\tau_{(t,t_{0})}$ (which is a linear symplectomorphism on $\Gamma$) provides a family of compatible complex structures $j_{t}=\tau_{(t,t_{0})}\, j\,\tau^{-1}_{(t,t_{0})}$ and inner products $\mu_{j_t}$, with $\mu_{j_t}(\tau_{(t,t_{0})} u,\tau_{(t,t_{0})}\tilde{u})={\mu_{j}}(u,\tilde{u})$. The complex structure $j_{{t}}$ is the complex structure generated by 
the time evolution, from time $t_{0}$ to time $t$, of the initial complex structure $j$, that we will rename $j_{t_0}$ from now on in order to emphasize the choice of initial time in the canonical formulation.

Alternatively, we can consider {the covariant perspective. Just as in the canonical phase space description, a complex structure} $J$ will equip the covariant phase space $(S,\Omega)$ with a real inner product space $\mu_{J}(\, \cdot\, , \, \cdot\, )=\Omega(J\, \cdot\, , \, \cdot\, )$. Linear symplectomorphisms on $S$ will define other compatible complex structures and, consequently, other real inner products as well. In particular, we have that time evolution in $S$ will induce a family of compatible complex structures, $J_{t}=T_{(t,t_{0})}\, J\, T^{-1}_{(t,t_{0})}$, and a family of real inner products $\mu_{J_t}$. Now, according to the discussion in Sec. \ref{subsec:cs-definitions}, a complex structure $J$ on $S$ (not necessarily {compatible}) endows the symplectic linear space $(S,\Omega)$ with a structure of complex vector space, and defines a field decomposition $\phi=\phi^{+}+\phi^{-}$, where $\phi^{+}$ is in the space of ``positive frequency'' solutions $S^{+}_{J}$, whereas $\phi^{-}$ is in the (complex conjugate) space of ``negative frequency'' solutions $S^{-}_{J}$ [see Eq. (\ref{pm-vect-sp})]. By requiring that $J$ be {compatible with $\Omega$}, we will get, apart from the real inner product $\mu_J$, the Hermitian inner products (\ref{herm-inner-prod-svp}) on $(S,\Omega)$ and (\ref{herm-inner-prod-pfs}) on $S^{+}_{J}$. The Cauchy completion of $S^{+}_{J}$ with respect to the norm associated with the Hermitian inner product  (\ref{herm-inner-prod-pfs}) yields the so-called ``one-particle Hilbert space'' ${\cal{H}}_{J}$. By repeating this construction for each  compatible {complex structure} $J_t$, we will obtain a family of (in general) distinct Hilbert spaces ${\cal{H}}_{J_t}$.

Let us briefly discuss how complex structures in $\Gamma$ and $S$ {are related} \cite{Cortez:2015mja}. Let $J$ be a complex structure on $S$, and consider the isomorphisms $I_t$ defined by the spacetime foliation. The complex structure on $\Gamma$ induced by $J$ at time $t$, via $I_t$, is $j_{t}=I_{t}JI^{-1}_{t}$. From this relation it immediately follows that $j_{t_2}=\tau_{(t_{2},t_{1})}j_{t_{1}}\tau^{-1}_{(t_{2},t_{1})}$. Once we have obtained the set of complex structures $j_{t}$ on $\Gamma$, we fix a bijection to identify $S$ with $\Gamma$; i.e. we chose a particular but arbitrary time and declare it as the initial reference time $t_0$. Thus, the complex structure 
\begin{equation}
\label{evolved-can-cs}
j_{t}=\tau_{(t,t_{0})}\, j_{t_{0}}\, \tau^{-1}_{(t,t_{0})} 
\end{equation}
is the complex structure generated by dynamical evolution of $j_{t_0}=I_{t_0}JI^{-1}_{t_0}$ from the initial time $t_0$ to time $t$. Now, since $I_{t_0}$ establishes a bijection between complex structures on $\Gamma$ and $S$, the family $j_t$  will provide a one-parameter family of complex structures on $S$, namely $J_{t}=I^{-1}_{t_0}j_{t}I_{t_0}$. Using that $j_t$ is the evolved complex structure of {$j_{t_0}$}, we then get
\begin{equation}
\label{evolved-cov-cs}
J_{t}=T_{(t,t_{0})}\,J\,T^{-1}_{(t,t_{0})}.
\end{equation}
{That is}, $J_t$ is the complex structure {obtained by} evolving $J$ in time.

Note that, although the introduction of a complex structure is a simple matter, the choice of a complex structure with physical content  is not. For instance, Poincar\'{e} invariance and time translation symmetry are exploited to select favored complex structures in Minkowski and stationary backgrounds, respectively. However, in the absence of stationarity {the issue becomes} more involved and extra requirements are needed in order to select a preferred family of complex structures. In the next subsection we will consider precisely a non-stationary system, concretely a scalar field propagating in an FLRW spacetime. Actually, the arguments that we will present apply equally well to more general, conformally ultrastatic spacetimes.

\subsection{Complex structures in FLRW spacetimes}
\label{subsec:CS-FLRW}

Let $\phi$ be a real scalar field with mass $m$ propagating in an FLRW spacetime. As it is well known, the 
FLRW cosmological models of homogeneous and isotropic universes {can be} described by the line element
\begin{equation}
\label{FLRW-line-element}
ds^{2}=a^{2}(t)\left[-dt^{2}+{\tilde{h}_{ab}dx^{a}dx^{b}}\right],
\end{equation}
where {$\tilde{h}_{ab}$ ($a,b=1,2,3$)} is the standard Riemannian metric of either a three-sphere, a three-dimensional Euclidean space, or a three-dimensional hyperboloid. The KG equation in this FLRW spacetime reads
\begin{equation}
\label{FLRW-eq-mot}
\ddot{\phi}+2\frac{\dot{a}}{a}\dot{\phi}-\Delta \phi+m^{2}a^{2}\phi=0.
\end{equation}
Here,  the dot stands for the derivative with respect to the conformal time $t$, and $\Delta$ denotes the LB operator associated with the spatial metric $\tilde{h}_{ab}$. 

The dynamics on phase space $\Gamma$ is dictated by the Hamiltonian
\begin{equation}
\label{Hamil-FLRW}
H=\frac{1}{2}\int_{\Sigma}d^{3}x\: a {\sqrt{h}\left(h^{-1}\pi^{2}+h^{ab}D_{a}\varphi D_{a}\varphi+m^{2}\varphi^{2}\right),}
\end{equation}
where $h$ stands for the determinant of ${h_{ab}=a^{2}\tilde{h}_{ab}}$ and $D_{a}$ is the derivative operator on $\Sigma$  associated with $h_{ab}$. A straightforward calculation shows that the equations of motion are given by
\begin{equation}
\label{FLRW-cano-eq-mot}
\begin{pmatrix}
 \dot{\varphi} \\
\dot{\pi} \end{pmatrix}
={\cal{T}} \begin{pmatrix}
 \varphi \\
\pi \end{pmatrix}, \qquad
{\cal{T}}=\begin{pmatrix}
  0 & ah^{-1/2}\\
 ah^{1/2}(D^{a}D_{a}-m^{2}) & 0
 \end{pmatrix}.
\end{equation}

By performing the polar decomposition of $\cal{T}$, one gets that the partial isometry $\vert {\cal{T}}\vert^{-1}{\cal{T}}$ provides a family of compatible complex structures on the space of Cauchy data \cite{ccq-annphys}. Specifically, the complex structure $j_{t}=\vert {\cal{T}}\vert^{-1}{\cal{T}}$ associated with the Cauchy surface $\Sigma_{t}$ is
\begin{equation}
\label{CS-FLRW}
j_{t}=  \begin{pmatrix}
  0 & -h^{-1/2}(-D^{a}D_{a} + m^{2})^{-1/2}\\
h^{1/2}(-D^{a}D_{a} + m^{2})^{1/2} & 0
 \end{pmatrix}.
\end{equation}
It is not difficult to check that $\mu_{j_{t}}((\varphi,\pi),(\dot{\varphi},\dot{\pi}))=0$, with $\mu_{j_{t}}(\,\cdot\, , \, \cdot\,)=\Omega(j_{t}\,\cdot\, , \, \cdot\,)$, so that{, for given $t$,} $j_{t}$ {can be considered} unique \cite{Ashtekar:1975zn}. In spite of this, it is worth remarking that for any two distinct times $t_1$ and $t_2$,  $j_{t_1}$ and $j_{t_2}$ give 
rise {in general to inequivalent} representations of the CCRs, implying that the time evolution cannot be represented by a unitary operator and, therefore, that we do not have at our disposal a Schr\"{o}dinger picture with an evolution that preserves the standard notion of probability. In an attempt to fix this drawback, we can use the freedom available in the choice of basic variables and, by applying a time dependent canonical transformation, redistribute the time dependence in an implicit part ({with evolution} generated by the corresponding Hamiltonian) and an explicit part (the factor of the transformation) {which varies in a way that is not necessarily unitary}. Let us hence introduce the time dependent scaling $\psi=a\phi$. By substituting $\phi=\psi/a$ in Eq. (\ref{FLRW-eq-mot}), we get that the dynamics in the new field description is dictated by
\begin{equation}
\label{eq-mot-scaled-field}
\ddot{\psi}-\Delta \psi+s(t)\psi=0,
\end{equation}
where $s(t)=m^{2}a^{2}-(\ddot{a}/{a})$. Thus, the system can be treated as a scalar field propagating in a fictitious static spacetime $ds^{2}=-dt^{2}+{\tilde{h}_{ab}}dx^{a}dx^{b}$, 
though now subject to a time varying potential $V(\psi)=s(t)\psi^{2}/2$ [or, equivalently, as a free scalar field with time dependent mass $\sqrt{s(t)}$ in a static background, provided
that $s(t)$ is a non-negative function]. The canonical equations of motion now are given by
\begin{equation}
\label{can-eq-mot-scaled-f}
\dot{\tilde{\varphi}}=\frac{1}{\sqrt{\tilde{h}}}\tilde{\pi},\quad \dot{\tilde{\pi}}=\sqrt{{\tilde{h}}}\left[\Delta \tilde{\varphi} -s(t)\tilde{\varphi}\right].
\end{equation}

Once an initial reference time $t_0$ is chosen, we introduce an initial complex structure on $\tilde{\Gamma}=\{(\tilde{\varphi},\tilde{\pi})\}$. The simplest complex structure guaranteeing an invariant Fock representation under the spatial symmetries is 
\begin{equation}
\label{CS-scaled-field}
j_{0}=  \begin{pmatrix}
  0 & -(-{\tilde{h}}\Delta)^{-1/2}\\
(-{\tilde{h}}\Delta)^{1/2} & 0
 \end{pmatrix}.
\end{equation}
Note that $j_0$ ignores the existence of the time varying potential (so, in particular, the existence of the mass) in the system. Thus, the Fock representation defined by the complex structure (\ref{CS-scaled-field}) can be referred to as the free massless field representation. As we will see in Sec. \ref{sec:FLRW}, {the $j_0$-Fock quantization is, up to unitary equivalence,} the unique invariant Fock representation {under spatial isometries that admits} a unitary implementation of the dynamics, both for closed \cite{CMV-PRD81,CMOV-JCAP10,CMOV-PRD83} and (compact) flat \cite{CCMMV-T3,MenaMarugan:2013tba,CCMMV-T3-ii} FLRW spacetimes.  

Before {we proceed to present our} uniqueness results about the quantization of {the KG field} with time varying mass in cosmological scenarios, {it may be helpful to analyze} in some detail the quantum theory of scalar fields in globally hyperbolic spacetimes. {This is the purpose of the next section.}

\section{Quantization}
\label{sec:q-theory}

{We now focus our discussion} on the quantization of real scalar fields in spacetimes {that admit a foliation by Cauchy surfaces}. We will first overview the program of canonical quantization
 on a Hilbert space for linear systems, {along the lines of Refs. \cite{Ashtekar-Tate,Ashtekar-book}}.  Next, we will apply the program to the case of scalar fields, and we will discuss Bogoliubov transformations and the unitary implementation of the time evolution. We will close this section with the presentation of the $j_0$-Fock quantization for scalar fields with time dependent mass. Throughout the section, special attention will be paid to the role of complex structures in the quantum theory. 

\subsection{Canonical quantization on a Hilbert space}
\label{subsec:canon-quant}

Consider a linear classical system (with a finite or infinite number of degrees of freedom), described by a symplectic vector space, {that we will call} $(X,\Omega)$. {The set of classical observables will hereby be denoted by ${\cal{O}}$.} Roughly speaking, by quantization {we will understand} the passage from a classical {description of a system to a quantum mechanical description}. In contrast to the situation in the classical theory, where states live in the phase space $(X,\Omega)$ and observables are real-valued functions on $(X,\Omega)$, in the quantum theory states belong to a Hilbert space ${\cal{H}}$, whereas observables are self-adjoint operators on $\cal{H}$. The basic {PB, that equip} the space of classical observables with an algebraic structure, {are replaced at the quantum level with} the canonical commutation relations (CCRs), that define an algebraic structure on the space of quantum observables. Thus, in very broad terms, the output of the quantization should be a Hilbert space $\cal{H}$ of quantum states, and quantum observables represented on $\cal{H}$ as self-adjoint operators, obeying the algebraic structure arising from the CCRs. For linear systems, the process of canonical quantization on a Hilbert space consists of (and it is accomplished by) three main steps: 

(i) A selection of basic (elementary, or fundamental) classical observables ${\cal{O}}_{0}\subset {\cal{O}}$. 

(ii) The construction of an abstract quantum algebra $\cal{A}$ of observables from ${\cal{O}}_{0}$, with the following two properties: (iia) for each basic observable $F\in {\cal{O}}_{0}$ there must be one, and only one, abstract quantum basic operator (observable) $\hat{F}\in {\cal{A}}$, and (iib) basic operators must satisfy the Dirac quantization condition, {relating their commutators with the corresponding PBs}. 

(iii) {The specification} of a Hilbert space $\cal{H}$ and a representation of the abstract basic observables as self-adjoint operators on $\cal{H}$. 

{For more details, we refer the reader, e.g., to Ref. \cite{Ashtekar-Tate}}.

These rules are far from determining a unique quantum description. Indeed, the process entails ambiguities at different stages, and a series of choices must be made in order to accomplish the quantization and arrive to a,  hopefully, well specified description. In fact, one has to face ambiguities from the very beginning of the process by making ``a judicious selection'' of fundamental observables ${\cal{O}}_{0}$. {This set of basic observables ${\cal{O}}_{0}$ is typically required to  be a vector subspace of $\cal{O}$, closed under PB, and such that every regular function on phase space can be obtained by (possibly a limit of) sums of products of its elements \cite{Ashtekar-Tate,Ashtekar-book}. These requirements are intended to achieve that observables in ${\cal{O}}_{0}$ will be appropriately promoted to quantum operators satisfying the CCRs, allowing to avoid ambiguities like e.g. the well-known problem of factor ordering. However, it is not uncommon that various distinct basic sets can be found for the same system. So, in general  there is not a unique canonical choice of elementary observables ${\cal{O}}_{0}$,  a fact which can give rise to non-equivalent quantum descriptions. This ambiguity is usually addressed by arguing ``naturalness and simplicity''  in favor of a particular classical canonical representation.}

Once the set of fundamental observables is specified, the next step in the quantization is to construct an abstract quantum algebra ${\cal{A}}$ of observables from the vector space ${\cal{O}}_{0}$. The algebra is constructed as follows. Let ${\cal{A}}_{0}$ be the free associative algebra over the complex numbers generated by ${\cal{O}}_{0}$, i.e. the free associative complex {algebra corresponding to} ${\cal{O}}_{0\,\mathbb{C}}={\cal{O}}_{0}\oplus i{\cal{O}}_{0}$.  Thus, every $F_{c}\in {\cal{O}}_{0\,\mathbb{C}}$ has a representative $\lambda(F_{c})$ in ${\cal{A}}_{0}$, where $\lambda$ is a linear mapping. Next, the algebra ${\cal{A}}_{0}$ is equipped with an involution operation $*$ which captures the {complex conjugation}; so, the representative $\lambda(F)\in{\cal{A}}^{*}_{0}$ of the real, basic observable $F\in {\cal{O}}_{0}$ is invariant under the involution operation, $[\lambda(F)]^{*}=\lambda(F)$. More generally, $[\lambda(F_{c})]^{*}=\lambda(G_{c})$ if and only if $\bar{F_{c}}=G_{c}$, where $F_{c},G_{c}\in {\cal{O}}_{0\,\mathbb{C}}$. Then, the algebraic structure on the space of classical observables, provided by the PB, is carried to an analogous algebraic structure on  quantum observables. {For this, one takes} the $*$-ideal ${\cal{I}}_{\rm{D}}$ of ${\cal{A}}^{*}_{0}$ generated by elements of the form $\lambda(-i\{F_{c},G_{c}\})+[\lambda(F_{c}),\lambda(G_{c})]\in {\cal{A}}^{*}_{0}$. This is precisely the Dirac quantization condition. The algebra of abstract quantum observables $\cal{A}$ is the quotient algebra of ${\cal{A}}^{*}_{0}$ by the ideal ${\cal{I}}_{\rm{D}}$. The associative algebras ${\cal{A}}^{*}_{0}$ and $\cal{A}$ are related by the homomorphism $\sigma(w)=w+{\cal{I}}_{\rm{D}}$, $w\in {\cal{A}}^{*}_{0}$. Let $\wedge$ be the mapping $\wedge=\sigma\circ\lambda$, and let us define $\hat{F_{c}}=\wedge(F_{c})$. Thus, in particular, we have that for each $F\in{\cal{O}}_{0}$, there is one and only one $*$-invariant operator $\hat{F}\in {\cal{A}}$. Given $F$, $G$, and $\{F,G\}$ in ${\cal{O}}_{0}$, their (abstract) $*$-invariant, basic operator counterparts $\hat{F}$, $\hat{G}$, and $\widehat{\{F,G\}}$ {in $\cal{A}$ satisfy the CCRs} $[\hat{F},\hat{G}]=i\widehat{\{F,G\}}$. By construction, any $\hat{A}\in\cal{A}$ can be expressed as a sum of products of elementary operators. 

The third and final step in the process is to find a Hilbert space $\cal{H}$ supporting a representation of the (abstract) fundamental quantum observables as self-adjoint operators. {This representation}, however, turns out to be not unique in general. There exist, typically, different (i.e. not unitarily equivalent) Hilbert space representations of the CCRs. So, one generally has to deal with the problem of determining a preferred representation. It should be noted that, in contrast with the ambiguity in the choice of basic observables, which affects both linear mechanical and linear field theory systems, the lack of uniqueness of {the representation is mainly an issue for field (i.e. infinite dimensional) systems. In fact, for} linear, finite  dimensional systems (i.e. linear mechanical systems), the specification of a unique preferred representation of the CCRs can be consistently and unambiguously established {under certain requirements}. Indeed, in view of the Stone-von Neumann uniqueness theorem, we can restrict our attention just to a single representation of the CCRs, namely the ordinary Schr\"{o}dinger representation of quantum mechanics. However, {the situation is quite different} for linear field theories. There are infinitely many inequivalent Hilbert space representations of the basic quantum observables as self-adjoint operators, and no analogue of the Stone-von Neumann theorem exists
 to confront the uniqueness issue. To {handle this} ambiguity in the representation of the CCRs, the usual procedure is to appeal to the spacetime symmetries of the field system and look for symmetry invariant representations. Though this strategy leads to a unique quantum theory for a certain class of field systems (for instance, linear field theories in both Minkowski and stationary spacetimes), it should be stressed that in more general cases (like e.g. non-stationary settings) symmetries will simply not be enough to pick out a preferred representation and, therefore, extra criteria must be imposed in order to set a unique quantum theory.

\subsection{Linear scalar field theory: Quantization}
\label{sec:quant-syst}

Let us now review the quantization of the scalar field theory introduced in Sec. \ref{subsec:classical-sft}.
We first consider the covariant phase space approach. As we have seen, linear functionals $\Omega(\phi,\, \cdot \,)$ provide a natural set of observables on $(S,\Omega)$ {with non-trivial  PB that are} proportional to the unit function  [see Eq. (\ref{PB-cov})]. Since observables on $(S,\Omega)$ can be obtained by taking linear combinations of products of natural observables $\Omega(\phi,\, \cdot \,)$ and the unit function $\rm{I}$ (which provides the constant functions on $S$), the subspace{\footnote{Here, $\{X\}_{\mathbb{F}}$ denotes the vector space given by the set $\{X\}$ over the field $\mathbb{F}$.}} $\{{\rm{I}},\Omega(\phi,\, \cdot \,)\, \vert\, \phi\in S\}_{\mathbb{R}}$ of $\cal{O}$ {qualifies as an} admissible set of basic observables. This, together with the ``naturalness and simplicity'' of our choice, leads us to {select the commented subspace}  as the set ${\cal{O}}_{0}$ of {fundamental (basic, or elementary)} classical observables.

By equipping the phase space $(S,\Omega)$ with a compatible complex structure $J$, the field $\phi$ can be  decomposed into the ``positive and negative frequency'' parts, $\phi^{+}$ and $\phi^{-}$, defined by $J$. In addition, the completion of the inner product space $(\,S^{+}_{J},\langle \,\cdot\, , \, \cdot\, \rangle_{S^{+}_J}=\Omega(\,\overline{J\,\cdot\,} , \, \cdot \,))$ in the norm $\|\cdot \|_{S^{+}_J}$ defines the ``one-particle'' Hilbert space ${\cal{H}}_J$. Notice that $\Omega(\,\overline{J\,\cdot\,} , \, \cdot \,)$  is an inner product not only for $S^{+}_J$, but also for ${\cal{S}}_{J}=S^{+}_J\oplus S^{-}_J$; in fact, $S^{+}_J$ and $S^{-}_J$ are orthogonal subspaces with respect to this product. Thus, the Cauchy completion of ${\cal{S}}_J$ gives a complex Hilbert space $H$, which decomposes into the orthogonal $(\pm\, i)$-eigenspaces of $J$, with the $(+i)$-eigenspace being precisely the so-called one-particle Hilbert space ${\cal{H}}_{J}$, whereas the $(-i)$-eigenspace is the complex conjugate of ${\cal{H}}_J$, $\overline{\cal{H}}_{J}$. Let $K_{J}:H\to {\cal{H}}_{J}$ and $\bar{K}_{J}:H\to \overline{\cal{H}}_{J}$ be the orthogonal projections arising from the inner product $\Omega(\,\overline{J\,\cdot\,} , \, \cdot \,)$. The restrictions of $K_J$ and $\bar{K}_{J}$ to $S$ are nothing but the real-linear bijections from $S$ to $S_{J}^{+}\subset {\cal{H}}_{J}$ and $S_{J}^{-}\subset \overline{\cal{H}}_{J}$, respectively. In terms of the restrictions of $K_J$ and $\bar{K}_{J}$ to $S$, the field decomposition defined by $J$ reads $\phi=K_{J}\phi+\bar{K}_{J}\phi$. Thus, basic observables can be written in the form $\Omega(\phi,\,\cdot\,)=ia(\bar{K}_{J}\phi)-i\bar{a}(K_{J}\phi)$, where 
\begin{equation}
\label{annihi-crealike-J}
a(\bar{K}_{J}\phi)=\Omega(J\bar{K}_{J}\phi,\, \cdot \,),\quad \bar{a}(K_{J}\phi)=\Omega(JK_{J}\phi,\, \cdot \,) 
\end{equation}
are, respectively, the annihilation and creation-like variables associated with the complex structure $J$. By using complex linearity and continuity, we get that 
\begin{equation}
\label{extended-basic-obs}
\Omega(\Phi,\,\cdot\,)=ia(\bar{\chi})-i\bar{a}(\xi),
\end{equation}
where $\chi , \xi \in {\cal{H}}_{J}$ and $\Phi\in H$, with $\Phi=\xi+\bar{\chi}$.

The next step in the process of quantization is to specify $\cal{A}$, the algebra of abstract quantum observables. According to the discussion in Sec. \ref{subsec:canon-quant}, this algebra is constructed from the complexification of ${\cal{O}}_{0}$, ${\cal{O}}_{0\, \mathbb{C}}$. However, notice that we have an enlarged vector space ${\mathscr{S}}=\{{\rm{I}},\Omega(\Phi,\,\cdot\,)\,\vert\, \Phi\in H\}_{\mathbb{C}}\supset{\cal{O}}_{0\, \mathbb{C}}$, perfectly valid to construct the algebra. So, we will take $\mathscr{S}$ to specify $\cal{A}$. Since every (complex) elementary variable $\Omega(\Phi,\,\cdot\,)$, with $\Phi\in H$, can be uniquely expressed in the form (\ref{extended-basic-obs}), the vector space $\mathscr{S}$ can be naturally rewritten as 
\begin{equation}
\label{Alg-gener-crea-annihi}
\mathscr{S}=\{{\rm{I}}, a(\bar{\chi}),\bar{a}(\xi)\, \vert \, \chi,\xi\in{\cal{H}}_J\ \}_{\mathbb{C}}.
\end{equation}
The only non-zero PB between the complex elementary variables (\ref{Alg-gener-crea-annihi})  are
\begin{equation}
\label{PB-annihi-cre}
\{a(\bar{\chi}),\bar{a}(\xi)\}=-i\big\langle \chi,\xi\big\rangle_{{\cal{H}}_{J}},
\end{equation}
where $\langle \,\cdot\, , \,\cdot\,\rangle_{{\cal{H}}_{J}}= \Omega(\,\overline{J\,\cdot\,} , \, \cdot \,)$ [i.e. the Hermitian inner product (\ref{herm-inner-prod-pfs})]. The quantum algebra $\cal{A}$ 
is {defined starting with} the complex vector space $\mathscr{S}$ (\ref{Alg-gener-crea-annihi}), 
exactly as we have explained in Sec. \ref{subsec:canon-quant} (with ${\cal{O}}_{0\, \mathbb{C}}$ replaced by $\mathscr{S}$). As a result of the construction, we get operators $\hat{a}(\bar{\chi})$ and $\hat{a}^{*}(\xi)$, satisfying $[\hat{a}(\bar{\chi})]^{*}=\hat{a}^{*}(\chi)$ and obeying the commutation relations
\begin{equation}
\label{CCR-annihi-cre}
[\hat{a}(\bar{\chi}),\hat{a}^{*}(\xi)]={\hat{\rm{I}}}\,\big\langle \chi,\xi \big\rangle_{{\cal{H}}_J}, \quad [\hat{a}(\bar{\chi}),\hat{a}(\bar{\xi})]=0, \quad [\hat{a}^{*}(\chi),\hat{a}^{*}(\xi)]=0,
\end{equation}
for all $\chi,\xi \in {\cal{H}}_{J}$. The abstract quantum counterparts of the basic observables $\Omega(\phi,\, \cdot \,)\in {\cal{O}}_{0}$ are given by the $*$-invariant elementary operators 
\begin{equation}
\label{fundam-op}
\hat{\Omega}(\phi,\,\cdot\,)=i\hat{a}(\bar{K}_{J}\phi)-i\hat{a}^{*}(K_{J}\phi),
\end{equation}
that fulfill the CCRs 
\begin{equation}
\label{CCR-fund-op}
[\,\hat{\Omega}(\phi,\, \cdot \,),\hat{\Omega}(\tilde{\phi},\, \cdot \,)\,]=-i\Omega(\phi,\tilde{\phi})\,\hat{\rm{I}}.
\end{equation}

In order to accomplish the quantization, we need to specify a Hilbert space supporting a representation of the fundamental quantum observables $\hat{\Omega}(\phi,\, \cdot \,)$ as self-adjoint operators. Note, however, that a Hilbert space structure has been already chosen from the introduction of a complex structure: the one-particle Hilbert space ${\cal{H}}_{J}$. It is from ${\cal{H}}_J$ that the Hilbert space of the quantum theory is constructed. Concretely,  the one-particle Hilbert space defines the symmetric Fock space  
\begin{equation}
\label{Fock-space}
{\cal{F}}_{J}=\oplus_{n=0}^{\infty}\left(\otimes^{n}_{(s)}{\cal{H}}_J\right),
\end{equation}
that is the desired Hilbert space. The structure of ${\cal{F}}_{J}$ allows for a natural representation of $\hat{a}(\bar{\chi})$ and $\hat{a}^{*}(\xi)$, subject to the commutation relations (\ref{CCR-annihi-cre}), as the annihilation and creation operators $\hat{a}(\bar{\chi})$ and $\hat{a}^{\dagger}(\xi)$ on ${\cal{F}}_{J}$. Thus, the fundamental observables $\hat{\Omega}(\phi,\, \cdot \,)$ are represented on the Fock space by the self-adjoint operators defined by Eq. (\ref{fundam-op}), obeying the CCRs (\ref{CCR-fund-op}). This is the standard procedure, in the covariant approach, for the Fock quantization of a linear scalar field 
{\emph{given} a complex structure $J$. Since} $J$ can be any compatible complex structure, what we really have is a family of Fock representations of the  CCRs parameterized by the set ${\cal{J}}(\Omega,S)$. This set splits naturally into equivalence classes $[J]$ of complex structures that lead to unitarily equivalent Fock representations, and it is well known that ${\cal{J}}(\Omega,S)$ {is formed by an infinite} number of them. That is, there are infinitely many inequivalent Fock representations of the CCRs. Therefore, in order to specify a unique quantum description, up to unitarity, a preferred complex structure $j_p$ (or, more generally, an equivalence class $[j_{p}]$) must be chosen.

Let us recall that, in general, there are no representations of the CCRs by bounded operators \cite{Wald-book,Putnam-book}. For the KG field, the quantum fundamental observables $\hat{\Omega}(\phi,\,\cdot\,)$ turn out to be all unbounded operators [except $\hat{\Omega}(0,\,\cdot\,)$]. So, questions concerning the domains of definition should be treated carefully. For instance, a proper definition of the elements of the quantum algebra on the Hilbert space becomes an intricate task, because $\cal{A}$ {contains polynomials}. In order to avoid this unwieldy situation, the usual procedure is to consider the exponentiated version of $\hat{\Omega}(\phi,\,\cdot\,)$, namely $W(\phi)=\exp[\,i\hat{\Omega}(\phi,\,\cdot\,)\,]$. Formally, the CCRs {are replaced with} the Weyl relations
\begin{equation}
\label{Weyl-relation1}
W(\phi)W(\tilde{\phi})=e^{\frac{i}{2}\Omega(\phi,\tilde{\phi})}W(\phi+\tilde{\phi}), 
\end{equation}
together with the adjoint relations
\begin{equation}
\label{Weyl-relation2}
W^{*}(\phi)=W(-\phi).
\end{equation}
By equipping the vector space spanned by all finite, complex linear combinations of the $W(\phi)$'s  with the product (\ref{Weyl-relation1}) --extended by linearity to the vector space-- and the involution operation (\ref{Weyl-relation2}), we get a complex associative $*$-algebra [with unit element ${\rm{I}}=W(0)$]. Given a Hilbert space representation of the CCRs (\ref{CCR-fund-op}), the $*$-algebra generated by the $W(\phi)$'s {becomes} a subalgebra, ${\cal{W}}_0$, of the $C^{*}$-algebra of all bounded linear operators on the Hilbert space. The closure of ${\cal{W}}_0$ thus leads to a $C^{*}$-(sub)algebra $\cal{W}$, which is known as the Weyl algebra. {Although one might think that} the Weyl algebra {defined in this way would be a} representation-dependent algebra, {actually this} is not the case: $\cal{W}$ is fully independent of the particular representation used \cite{Segal,Slawny}. Thus, in order to avoid domain problems, one can consider the Weyl algebra $\cal{W}$ and look for a unique preferred representation of the relations (\ref{Weyl-relation1}) and (\ref{Weyl-relation2}).

From the definition of $W(\phi)$ and Eq. (\ref{fundam-op}), we can write the Weyl generators in terms of the annihilation and creation operators on ${\cal{F}}_{J}$, 
\begin{equation}
\label{Weyl-crea-annihi}
W(\phi)=\exp\big[\,\hat{a}^{\dagger}(K_{J}\phi)-\hat{a}(\bar{K}_{J}\phi)\,\big].
\end{equation}
By using the commutation relations (\ref{CCR-annihi-cre}), the relationship $[\hat{a}(\bar{K}_{J}\phi)]^{\dagger}=\hat{a}^{\dagger}(K_{J}\phi)$, and the Baker-Campbell-Hausdorff (BCH) formula, it is not difficult to see that the generators in Eq. (\ref{Weyl-crea-annihi}) satisfy indeed the relations (\ref{Weyl-relation1}) and (\ref{Weyl-relation2}). In this way, we can construct the (concrete) Weyl algebra ${\cal{W}}_{J}$, which is a subalgebra of $L({\cal{F}}_{J})$, the $C^{*}$-algebra of all bounded linear operators on ${\cal{F}}_{J}$. 
{Let us now consider the vacuum state $\vert 0\rangle\in {\cal{F}}_{J}$, i.e. the unique normalized state
$\vert 0\rangle$ that is annihilated by all the  annihilation  operators $\hat{a}(\bar{\chi})$.}
By using the BCH formula and Eq. (\ref{CCR-annihi-cre}), a direct calculation shows that the vacuum expectation value of $W(\phi)$ in ${\cal{F}}_{J}$ is given by
\begin{equation}
\label{algstate-mechsyst}
\langle W(\phi) \rangle_{\rm{vac}}=e^{-\frac{1}{4}\| \phi\|^{2}_{J}}.
\end{equation}
Here, $\| \phi\|_{J}$ is the norm of $\phi\in S$ with the real inner product  $\mu_{J}$  defined by $J\in {\cal{J}}(\Omega,S)$ [see Eq. (\ref{inner-prod-svp})]. 

The relationship (\ref{algstate-mechsyst}) defines a quasi-free algebraic state $\omega_{J}[W(\phi)]=\exp(-\tfrac{1}{4}\| \phi\|^{2}_{J})$. The triple $({\cal{F}}_{J},{\cal{W}}_{J},\vert 0\rangle)$ is, in fact, the same that would be obtained by employing $\omega_{J}$ on the (abstract) Weyl algebra $\cal{W}$ in the so-called {Gelfand-Naimark-Segal (GNS)} construction \cite{GN,Seg}.  The representation ${\cal{W}}_{J}$ of the Weyl algebra $\cal{W}$, defined by the complex structure $J$, is moreover irreducible, which is tantamount to saying that the state $\omega_{J}$ is pure. Conversely, pure quasi-free states of the Weyl algebra are associated with complex structures, and give rise to Fock representations as above \cite{MV,Wald-book}.

Let us consider now the quantization in the canonical phase space approach. Our choice of a natural set of elementary classical observables on $(\Gamma,\Omega)$ leads to the real vector space ${\cal{O}}_{0}=\{{\rm{I}},\Omega((g,f),\,\cdot\,)\,\vert \, (g,f)\in\Gamma\}_{\mathbb{R}}=\{{\rm{I}},\varphi[f],\pi[g]\,\vert \, (g,f)\in\Gamma\}_{\mathbb{R}}$ equipped with the PB (\ref{PB-cano}). In addition, let us introduce a complex structure $j\in {\cal{J}}(\Omega,\Gamma)$. The abstract algebra $\cal{A}$ is constructed from the complex vector space ${\mathscr{S}}=\{{\rm{I}},\varphi[F],\pi[G]\,\vert \, (G,F)\in\gamma_{j}\}_{\mathbb{C}}$, where $\gamma_{j}$ is the Cauchy completion {of} $\Gamma\oplus i\Gamma$ with respect to $\mu_{j}(\overline{\, \cdot\,},\,\cdot\,)=\Omega(\overline{j\, \cdot\,},\,\cdot\,)$. The fundamental quantum operators $\hat{\varphi}[f]$ and $\hat{\pi}[g]$ in $\cal{A}$ satisfy the CCRs: $[\hat{\varphi},\hat{\pi}]=i\Omega((0,f),(g,0))\,\hat{\rm{I}}$. {Schr\"{o}dinger-like representations of the CCRs are naturally available in the canonical approach, as follows. The} CCRs are represented on a Hilbert space ${\mathscr{H}}_{j}=L^{2}(\bar{C},d\varrho)$ of wave functionals on a quantum configuration space $\bar{C}$, with the basic operators of configuration and momentum, $\hat{\varphi}[f]$ and $\hat{\pi}[g]$, acting on {the} wave functionals by multiplication and {by} derivation plus multiplication, respectively. The measure $\varrho$, {that} is of Gaussian type, is determined by the complex structure $j$. However, it does not encode the full information about the complex structure, in general. Apart from a derivative, the momentum operator $\hat{\pi}[g]$ contains, in general, two multiplicative terms, namely a factor associated with the Gaussian character of $\varrho$, and {possibly another (non-trivial) multiplicative term that contains further} information about the complex structure $j$. To be more specific, the general form of a complex structure $j\in {\cal{J}}(\Omega,\Gamma)$ is given by $-j(\varphi,\pi)=(\frak{a}\varphi+\frak{b}\pi,\frak{c}\pi+\frak{d}\varphi)$, where $\frak{a,b,c}$, and $\frak{d}$ are linear operators satisfying
\begin{equation}
\label{prop-j}
\frak{a}^{2}+\frak{bd}=-{\rm{I}},\quad \frak{c}^{2}+\frak{db}=-{\rm{I}},\quad \frak{ab}+\frak{bc}=0,\quad \frak{da}+\frak{cd}=0,
\end{equation}
and
\begin{eqnarray}
\label{prop-j-comp}
\int_{\Sigma}f\frak{b}f'=\int_{\Sigma}f'\frak{b}f, \quad \int_{\Sigma}g\frak{d}g'=\int_{\Sigma}g'\frak{d}g,\quad \int_{\Sigma}f\frak{a}g=-\int_{\Sigma}g\frak{c}f, \nonumber \\ 
\int_{\Sigma}f\frak{b}f'>0, \quad \int_{\Sigma}g\frak{d}g'<0,
\end{eqnarray}
for all unit weight scalar densities $f,f'\in C^{\infty}_{0}(\Sigma)$ and scalars $g,g'\in C^{\infty}_{0}(\Sigma)$. Relationships (\ref{prop-j}) {come} from {the condition} $j^{2}=-{\rm{I}}$, whereas {restrictions} (\ref{prop-j-comp}) {follow} from requiring that $\mu_{j}(\, \cdot\,,\,\cdot\,){=}\Omega(j\, \cdot\,,\,\cdot\,)$ be a symmetric and positive definite bilinear form. The measure and the basic operators of configuration and momentum {are} \cite{ccq-annphys,ccq-prd66}
\begin{equation}
\label{the-gaussian-measure}
d\varrho=\exp\left(-\int_{\Sigma}\varphi{\frak{b}}^{-1}\varphi\right){\cal{D}}\varphi,
\end{equation}
\begin{equation}
\label{the-conf-mom-rep}
\hat{\varphi}[f]\Psi=\varphi[f]\Psi,\quad \hat{\pi}[g]\Psi=-i\int\left(g\frac{\delta}{\delta \varphi}-\varphi({\frak{b}}^{-1}-i{\frak{c}}{\frak{b}}^{-1})g\right)\Psi.
\end{equation}

{Note that the  representation defined by Eq. (\ref{the-conf-mom-rep}), in the Hilbert space ${\mathscr{H}}_{j}$, is a representation of the Fock type,  i.e. corresponds to a pure quasi-free state of the Weyl algebra.To see this explicitly, let us introduce the Weyl operators $W(g,f)=\exp[i\hat{\Omega}((g,f),\,\cdot\,)]$. For an initial reference time $t_0$, the map $I_{t_{0}}$, where $(g,f)=I_{t_{0}}\phi$, naturally induces a bijection between the algebra generated by the objects $W(\phi)$  and the corresponding one generated by the operators $W(g,f)$, 
that we will call $W_j$. Consider now the unit constant functional
$\psi_0\in {\mathscr{H}}_{j}$. It can again be shown that the expectation values of the Weyl generators read $\langle\psi_0,W(g,f)\psi_0\rangle=\exp(-\tfrac{1}{4}\|(g,f)\|^{2}_{j})$,where $\| \, \cdot \,\|_{j}$ is the norm associated with the real inner product $\mu_{j}(\, \cdot\, , \, \cdot\,)=\Omega(j\,\cdot\, , \, \cdot\,)$ on {the} phase space $(\Gamma,\Omega)$. Using the bijection $I_{t_{0}}$, one concludes that $\omega_{J}[W(\phi)]=\exp(-\tfrac{1}{4}\| \phi\|^{2}_{J})$ indeed defines a pure quasi-free state of the Weyl algebra, associated with the complex structure $J=I^{-1}_{t_{0}}jI_{t_0}\in {\cal{J}}(\Omega,S)$. Since a state (in fact the evaluation of a state on the generators) uniquely characterizes a unitary equivalence class of the Weyl algebra, it follows that $({\cal{F}}_{J},{\cal{W}}_{J},\vert 0\rangle)$ and $({\mathscr{H}}_{j},{\cal{W}}_{j},\psi_{0})$ are just different realizations of the same representation of the Weyl relations, i.e. there is a unitary map $U:{\cal{F}}_{J}\to {\mathscr{H}}_{j}$, with $U\vert 0\rangle=\psi_{0}$, that intertwines ${\cal{W}}_{J}$ with ${\cal{W}}_{j}$. To make this 
relationship fully explicit,  let us display the form of the annihilation and creation operators on ${\mathscr{H}}_j$, that are readily seen to be
\begin{equation}
\label{anni-crea-L2}
\hat{a}(\overline{\gamma^{+}})=\tfrac{1}{2}\left(\hat{\varphi}[\bar{\sigma}]+\hat{\pi}[\bar{\rho}]\right),\quad \hat{a}^{\dagger}(\gamma^{+})=\tfrac{1}{2}\left(\hat{\varphi}[\sigma]+\hat{\pi}[\rho]\right),
\end{equation}
where $\sigma={\frak{d}}g-{\frak{c}}f+if$, $\rho={\frak{b}}f-{\frak{a}}g+ig$ and $\gamma^{+}=(-g,f)^{+}\in\gamma_{j}$ is the ``positive frequency'' part of the Cauchy data $\gamma=(-g,f)\in \Gamma$, that is to say $\gamma^{+}=(\gamma - ij\gamma)/2$.
(For a comprehensive discussion on the Schr\"{o}dinger representation for a linear scalar field  in flat and curved spacetime, including  the relationship between the covariant and the canonical approaches to quantization, as well as  measure theoretical aspects, see Refs. \cite{ccq-annphys,ccq-prd66,Ve}).}

{Representations of the type $({\cal{F}}_{J},{\cal{W}}_{J},\vert 0\rangle)$, determined by a complex structure $J$ in the covariant phase space, will hereafter be called $J$-Fock representations, whereas the corresponding representations of the form $({\mathscr{H}}_{j},{\cal{W}}_{j},\psi_{0})$, constructed from the canonical perspective, will be called $j$-Fock representations.}

In the rest of our discussion, the domain of definition of the different quantum observables will not play a relevant role. Hence, in what follows we will consider representations of the CCRs only.

\subsection{Bogoliubov transformations and unitary implementability}
\label{subsec:remarks-quant}

Let us {take} two {compatible} complex structures on {the} phase space $(S,\Omega)$, say $J_1$ and $J_2$, {and} assume that their associated inner products, $\mu_{i}(\, \cdot\, , \, \cdot\, )=\Omega(J_{i}\,\cdot\, , \,\cdot\, )$ for $i=1,2$, define equivalent norms on $(S,\Omega)$. Then, the corresponding Hilbert spaces $H_{1}={\cal{H}}_{1}\oplus \overline{\cal{H}}_{1}$ and $H_{2}={\cal{H}}_{2}\oplus \overline{\cal{H}}_{2}$ may be identified, and can be viewed as two distinct splittings {of} the same Hilbert space $H$ \cite{Wald-book}. {Consider also} the orthogonal projections $K_{J_{i}}:H\to {\cal{H}}_{i}$ and $\bar{K}_{J_{i}}:H\to \overline{{\cal{H}}_{i}}$ defined by the inner product $\Omega(\overline{J_{i}\,\cdot}\, , \,\cdot\, )$ on $H$. {Then,} let $A:{\cal{H}}_{2}\to {\cal{H}}_{1}$ and $B:{\cal{H}}_{2}\to \overline{\cal{H}}_{1}$ be the restrictions of $K_{J_1}$ and $\bar{K}_{J_1}$, respectively, to ${\cal{H}}_{2}$. Similarly, {let $C:{\cal{H}}_{1}\to {\cal{H}}_{2}$ and $D:{\cal{H}}_{1}\to \overline{\cal{H}}_{2}$ be the respective} restrictions of $K_{J_2}$ and $\bar{K}_{J_2}$ to ${\cal{H}}_{1}$. {In this setting}, it can be shown that \cite{Wald-book}
\begin{equation}
\label{Bogo-ops-relations-1}
A^{\dagger}A-B^{\dagger}B={\rm{I}}, \quad A^{\dagger}\bar{B}=B^{\dagger}\bar{A},
\end{equation}
\begin{equation}
\label{Bogo-ops-relations-2}
C^{\dagger}C-D^{\dagger}D={\rm{I}}, \quad C^{\dagger}\bar{D}=D^{\dagger}\bar{C},
\end{equation}
and
\begin{equation}
\label{Bogo-ops-relations-3}
C=A^{\dagger}, \quad D=-\bar{B}^{\dagger} .
\end{equation}

For an element $\varsigma$  of $H$, let $\psi\in {\cal{H}}_{1}$ and $\bar{\xi}\in \overline{{\cal{H}}_{1}}$ be the components of $\varsigma$ with respect to the splitting $H_1$ of $H$, and $\chi\in {\cal{H}}_{2}$ and $\bar{\eta}\in \overline{{\cal{H}}_{2}}$ their components with respect to the splitting $H_2$ of $H$. In short, $\varsigma=(\psi,\bar{\xi})_{H_{1}}\in {\cal{H}}_{1}\oplus \overline{\cal{H}}_{1}$ and $\varsigma=(\chi,\bar{\eta})_{H_{2}}\in {\cal{H}}_{2}\oplus \overline{\cal{H}}_{2}$. We know, in particular, that $(\psi,\bar{\xi})_{H_{1}}=(\psi,0)_{H_{1}}+(0,\bar{\xi})_{H_{1}}$. Since the orthogonal projections of $(\psi,0)_{H_{1}}$ and $(0,\bar{\xi})_{H_{1}}$ onto ${\cal{H}}_{2}$ are $(C\psi,0)_{H_2}$ and $(\bar{D}\bar{\xi},0)_{H_2}$, we get that $(C\psi+\bar{D}\bar{\xi},0)_{H_2}$ is the orthogonal projection of $(\psi,\bar{\xi})_{H_{1}}$ onto ${\cal{H}}_{2}$. A similar calculation shows that the orthogonal projection of $(\psi,\bar{\xi})_{H_{1}}$ onto $\bar{\cal{H}}_{2}$ is given by $(0,D\psi+\bar{C}\bar{\xi})_{H_{2}}$. {So, we have}
\begin{equation}
\label{Bogo-T-general}
\chi=C\psi+\bar{D}\bar{\xi}, \quad \bar{\eta} = D\psi+\bar{C}\bar{\xi}.
\end{equation}
This transformation, with $C$ and $D$ satisfying relationships (\ref{Bogo-ops-relations-2}), is known as a Bogoliubov transformation. {Note that, from Eqs.} (\ref{Bogo-ops-relations-2}) and (\ref{Bogo-ops-relations-3}), {the} inverse of (\ref{Bogo-T-general}) {is}
\begin{equation}
\label{Bogo-Tinv-general}
\psi=A\chi+\bar{B}\bar{\eta}, \quad \bar{\xi} = \bar{A}\bar{\eta}+B\chi.
\end{equation}

Associated to each  of the complex structures $J_1$ and $J_2$, there is a set of elementary variables [{see Eq.} (\ref{Alg-gener-crea-annihi})], 
\begin{equation}
{\mathscr{S}}_{1}=\{{\rm{I}}, a_{1}(\overline{\xi}),\bar{a}_{1}(\psi)\, \vert \, \xi,\psi\in{\cal{H}}_1\ \}_{\mathbb{C}}, \quad {\mathscr{S}}_{2}=\{{\rm{I}}, a_{2}(\overline{\eta}),\bar{a}_{2}(\chi)\, \vert \, \eta,\chi\in{\cal{H}}_2\ \}_{\mathbb{C}}.
\end{equation}
{Nonetheless,} the vector spaces ${\mathscr{S}}_{1}$ and ${\mathscr{S}}_{2}$ are, in fact, the same vector space ${\mathscr{S}}$ [{recall} that ${\mathscr{S}}=\{{\rm{I}},\Omega(\Phi,\,\cdot\,)\,\vert\, \Phi\in H\}_{\mathbb{C}}$, so that ${\mathscr{S}}_{1}$ and ${\mathscr{S}}_{2}$ are simply two different decompositions of the linear space ${\mathscr{S}}$]. Let us be more precise. It follows from {Eq.} (\ref{extended-basic-obs}) that the linear space $L_{0}{=}\{\Omega(\Phi,\,\cdot\,)\,\vert\, \Phi\in H\}_{\mathbb{C}}$ is decomposed by a complex structure $J$ into the direct sum of $A^{+}_{J}=\{\bar{a}(\rho)\, \vert\, \rho\in {\cal{H}}_{J}\}_{\mathbb{C}}$ and $A^{-}_{J}=\{a(\bar{\sigma})\, \vert\, \sigma\in {\cal{H}}_{J}\}_{\mathbb{C}}$. Hence, {the} complex structures $J_1$ and $J_2$ decompose the vector space $L_{0}$ as $A^{+}_{1}\oplus A^{-}_{1}$ and $A^{+}_{2}\oplus A^{-}_{2}$, respectively. Since ${\mathscr{S}}=\mathbb{C}\oplus L_{0}$, we get that ${\mathscr{S}}_{1}$ and ${\mathscr{S}}_{2}$ are nothing but two different decompositions of ${\mathscr{S}}$, {as we had commented}. The explicit relationship between the annihilation and creation-like variables associated with $J_1$ and $J_2$ {are}
\begin{equation}
\label{annihi-crea-diffe-CS}
a_{2}(\bar{\eta})=a_{1}\big(\bar{A}\bar{\eta}\big)-\bar{a}_{1}\big(\bar{B}\bar{\eta}\big), \quad \bar{a}_{2}(\chi)=
\bar{a}_{1}\big(A\chi\big)-a_{1}\big(B\chi).
\end{equation}
{In order to get these identities}, we have used the definition of the annihilation and creation-like variables (\ref{annihi-crealike-J}), the action of {the} complex structures $J_1$ and $J_2$ on their corresponding eigenvectors, and the linearity of $\Omega$, as well as the decomposition of $\chi\in {\cal{H}}_{2}$ and $\bar{\eta}\in \overline{\cal{H}}_2$ with respect to $H_1$, namely $\chi=A\chi+B\chi$ and $\bar{\eta}=\bar{A}\bar{\eta}+\bar{B}\bar{\eta}$. Relationships (\ref{annihi-crea-diffe-CS}) {give the form in ${\mathscr{S}}_{1}$ of} the annihilation-like varibles $a_2$ and the creation-like variables $\bar{a}_2$, defined by the complex structure $J_2$. It is not difficult to see that the PB between {the} variables $a_{2}(\bar{\eta})$ and $\bar{a}_{2}(\chi)$, given in {Eq.} (\ref{annihi-crea-diffe-CS}), {satisfy} indeed {Eq.} (\ref{PB-annihi-cre}). 

We emphasize that different complex structures{\footnote{More precisely, we refer to complex structures {compatible with $\Omega$ that give} rise to equivalent norms on $S$.}} provide different generators for the (same) abstract quantum algebra $\cal{A}$ (this is so because {different} complex structures just introduce different splittings in $\mathscr{S}$, the space from which $\cal{A}$ is constructed). Let us denote the abstract algebra $\cal{A}$ by ${\cal{A}}_{i}$ in the basis provided by the annihilation and creation-like variables defined by the complex structure $J_i$; i.e. ${\mathscr{S}}={\mathscr{S}}_{i}$ with ${\mathscr{S}}_{i}=\{{\rm{I}}, a_{i}(\overline{\sigma}_{i}),\bar{a}_{i}(\rho_{i})\, \vert \, \sigma_{i},\rho_{i}\in{\cal{H}}_i\ \}_{\mathbb{C}}$, where $a_{i}(\overline{\sigma}_{i})=\Omega(J_{i}\overline{\sigma}_{i},\,\cdot\,)$ and $\bar{a}_{i}(\rho_{i})=\Omega(J_{i}\rho_{i},\,\cdot\,)$. The abstract quantum counterparts of $a_{i}(\bar{\sigma}_{i})$ and $\bar{a}_{i}(\rho_{i})$ are the operators $\hat{a}_{i}(\bar{\sigma}_{i})$ and $\hat{a}^{*}_{i}(\rho_{i})$, satisfying $[\hat{a}_{i}(\bar{\sigma}_{i})]^{*}=\hat{a}^{*}_{i}({\sigma}_{i})$ and the CCRs (\ref{CCR-annihi-cre}). {When  $a_{i}$ and $\bar{a}_{i}$ are replaced}, respectively, {with} $\hat{a}_{i}$ and $\hat{a}^{*}_{i}$ (for $i=1,2$) in {Eq.} (\ref{annihi-crea-diffe-CS}) we get {expressions for}  $\hat{a}_{2}$ and $\hat{a}^{*}_{2}$ {in} ${\cal{A}}_{1}$. According to the discussion in Sec. \ref{sec:quant-syst}, the algebra ${\cal{A}}_{i}$ (for $i=1,2$) is {then} represented on the Fock space ${\cal{F}}_{i}$ (constructed from the one-particle Hilbert space ${\cal{H}}_i$) by declaring (representing) $\hat{a}_{i}$ and $\hat{a}^{*}_{i}$ as the annihilation and creation operators on ${\cal{F}}_{i}$, renaming then $\hat{a}_{i}$ and $\hat{a}^{\dagger}_{i}$. Thus, in spite of the $J$-independence of $H$ and $\cal{A}$, the representation of the algebra {on the} Hilbert space turns out to be a decomposition-dependent process: every complex structure ({or, equivalently,} decomposition) {gives} rise to a different Fock space representation of $\cal{A}$. The annihilation and creation operators on the Fock space ${\cal{F}}_{2}$, $\hat{a}_{2}$ and $\hat{a}^{\dagger}_{2}$, are represented on ${\cal{F}}_{1}$ {as}
\begin{equation}
\label{ac-ops-diff-fock}
\hat{a}'_{2}(\bar{\eta})=\hat{a}_{1}\big(\overline{A\eta}\big)-\hat{a}^{\dagger}_{1}\big(\overline{B\eta}\big), \quad \hat{a}^{\prime\,\dagger}_{2}(\chi)=\hat{a}^{\dagger}_{1}\big(A\chi\big)-\hat{a}_{1}\big(B\chi).
\end{equation}
Hence, in general, $\hat{a}'_{2}(\bar{\eta})$ {does} not annihilate the ${\cal{F}}_1$-vacuum state $\vert 0\rangle_{1}$. A direct calculation shows that the ${\cal{F}}_2$-number operator $\hat{a}^{\dagger}_{2}(\chi)\, \hat{a}_{2}(\bar{\eta})$, represented on ${\cal{F}}_1$ , has the following expectation value in the vacuum state $\vert 0\rangle_{1}\in{\cal{F}}_1$:
\begin{equation}
\label{numb-op}
\big\langle\, \hat{a}^{\prime\, \dagger}_{2}(\chi)\, \hat{a}'_{2}(\bar{\eta})\,\big\rangle_{{\rm{vac}}}=\langle \eta , B^{\dagger}B\chi \rangle_{{\cal{H}}_2}.
\end{equation}
The vacuum state $\vert 0\rangle_{2}\in {\cal{F}}_2$ in the Fock representation ${\cal{F}}_1$ corresponds to a state $\vert 0'\rangle_{2}$ {satisfying}
\begin{equation}
\label{vac-diff-fock}
\hat{a}_{1}(\bar{A}\bar{\eta})\,\vert 0'\rangle_{2}=\hat{a}^{\dagger}_{1}(\bar{B}\bar{\eta})\,\vert 0'\rangle_{2}.
\end{equation} 

Actually, provided that $\mu_1$ and $\mu_2$ define equivalent norms, it can be shown \cite{Wald-book} that the necessary and sufficient condition for {the} unitary equivalence of the Fock representations $({\cal{F}}_{1},\hat{a}_{1}, \hat{a}^{\dagger}_{1})$ and $({\cal{F}}_{2},\hat{a}_{2}, \hat{a}^{\dagger}_{2})$ is that $B$ {fulfills} the Hilbert-Schmidt condition 
\begin{equation}
\label{HS-condition}
{\rm{tr}}(B^{\dagger}B)<\infty.
\end{equation} 
In that case, there exists a unitary map ${\cal{U}}:{\cal{F}}_{1}\to {\cal{F}}_{2}$ such that 
\begin{equation}
\label{unit-between-fock-sp}
{\cal{U}}^{-1}\,\hat{a}_{2}(\bar{\eta})\,{\cal{U}}=\hat{a}'_{2}(\bar{\eta}),\quad  {\cal{U}}^{-1}\,\hat{a}^{\dagger}_{2}(\chi)\,{\cal{U}}=\hat{a}^{\prime\, \dagger}_{2}(\chi), \quad \vert 0'\rangle_{2}={\cal{U}}^{-1}\vert 0\rangle_{2},
\end{equation}
with $\hat{a}'_{2}(\bar{\eta})$ and $\hat{a}^{\prime\, \dagger}_{2}(\chi)$ given by relationships (\ref{ac-ops-diff-fock}), and $\vert 0'\rangle_{2}$ {solving Eq.} (\ref{vac-diff-fock}). We also note that the requirement (\ref{HS-condition}) on $B$ is equivalent to {impose} the Hilbert-Schmidt condition on $(J_{2}-J_{1})$. Indeed, {since} $\chi=A\chi +B\chi$, it follows that $(J_{2}-J_{1})\chi=2iB\chi$ for all $\chi\in {\cal{H}}_{2}$. Similarly, we have that $(J_{2}-J_{1})\psi=2i\bar{B}^{\dagger}\psi$ for all $\psi\in {\cal{H}}_{1}$. Thus, {the two} complex structures lead to unitary equivalent representations of the CCRs if and only if $(J_{2}-J_{1})$ defines a Hilbert-Schmidt operator, either on ${\cal{H}}_1$ or on ${\cal{H}}_2$. 

Let us now discuss the issue of dynamics. {Consider} a {compatible} complex structure {$J$ on} phase space $(S,\Omega)$. As we have seen in Sec. \ref{subsec:classical-sft}, $J$ {evolves} according to $J_{t}=T_{(t,t_{0})}\,J\,T^{-1}_{(t,t_{0})}$ [{see Eq.} (\ref{evolved-cov-cs})], where  $T_{(t,t_{0})}:S\to S$ is the linear symplectic transformation {corresponding to} {the} time evolution from $t_0$ to $t$. \footnote{Here, $J$ plays the role of an initial complex structure. Accordingly, all objects defined by $J$, such as the annihilation and creation operators or the associated Hilbert space, will be labelled with a subscript, or superscript, $t_0$.} Thus, every $J_t$ {belongs to} ${\cal{J}}(S,\Omega)$ and, consequently, we {get} a family of real inner products, $\mu_{t}{=}\Omega(J_{t}\,\cdot\, , \, \cdot\,)$, on $S$. {Assume} that, for each {time $t$}, the linear symplectic bijections $T_{(t,t_{0})}$ and $T^{-1}_{(t,t_{0})}$ are both continuous mappings on $S_{\mu}$ (the Hilbert completion of $S$ with respect to the norm $\|\cdot\|_{J}$ defined by the inner product $\mu_{t_0}=\mu_J$). Then, $\mu_{t_0}$ and $\mu_{t}$ define equivalent norms{\footnote{Let $S_{\mu}$ be the Cauchy completion of $S$ with respect to $\|\cdot\|_{J}$. Suppose that the linear {symplectomorphism} $R:S_{\mu}\to S_{\mu}$ and (its inverse) $R^{-1}:S_{\mu}\to S_{\mu}$ are continuous. Then, $R$ and $R^{-1}$ are bounded in the norm $\|\cdot\|_{J}$. {Hence,} using that $\mu_{R}(R\phi,R\phi)=\mu(\phi,\phi)$, {it} follows that $\mu_{J}=\Omega(J\,\cdot\, , \, \cdot\,)$ and $\mu_{J_{R}}=\Omega(J_{R}\,\cdot\, , \, \cdot\,)$, with $J_R=R J R^{-1}$, define equivalent norms.}} for all $t\in {\mathbb{I}\subset \mathbb{R}}$. The annihilation and creation operators induced by time evolution $T_{(t,t_{0})}$ on ${\cal{F}}_{t_{0}}{=}{\cal{F}}_{J}$, namely $\hat{a}'_{t}$ and $\hat{a}^{\prime\,\dagger}_{t}$, are given by Bogoliubov transformations of the form (\ref{ac-ops-diff-fock}),
\begin{equation}
\label{ac-prime-ops-diff-fock}
\hat{a}'_{t}(\bar{\eta}_{t})=\hat{a}_{t_0}\big(\overline{{A}_{(t,t_{0})}\eta_{t}}\big)-\hat{a}^{\dagger}_{t_0}\big(\overline{{B}_{(t,t_{0})}\eta_{t}}\big), \quad \hat{a}^{\prime\,\dagger}_{t}(\chi_{t})=\hat{a}^{\dagger}_{t_0}\big(A_{(t,t_{0})}\chi_{t}\big)-\hat{a}_{t_0}\big(B_{(t,t_{0})}\chi_{t}),
\end{equation}
for each $t\in {\mathbb{I}\subset \mathbb{R}}$. Here, both $\eta_{t}$ and $\chi_{t}$ are in ${\cal{H}}_{t}$, whereas the orthogonal projections (with respect to the $H_{t_0}$-decomposition) $A_{(t,t_{0})}:{\cal{H}}_{t}\to {\cal{H}}_{t_0}$ and $B_{(t,t_{0})}:{\cal{H}}_{t}\to \overline{\cal{H}}_{t_0}$ satisfy relationships (\ref{Bogo-ops-relations-1}). So, $\hat{a}'_{t}(\bar{\eta}_{t})$ and $ \hat{a}^{\prime\,\dagger}_{t}(\chi_{t})$ fulfill the CCRs (\ref{CCR-annihi-cre}). Clearly, $A_{(t_{0},t_{0})}$ and $B_{(t_{0},t_{0})}$ are the identity and the zero maps, respectively. 

Since classical observables evolve according to{\footnote{Indeed, under time evolution, $\Omega(\phi,\phi') \mapsto\Omega(\phi,T_{(t,t_{0})}\phi')=\Omega(T_{(t,t_{0})}^{-1}\phi,\phi')$.}} $\Omega(\phi,\,\cdot\,)\mapsto \Omega(T^{-1}_{(t,t_{0})}\phi,\,\cdot\,)$, we have that $a'_{t}(\bar{\eta}_{t})=\Omega(T^{-1}_{(t,t_{0})}J\bar{\eta}_{t_0},\, \cdot\,)$ and $\bar{a}'_{t}(\chi_{t})=\Omega(T^{-1}_{(t,t_{0})}J\chi_{t_0},\, \cdot\,)$. On the other hand, the symplectic transformations $T_{(t,t_{0})}$ induce a two-parameter family of $*$-{automorphisms on $\cal{A}$,} $\hat{\Omega}(\phi,\,\cdot\,)\mapsto \zeta_{\, (t,t_{0})}\cdot \hat{\Omega}(\phi,\,\cdot\,)=\hat{\Omega}(T_{(t,t_{0})}\phi,\,\cdot\,)$. Thus, {the} time evolution of the (abstract) elementary quantum observables is given by $\hat{\Omega}(\phi,\,\cdot\,)\mapsto \zeta_{\, (t,t_{0})}^{-1}\cdot \hat{\Omega}(\phi,\,\cdot\,)$. In particular, we have that $\hat{a}'_{t}(\bar{\eta}_{t})=\zeta_{\, (t,t_{0})}^{-1}\cdot \hat{a}_{t_{0}}(\bar{\eta}_{t_{0}})$ and $\hat{a}^{\prime\,\dagger}_{t}(\chi_{t})=\zeta_{\, (t,t_{0})}^{-1}\cdot \hat{a}^{\dagger}_{t_0}(\chi_{t_0})$, {on} ${\cal{A}}_{t_0}$.

{The question of unitary implementability of the dynamics in the $J$-Fock representation is whether or not there exist unitary operators $U_{(t,t_0)}:{\cal{F}}_{t_0}\to {\cal{F}}_{t_0}$ such that}
\begin{equation}
\label{zz1}
\zeta_{\, (t,t_{0})}\cdot \hat{\Omega}(\phi,\,\cdot\,)=U^{\dagger}_{(t,t_0)}\hat{\Omega}(\phi,\,\cdot\,)\,U_{(t,t_0)}.
\end{equation}
{If such operators exist, it also means that, within the Heisenberg picture, the evolution expressed by Eqs. \eqref{ac-prime-ops-diff-fock} is unitary, i.e.}
\begin{equation}
\label{transf-annihi-op}
U_{(t,t_0)}\,\hat{a}_{t_0}(\bar{\eta}_{t_0})\,U^{\dagger}_{(t,t_0)}=\hat{a}_{t_0}\big(\overline{{A}_{(t,t_{0})}\eta_{t}}\big)-\hat{a}^{\dagger}_{t_0}\big(\overline{{B}_{(t,t_{0})}\eta_{t}}\big),
\end{equation}
\begin{equation}
\label{transf--crea-op}
U_{(t,t_0)}\,\hat{a}^{\dagger}_{t_0}(\chi_{t_0})\,U^{\dagger}_{(t,t_0)}=\hat{a}^{\dagger}_{t_0}\big(A_{(t,t_{0})}\chi_{t}\big)-\hat{a}_{t_0}\big(B_{(t,t_{0})}\chi_{t}).
\end{equation}
{It follows from the discussion above that the unitary operators $U_{(t,t_0)}$ exist, i.e. the classical dynamics dictated by $T_{(t,t_{0})}$ is unitarily implementable in the $J$-Fock representation,}  if and only if $B_{(t,t_{0})}$ satisfies the Hilbert-Schmidt condition (\ref{HS-condition}) for all $t$. This {is tantamount to requiring} that $(J-J_{t})$ be {Hilbert-Schmidt} on ${\cal{H}}_{t_0}$ for all $t$, as we have seen. {Another way to formulate this condition is to say that} the antilinear part of $T_{(t,t_{0})}$ {must} be {Hilbert-Schmidt} on ${\cal{H}}_{t_0}$ for all $t$ ({indeed}, a symplectic transformation $R$ is unitarily implementable on a Fock space ${\cal{F}}_{J}$ if and only if its antilinear {part} with respect to the complex structure $J$, namely $R_{J}=(R+JRJ)/2$, {is} Hilbert-Schmidt {on} the one-particle space $\cal{H}$ defined by $J$ \cite{Shale,Hone-Rieck}).

{Turning to the more algebraic perspective, the $*$-automorphisms $\zeta_{\, (t,t_{0})}$ of 
the algebra $\cal{A}$ define $*$-automorphisms $\zeta'_{(t,t_0)}$ of the Weyl algebra $\cal{W}$ via $\zeta'\cdot W(\phi)=\exp[i\zeta\cdot \hat{\Omega}(\phi, \, \cdot\,)]$, i.e. $\zeta'_{\, (t,t_{0})}\cdot W(\phi)=W(T_{(t,t_{0})}\phi)$. A simple calculation shows that $\omega_{J_{t}}[W(\phi)]=\omega_{J}[\zeta_{\, (t,t_{0})}^{\prime\,-1}\cdot W(\phi)]$; that is to say, {the} time evolution of observables in the Heisenberg picture is represented by the inverse of {the} automorphisms $\zeta'_{\, (t,t_{0})}$, {related to} the inverse of $\zeta_{\, (t,t_{0})}$, of course ({for} details about symplectic transformations and automorphisms in the Weyl algebra see, for instance, {Ref.} \cite{Torre-Varadarajan-cqg}).  Again, the family of automorphisms $\zeta'_{\, (t,t_{0})}$ of the abstract Weyl algebra corresponds to unitary transformations in the $J$-Fock representation if and only if $B_{(t,t_{0})}$ is Hilbert-Schmidt for all $t$. Note also that  the relation $\omega_{J_t}=\omega_{J}\circ \zeta^{\prime\, -1}_{\, (t,t_{0})}$ between {the} algebraic states can be interpreted as the  time evolution of the ``initial'' algebraic state $\omega_{J}$. So, in the Schr\"{o}dinger picture, the issue of a unitary quantum dynamics becomes the question of whether or not the family of algebraic states $\omega_{J}\circ \zeta^{\prime\,-1}_{\, (t,t_{0})}$ provide unitary equivalent  representations of the (adjoint and) Weyl relations (\ref{Weyl-relation1}) {and} (\ref{Weyl-relation2}) [or, equivalently, of the CCRs (\ref{CCR-fund-op})].}

{Let us now focus in particular on the case} of a free scalar field propagating in a spatially compact spacetime. Because of spatial compactness, {every} $\psi\in{\cal{H}}_{t}$ and $\bar{\xi}\in\overline{\cal{H}}_{t}$ can be written as
\begin{equation}
\psi=\sum_{k}c_{k}u^{t}_{k}, \quad \bar{\xi}=\sum_{k}d_{k}\bar{u}^{t}_{k}.
\end{equation}
Here, $\{u^{t}_{k}\}$ and $\{\bar{u}^{t}_k\}$ are orthonormal bases with respect to the Hermitian inner product $\mu_{(H,J_{t})}(\, \cdot\,, \, \cdot\,){=}\Omega(\overline{J_{t}\, \cdot\, }, \, \cdot\,)$ for, respectively, ${\cal{H}}_{t}$ and $\overline{\cal{H}}_{t}$, whereas $c_{k}$ and $d_{k}$ are complex constant numbers. Clearly, $\{(u^{t}_{k},\bar{u}^{t}_{k})\}$ is an orthonormal basis for the $J_t$-decomposition of $H$, namely ${\cal{H}}_{t}\oplus\overline{\cal{H}}_{t}$. As before, let us denote by $\langle\,\cdot\, ,\, \cdot\, \rangle_{{\cal{H}}_{t}}$ and $\langle\,\cdot\, ,\, \cdot\, \rangle_{\overline{\cal{H}}_{t}}$ the restriction of the Hermitian inner product $\mu_{(H,J_{t})}$ to, respectively, ${\cal{H}}_{t}$ and $\overline{\cal{H}}_{t}$. The  projection of $u^{t}_{m}\in {\cal{H}}_{t}$ onto $u^{t_0}_{k}\in {\cal{H}}_{t_0}$ gives the vector $\langle u^{t_0}_{k} , u^{t}_{m}\rangle_{{\cal{H}}_{t_0}}u^{t_0}_{k}$. Similarly, the projection of $u^{t}_{m}$ onto $\bar{u}^{t_0}_{k}\in \overline{\cal{H}}_{t_0}$, gives the vector $\langle \bar{u}^{t_0}_{k} , u^{t}_{m}\rangle_{\overline{\cal{H}}_{t_0}}\bar{u}^{t_0}_{k}$. Hence, we have that
\begin{equation}
\label{AB-proj}
A_{(t,t_{0})}u^{t}_{m}=\sum_{k}A_{km}(t,t_0)u^{t_0}_{k}, \quad B_{(t,t_{0})}u^{t}_{m}=\sum_{k}\bar{B}_{km}(t,t_0)\bar{u}^{t_0}_{k},\end{equation}
with
\begin{equation}
\label{AB-coeff}
A_{km}(t,t_0)=\langle u^{t_0}_{k} , u^{t}_{m}\rangle_{{\cal{H}}_{t_0}}, \quad \bar{B}_{km}(t,t_0)=\langle \bar{u}^{t_0}_{k} , u^{t}_{m}\rangle_{\overline{\cal{H}}_{t_0}}.
\end{equation}
By performing an analogous calculation, one gets $C_{(t,t_{0})}u^{t_0}_{m}$ and $D_{(t,t_0)}u^{t_0}_{m}$ [{in} fact, we can obtain them by simply switching the $t$ and $t_0$ parameters in {Eqs.} (\ref{AB-proj}) {and} (\ref{AB-coeff})]. Since $C_{(t,t_{0})}=A^{\dagger}_{(t,t_{0})}$ and $D_{(t,t_{0})}=-\bar{B}^{\dagger}_{(t,t_0)}$ [{see Eq.} (\ref{Bogo-ops-relations-3})], we thus get that
\begin{equation}
\label{AB-dagg-proj}
A^{\dagger}_{(t,t_{0})}u^{t_0}_{m}=\sum_{k}\bar{A}_{mk}(t,t_0)u^{t}_{k}, \quad B^{\dagger}_{(t,t_{0})}\bar{u}^{t_0}_{m}=\sum_{k}B_{mk}(t,t_0)u^{t}_{k}
\end{equation}
{where we have used} 
\begin{equation}
\langle u^{t}_{k} , u^{t_0}_{m}\rangle_{{\cal{H}}_{t}}=\overline{\langle u^{t_0}_{m} , u^{t}_{k}\rangle_{{\cal{H}}_{t_0}}}=\bar{A}_{mk}(t,t_{0}), \quad\quad -\langle u^{t}_{k} , \bar{u}^{t_0}_{m}\rangle_{{\cal{H}}_{t}}=\overline{\langle \bar{u}^{t_0}_{m} , u^{t}_{k}\rangle_{\overline{\cal{H}}_{t_0}}}=B_{mk}(t,t_{0}).
\end{equation}
The first {relation} in {Eq.} (\ref{Bogo-ops-relations-1}), together with {Eqs.} (\ref{AB-proj}) and (\ref{AB-dagg-proj}),  implies that the Bogoliubov coefficients $A_{km}(t,t_{0})$ and $\bar{B}_{km}(t,t_{0})$ satisfy
\begin{equation}
\label{Bogo-coeff-rel}
\sum_{k}(A_{km}\bar{A}_{kn}-\bar{B}_{km}B_{kn})=\delta_{nm}.
\end{equation}

From $u^{t}_{m}=A_{(t,t_{0})}u^{t}_{m}+B_{(t,t_{0})}u^{t}_{m}$ and {Eq.} (\ref{AB-proj}), it follows that the bases $\{(u^{t_0}_{m},\bar{u}^{t_0}_{m})\}$ and $\{(u^{t}_{m},\bar{u}^{t}_{m})\}$ of $H$ are related by 
\begin{equation}
\label{Bogo-Transf-Dyn}
u^{t}_{m}=\sum_{k} \left(A_{km}(t,t_0)u^{t_0}_{k}+\bar{B}_{km}(t,t_0)\bar{u}^{t_0}_{k}\right),
\end{equation}
\begin{equation}
\label{Bogo-Transf-Dyn-cc}
\bar{u}^{t}_{m}=\sum_{k}\left(\bar{A}_{km}(t,t_0)\bar{u}^{t_0}_{k}+B_{km}(t,t_0)u^{t_0}_{k}\right),
\end{equation}
where $A_{km}(t,t_{0})$ and $\bar{B}_{km}(t,t_{0})$ are given by {Eq.} (\ref{AB-coeff}), and satisfy the {relation} (\ref{Bogo-coeff-rel}). {Equations} (\ref{Bogo-Transf-Dyn}) and (\ref{Bogo-Transf-Dyn-cc}) are the Bogoliubov transformations between basis vectors.

Let us consider the expansion of the field in terms of the basis modes $\{(u^{t_0}_{k},\bar{u}^{t_0}_{k})\}$ associated with the initial complex structure $J$, 
\begin{equation}
\label{field-decom-u}
\phi=\sum_{k}(a^{0}_{k}u^{t_0}_{k}+\bar{a}^{0}_{k}\bar{u}^{t_0}_{k}).
\end{equation}
It is straightforward to check that the annihilation and creation-like observables $a_{t_{0}}(\bar{u}_{k}^{t_0})$ and $\bar{a}_{t_{0}}(u_{k}^{t_0})$ evaluated at $\phi\in S$ give $a_{t_{0}}(\bar{u}_{k}^{t_0})[\phi]=a^{0}_{k}$ and $\bar{a}_{t_{0}}(u_{k}^{t_0})[\phi]=\bar{a}^{0}_{k}$. That is, $a_{t_{0}}(\bar{u}_{k}^{t_0})$ and $\bar{a}_{t_{0}}(u_{k}^{t_0})$ can be viewed {as} coordinate functions{\footnote{The space $S$ can be identified with the space of coefficients $\{(a^{0}_{k},\bar{a}^{0}_{k})\}$.}} on $S$, so {that} we can write $\hat{a}_{t_0}(\bar{u}_{k}^{t_0})=\hat{a}^{0}_{k}$ and $\hat{a}^{\dagger}_{t_{0}}(u_{k}^{t_0})=\hat{a}^{0\, \dagger}_{k}$. From {Eq.} (\ref{CCR-annihi-cre}) we get that $\hat{a}^{0}_{k}$ and $\hat{a}^{0\, \dagger}_{k}$ satisfy the standard CCRs $[\hat{a}^{0}_{k},\hat{a}^{0\, \dagger}_{m}]=\hat{{\rm{I}}}\delta_{km}$. The Fock space of quantum states ${\cal{F}}_{t_0}$ is generated by repeatedly applying the creation operators $\hat{a}^{0\, \dagger}_{m}$ on $\vert 0 \rangle$, the state annihilated by all $\hat{a}^{0}_{k}$. Employing {Eq.} (\ref{transf-annihi-op}) {on} the basis modes $u^{t}_{m}$ (i.e. $\eta_{t}=u^{t}_{m}$) and using {Eq.} (\ref{AB-proj}), we get that  {-- if it turns out to be unitary --} the ``evolution''  of $\hat{a}^{0}_{m}$ from $t$ to $t_0$ would be given by
\begin{equation}
U_{(t,t_0)}\,\hat{a}^{0}_{m}\,U^{\dagger}_{(t,t_0)}=\sum_{k}\left(\bar{A}_{km}(t,t_0)\,\hat{a}^{0}_{k}-B_{km}(t,t_0)\,\hat{a}^{0\, \dagger}_{k}\right).
\end{equation}
The time evolution from the initial time  $t_{0}$ to an arbitrary final time $t$ {is obtained  simply by interchanging $t$ with $t_{0}$} in the above equation. Note that $\bar{A}_{km}(t_0,t)=A_{mk}(t,t_{0})$ and that $B_{km}(t_0,t)=-B_{mk}(t,t_{0})$. Thus, the evolution of the annihilation operator $\hat{a}^{0}_{m}$ associated with $u^{t_0}_{k}$, from $t_0$ to $t$, would be given by
\begin{equation}
\label{evol-annihi-op}
U^{\dagger}_{(t,t_0)}\,\hat{a}^{0}_{m}\,U_{(t,t_0)}=\sum_{k}\left(A_{mk}(t,t_0)\,\hat{a}^{0}_{k}+B_{mk}(t,t_0)\, \hat{a}^{0\, \dagger}_{k}\right).
\end{equation}
Analogously, we obtain that the evolution of the creation operator $\hat{a}^{\dagger}_{m}$, from $t_0$ to $t$, would be dictated by the unitary transformation
\begin{equation}
\label{evol-crea-op}
U^{\dagger}_{(t,t_0)}\,\hat{a}^{0\, \dagger}_{m}\,U_{(t,t_0)}=\sum_{k}\left(\bar{A}_{mk}(t,t_0)\,\hat{a}^{0\, \dagger}_{k}+\bar{B}_{mk}(t,t_0)\, \hat{a}^{0}_{k}\right).
\end{equation}

A direct calculation shows that the {unitarity condition, i.e. the Hilbert-Schimdt  condition on $B_{(t,t_{0})}$,} turns out to be the requirement that the Bogoliubov coefficients $B_{km}(t,t_{0})$ be square summable, 
\begin{equation}
\label{sqs-general}
\sum_{km} \vert B_{km}(t,t_{0})\vert^{2}<\infty ,\: \forall t.
\end{equation}
It is worth remarking that unitarity (or not) of $U_{(t,t_{0})}$ is a basis-independent issue. Indeed, given any other orthonormal basis $\{\tilde{u}^{t}_{k}\}$ in ${\cal{H}}_{t}$, it is not  difficult to see that $\sum_{km}\vert \tilde{B}_{km}(t,t_{0})\vert^{2}$ is equal to $\sum_{km}\vert B_{km}(t,t_{0})\vert^{2}$, i.e. the result of ${\rm{tr}}\big(B_{(t,t_{0})}^{\dagger}B_{(t,t_{0})}\big)$ does not depend on the specific choice of basis considered to perform the calculation.

By using the isomorphism $I_{t_0}$ between the linear spaces $S$ and $\Gamma$, {one can obtain the counterpart of the above quantization in the canonical approach. The} configuration and momentum of the field $\phi$, expanded in the positive and negative frequency mode solutions associated with $J$ [{see Eq.} (\ref{field-decom-u})], are given by
\begin{equation}
\label{confmom}
\varphi=\sum_{k}(a^{0}_{k}g_{k}+\bar{a}^{0}_{k}\bar{g}_{k}),\quad
\pi=\sum_{k}(a^{0}_{k}f_{k}+\bar{a}^{0}_{k}\bar{f}_{k}),
\end{equation}
where $g_{k}=u^{t_0}_{k}\vert_{t_0}$ and $f_{k}=\sqrt{h}L_{n}u^{t_0}_{k}\vert_{t_0}$. In terms of the {complex structure induced on $\Gamma$, i.e.} $j_{t_0}=I_{t_0}JI_{t_0}^{-1}$, the annihilation and creation-like variables read $a^{0}_{k}=\Omega(j_{t_0}(\bar{g}_{k},\bar{f}_{k}),(\varphi,\pi))$ and $\bar{a}^{0}_{k}=\Omega(j_{t_0}(g_{k},f_{k}),(\varphi,\pi))$, respectively. 
The promotion of these variables to quantum operators corresponds to the annihilation and creation operators (\ref{anni-crea-L2}) in the Schr\"{o}dinger representation, with label $\gamma^{+}_{k}=(g_{k},f_{k})$. The time evolved operators of annihilation and creation, $\hat{a}_{k}(t)=\mathscr{U}^{\dagger}_{(t,t_0)}\hat{a}^{0}_{k}\mathscr{U}_{(t,t_0)}$ and $\hat{a}^{\dagger}_{k}(t)=\mathscr{U}^{\dagger}_{(t,t_0)}\hat{a}^{0\,\dagger}_{k}\mathscr{U}_{(t,t_0)}$, are respectively given by the right-hand side of {Eqs.} (\ref{evol-annihi-op}) and (\ref{evol-crea-op}), that define the mapping $\mathscr{U}_{(t,t_0)}$ in the current representation.

{Let us conclude with the following  remark concerning unitarity}.  It follows from {Eqs.} (\ref{unit-between-fock-sp}), (\ref{ac-prime-ops-diff-fock}), (\ref{transf-annihi-op}), and (\ref{transf--crea-op}) that if $U$ is a unitary map, then so is $\cal{U}$ (and vice versa). For unitary $U$, we have in particular that
\begin{equation}
\label{consist-rel}
U\, \left(\hat{a}_{t_{0}}^{\dagger}(\eta_{t_0})-\hat{a}_{t_0}(\bar{\eta}_{t_0})\right)\,U^{-1}={\cal{U}}^{-1}\, \left(\hat{a}_{t}^{\dagger}(\eta_{t})-\hat{a}_{t}(\bar{\eta}_{t})\right)\,{\cal{U}}.
\end{equation}
Let us now suppose, without any {further  assumptions}, that {Eq.} (\ref{consist-rel}) is satisfied. Then, a calculation along the lines of {Ref.} \cite{Asht-Agu} shows that
\begin{equation}
\label{non-necess-unit}
-2i\,U\,\hat{a}_{t_{0}}(\bar{\eta}_{t_0})\,U^{-1}={\cal{U}}^{-1}\, \left(\hat{a}_{t}^{\dagger}[(J_{t}-TJT^{-1})\eta_{t}]+\hat{a}_{t}[(J_{t}+TJT^{-1})\bar{\eta}_{t}]\right)\,{\cal{U}}.
\end{equation}
Let $\tilde{U}$ be the composition ${\cal{U}}U:{\cal{F}}_{t_{0}}\to {\cal{F}}_{t}$. Thus, we obtain from {Eq.} (\ref{non-necess-unit}) that
\begin{equation}
\label{a-op-interms-others}
-2i\,\hat{a}_{t_{0}}(\bar{\eta}_{t_0})={\tilde{U}}^{-1}\, \left(\hat{a}_{t}^{\dagger}[(J_{t}-T
JT^{-1})\eta_{t}]+\hat{a}_{t}[(J_{t}+TJT^{-1})\bar{\eta}_{t}]\right)\,{\tilde{U}}.
\end{equation}
By applying {Eq.} (\ref{a-op-interms-others}) to the vacuum state $\vert 0\rangle$ of ${\cal{F}}_{t_0}$, we get that the state $\vert \Psi_{t}\rangle={\tilde{U}}\vert 0\rangle$ in ${\cal{F}}_{t}$ must satisfy the relationship
\begin{equation}
\label{cond-on-psi-t}
\hat{a}_{t}\left[(J_{t}+TJT^{-1})\bar{\eta}_{t}\right]\vert \Psi_{t}\rangle=\hat{a}_{t}^{\dagger}\left[(TJT^{-1}-J_{t})\eta_{t}\right]\vert \Psi_{t}\rangle.
\end{equation}
{Therefore}, the maps $\tilde{U}$ are unitary mappings if and only if $(TJT^{-1}-J_{t})$ is {Hilbert-Schimdt} on ${\cal{H}}_{t}$. However, since $J_t$ is precisely the complex structure resulting from evolving $J$ in time, we have that the {Hilbert-Schimdt} condition is {\emph{trivially}} satisfied and, therefore, $\tilde{U}$ is {always} a unitary map {for all $t_{0}$ and $t$}. Note, {nonetheless}, that unitarity of $\tilde{U}$ does not imply that $U$ ({nor} $\cal{U}$) must be necessarily unitary. 

{Let} us consider the above condition (\ref{cond-on-psi-t}) with $J_t$ replaced {with} some $J'_{t}\neq TJT^{-1}$. {Then,} unitarity of $\tilde{U}$ means that complex structures $J'_t$ differing from $TJT^{-1}$ can be consistently considered at time $t$ only if $(J'_{t}-TJT^{-1})$ is a {Hilbert-Schmidt} operator. {More specifically, from the} unitarity of $\tilde{U}$, it follows that $\langle 0\vert W(\phi) \vert0\rangle_{J}=\langle \Psi'_{t}\vert W'(T\phi)\vert \Psi'_{t} \rangle_{J'_t}$, where $W'(T\phi)=\tilde{U}W(\phi)\tilde{U}^{-1}$ and $\vert \Psi'_{t}\rangle =\tilde{U}\vert 0 \rangle$ is a normalizable state satisfying {Eq.} (\ref{cond-on-psi-t}). Since the expectation value of $W(\phi)$ at final time is given by  $\langle W(\phi)\rangle_{TJT^{-1}}=\langle W(T^{-1}\phi)\rangle_{J}$ (see for instance {Ref.} \cite{Torre-Varadarajan-cqg}), {we} have that $\langle W(\phi)\rangle_{TJT^{-1}}=\langle \Psi'_{t}\vert W'(\phi)\vert \Psi'_{t} \rangle_{J'_t}$, which certainly holds only if $(J'_{t}-TJT^{-1})$ is {Hilbert-Schmidt}\footnote{{The consistency condition that $(J'_{t}-TJT^{-1})$ be Hilbert-Schmidt was introduced and considered in Refs. \cite{Asht-Agu, Much:2018ehc} (within the canonical space approach) as a general condition  of unitary evolution.}} either on ${\cal{H}}_{J'_{t}}$ or {on} ${\cal{H}}_{TJT^{-1}}$ . For a thorough discussion on quantum unitary dynamics in cosmological scenarios see {Ref.} \cite{Cortez:2015mja}. 

\subsection{The scalar field with time dependent mass}
\label{subsec:KG-time-mass}

As we pointed out in Sec. \ref{subsec:CS-FLRW}, the $0$-spin boson field $\phi$ propagating in a spatially compact {FLRW} spacetime can be treated, after the time dependent scaling $\psi=a\phi$, as a free scalar field with time dependent mass (or, equivalently, as a scalar field subject to a time dependent potential) propagating in a static background, obeying the equation of motion (\ref{eq-mot-scaled-field}). Here, we {will} consider the same class of system, but adding {also} the case {of a background with} one-dimensional {spatial sections} with the topology of a circle. Besides, the  time dependent function $s(t)$ in the potential $V(\psi)=s(t)\psi^{2}/2$ will be considered (except for very mild conditions that will be specified below) as a general {real} function. Let us remark that for non-negative $s(t)$, the function can be interpreted as a squared time dependent mass. 

More concretely, we consider here a real scalar field $\psi$ governed by the equation
\begin{equation}
\label{eq-mot-scaled-field-ii}
\ddot{\psi}-\Delta \psi+s(t)\psi=0,
\end{equation}
in a static background
\begin{equation}
\label{static-metric}
{g_{\alpha\beta}dx^{\alpha}dx^{\beta}=-dt^{2} + h_{ab} dx^a dx^b,}
\end{equation}
where $h_{ab}$ is the standard Riemannian metric of {a spatial manifold $\Sigma$ that we will allow to be} either a circle $S^1$, a three-sphere $S^{3}$, or a three-dimensional torus $T^{3}$. {Besides,} $\Delta$ is the {LB} operator associated {to} $h_{ab}$. According to our general discussion in Sec. \ref{subsec:classical-sft}, the canonical phase space is the real linear space $\Gamma=\{(\varphi,\pi)\vert \varphi,\pi \in C^{\infty}(\Sigma)\}$ equipped with the standard symplectic structure (\ref{sympl-struc-KG-field}).\footnote{With respect to Eq. \eqref{can-eq-mot-scaled-f}, we now rename $\tilde{\varphi}\rightarrow \varphi$ and $\tilde{\pi}\rightarrow\pi$ to simplify our notation.} The covariant phase space is the linear space $S$ of smooth solutions to {Eq.} (\ref{eq-mot-scaled-field-ii}) arising from initial data on $\Gamma$, $\varphi=\psi\vert_{t_0}$ and $\pi=\sqrt{h}\dot{\psi}\vert_{t_0}$, equipped with the symplectic structure (\ref{sympl-struc-cov}) {(with the identification} $\Sigma_{t_0}\approx S^{1}$, $\Sigma_{t_0}\approx S^{3}$, or $\Sigma_{t_0}\approx T^{3}${)}. {We recall that} $t_0$ stands for {the} fixed (but) arbitrary initial reference time. {The} PB between the canonically conjugate variables of configuration and momentum {are} given by $\{\varphi(x),\pi(x')\}=\delta(x-x')$, where $x$ {denotes abstractly the coordinates of a point on} $\Sigma$. 

Scalar functions on $\Sigma$ can be expanded in terms of harmonics, {i.e.} in terms of solutions of the eigenvalue equation for the LB operator on $\Sigma$: $-\Delta X_{\mathbf{n}}=\omega^{2}_{n}X_{\mathbf{n}}$, where (1) $\omega^{2}_{n}=n^2$ for $S^{1}$, with $n\in \mathbb{Z}$, (2) $\omega^{2}_{n}=n(n+2)$ for $S^{3}$, with $n\in \mathbb{N}$, and (3) $\omega^{2}_{n}=\vec{n}\cdot \vec{n}$ for $T^{3}$, with $\vec{n}=(n_{1},n_{2},n_{3})$ and $n_{i}\in \mathbb{Z}$ ($i=1,2,3$). The eigenfunctions $X_{\mathbf{n}}$ {can be chosen as} the complex exponential functions $\exp(inx)/(2\pi)^{1/2}$ and $\exp(i\vec{n}\cdot \vec{x})/(2\pi)^{3/2}$ for the $S^1$ and the $T^{3}$ cases
 [${\mathbf{n}}$ denotes the integer $n$ and the {triple} $\vec{n}=(n_{1},n_{2},n_{3})$, respectively], whereas for the $S^3$ case $X_{\mathbf{n}}$ stands for the (hyper)spherical harmonics{\footnote{For a description of the harmonics in non-vanishing spatial curvature see, for instance, Ref. \cite{Harrison-harmonics}.}} $Q_{n\ell m}(x)$ on $S^{3}$ [{here} ${\mathbf{n}}$ denotes collectively the {set of indices {$(n,\ell,m)$}}, with $n\in \mathbb{N}$, $0\leq \ell \leq n$, and $-\ell\leq m \leq \ell$]. The functions $X_{\mathbf{n}}$ are orthonormal with respect to the $L^{2}$-product on $\Sigma$, namely $(X_{\mathbf{n}},X_{\mathbf{m}})=\delta_{\mathbf{nm}}$, where $(X_{\mathbf{n}},X_{\mathbf{m}})=\int \bar{X}_{\mathbf{n}}X_{\mathbf{m}}\sqrt{h}d^{3}x$. {The} configuration and momentum of the field can be expressed as 
\begin{equation}
\label{decomp-FLRW-S3-T3}
\varphi(x)=\sum_{\mathbf{n}}\varphi_{\mathbf{n}}X_{\mathbf{n}}(x),\quad \pi(x)=\sqrt{h}\sum_{\mathbf{n}}\pi_{\mathbf{n}}X_{\mathbf{n}}(x),
\end{equation}
where $\varphi_{\mathbf{n}}$ and $\pi_{\mathbf{n}}$ are the complex Fourier coefficients of the expansion in the complete set $\{X_{\mathbf{n}}(x)\}$. Since the field is a real one, these Fourier coefficients satisfy the following reality conditions:{\footnote{The reality conditions are obtained by using {that} $\varphi_{\mathbf{n}}=\int_{\Sigma}\sqrt{h}\varphi\bar{X}_{\mathbf{n}}$ and $\pi_{\mathbf{n}}=\int_{\Sigma}\pi\bar{X}_{\mathbf{n}}$, that the configuration $\varphi$ and the momentum $\pi$ of the field are real functions, and by employing the specific relationship between $X_{\mathbf{n}}$ and its complex conjugate $\bar{X}_{\mathbf{n}}$:  $\bar{X}_{k}=X_{-k}$ on $S^1$, $\bar{X}_{\vec{k}}=X_{-\vec{k}}$ on $T^3$, and $\bar{X}_{n\ell m}=(-1)^{m}X_{n\ell -m}$ on $S^3$.}} $\bar{\eta}_{k}=\eta_{-k}$ for the circle case, $\bar{\eta}_{\vec{k}}=\eta_{-\vec{k}}$ for the three-torus case, and $\bar{\eta}_{n\ell m}=(-1)^{m}\eta_{n\ell -m}$ for the three-sphere case. Here, $\eta_{\mathbf{n}}={(\varphi_{\mathbf{n}},\pi_{\mathbf{n}})}$.

{Incorporating the time dependence in our field, and recalling that it is real, we can decompose it in a Fourier expansion of the form
\begin{equation}
\label{zz3}
\psi(t,x)=\sum_{\mathbf{n}}\left(\xi_{\mathbf{n}}(t)X_{\mathbf{n}}(x)+{\rm{c.c.}}\right),
\end{equation}
where the functions of time $\xi_{\mathbf{n}}$ are solutions to the second-order differential equations 
\begin{equation}
\label{eq-of-mot-Tn}
\ddot{f}=-(\omega^{2}_{n}+s)f.
\end{equation}}
{This equation follows from the field equation \eqref{eq-mot-scaled-field-ii} when the spatial part is evaluated in the harmonic $X_{\mathbf{n}}$. We note that the equation \eqref{eq-of-mot-Tn} is real. Therefore if $\xi_{\mathbf{n}}$ provides a solution, so does its complex conjugate $\bar\xi_{\mathbf{n}}$. The relation between the functions $\xi_{\mathbf{n}}$ and the coefficients $\eta_{\mathbf{n}}$ above depend on the complex conjugation properties of the eigenfunctions $X_{\mathbf{n}}$ of the LB operator. For instance, in the $S^3$ case we get that the Fourier coefficients of the configuration field are given by $\varphi_{\mathbf{n}}= \xi_{\mathbf{n}}(t_0)+\bar\xi_{\mathbf{-n}}(t_0)$, where $t_0$ is the initial time. On the other hand, it is worth remarking that the dynamical equation \eqref{eq-of-mot-Tn}  depends exclusively on the eigenvalue of the LB operator, $-\omega_n^2$, rather than on the label of the harmonic, $\mathbf{n}$. As a consequence, except for the dependence on $\mathbf{n}$ that the initial conditions determined by  $\eta_{\mathbf{n}}$ may impose at $t_0$, the functions $\xi_{\mathbf{n}}(t)$ vary only with the value of $\omega_n^2$. Indicating the dependence on this eigenvalue with a subscript $n$, we can then rewrite the field \eqref{zz3} in the following manner:
\begin{equation}
\label{zz4}
\psi(t,x)=\sum_{\mathbf{n}}\left(a_{\mathbf{n}}T_n(t) X_{\mathbf{n}}(x)+{\rm{c.c.}}\right).
\end{equation}
This field decomposition respects the symmetries of the field equations. Here, $a_{\mathbf{n}}$ is a set of arbitrary complex constants, and the functions $T_n$  are conveniently normalized solutions to Eq. \eqref{eq-of-mot-Tn} (as we explain below). The subscript $n$ can be chosen to correspond to the absolute value of the harmonic label $n$ for the case of the circle, to the Euclidean norm of $\vec{n}$ for the three-torus, and to the first index in the set $\mathbf{n}\equiv\{n,\ell,m\}$ for the three-sphere}.

{Most important for the quantization it is the fact that, given that Eq. \eqref{eq-of-mot-Tn} is real and of second-order, as we have commented, we can choose the complex solution $T_n$ so that $\bar{T}_n$ is an independent solution. In this way, we obtain a splitting of the space of solutions between ``positive and negative'' frequency modes, namely $\psi_{\mathbf{n}}(t,x)=T_{n}(t){X}_{\mathbf{n}}(x)$ and $\bar{\psi}_{\mathbf{n}}(t,x)=\bar{T}_{n}(t)\bar{{X}}_{\mathbf{n}}(x)$. According to the discussion in Sec. \ref{sec:quant-syst}, there is an associated complex structure $J$, with corresponding annihilation-like variables given by $a_{\mathbf{n}}=\Omega(J\bar{\psi}_{\mathbf{n}}, \psi)$ and creation-like variables provided by their complex conjugates. From the orthonormality of the field solutions $\psi_{\mathbf{n}}$ with respect to the Hermitian inner product $\Omega(\overline{J\,\cdot\,},\,\cdot\,)$, and of the {eigenfunctions} $X_{\mathbf{n}}$ with respect to the $L^2$-product on $\Sigma$, it follows that 
\begin{equation}
\label{zz2}
T_{n}\dot{\bar{T}}_{n}-\bar{T}_{n}\dot{T}_{n}=i, \ \forall n.
\end{equation}}

In this perspective, $J$ is  ultimately defined by the functions $T_n$, and thus the choice of a complex structure is equivalent to the choice of a set of complex solutions $T_n$ to Eq. \eqref{eq-of-mot-Tn} for every of the LB eigenspaces, satisfying Eq. \eqref{zz2} (see Ref. \cite{LR} for details). The field decomposition \eqref{zz4} is fully adapted to this perspective, and immediately gives an expression for the field operator in the Heisenberg picture, when the constants $a_{\mathbf{n}}$ and $\bar a_{\mathbf{n}}$ are replaced  with annihilation and creation operators, acting on the Hilbert space constructed from $J$, as described in Sec. \ref{sec:quant-syst}.

Making contact with the canonical perspective, and since the solutions $T_n$ are determined by
the initial conditions, we have that, in terms of the Cauchy data at the initial reference time $t_0$, the annihilation-like variables are given by
\begin{equation}
\label{S3T3-annhi-var-gen-T}
a_{\mathbf{n}}=\Omega\big(j_{t_0}(\bar{g}_{\mathbf{n}},\bar{f}_{\mathbf{n}}\big), (\varphi,\pi));\quad g_{\mathbf{n}}(x)=T_{n}(t_{0})X_{\mathbf{n}}(x),\quad f_{\mathbf{n}}=\sqrt{h}\dot{T}_{n}(t_{0})X_{\mathbf{n}}(x),
\end{equation}
where $j_{t_0}$ is the initial complex structure on $\Gamma$ induced by $J$, namely  $j_{t_0}=I_{t_0}JI^{-1}_{t_{0}}$.

Clearly, the group of {spatial} symmetries of the metric $h_{ab}$, say $G_{h}$, is a group of symmetries of $\Delta$ and, consequently, of the equation of motion (\ref{eq-mot-scaled-field-ii}). {In the same spirit of demanding invariance under such symmetries that we adopted above, we note} that a $G_h$-invariant complex structure {does} not only allow for a unitary implementation of {the spatial isometries corresponding to $h_{ab}$, but furthermore} for a $G_h$-invariant representation of the CCRs. A simple obvious choice, that ensures a $G_h$-invariant Fock representation, is the massless free field representation provided by the complex structure (\ref{CS-scaled-field}),
\begin{equation}
\label{CS-scaled-field-ii}
j_{0}(\varphi,\pi)=  \big( -[-h\Delta]^{-1/2}\pi,[-h\Delta]^{1/2}\varphi\big).
\end{equation}
This complex structure defines the annihilation-like variables{\footnote{{We exclude in principle the zero mode. This does not affect} the field properties of the system. {Besides, the} zero mode can be quantized separately as a mechanical system.}} $a^{0}_{\mathbf{n}}=[\omega_{n}\varphi_{\mathbf{n}}+i\pi_{\mathbf{n}}]/\sqrt{2\omega_{n}}$.  {Notice that $j_0$ is determined by the initial conditions $T_{n}(t_0)=1/\sqrt{2\omega_{n}}$ and $\dot{T}_{n}(t_{0})=-i\sqrt{\omega_{n}/2}$} for {Eq.} \eqref{eq-of-mot-Tn} ({these can be checked to provide valid initial conditions; see, for instance, Refs.} \cite{CMSV,CMV-PRD81}). {Indeed, substituting the complex structure (\ref{CS-scaled-field-ii}) into Eq. (\ref{S3T3-annhi-var-gen-T})} we get
\begin{equation}
\label{S3-T3-annhi-var-inter-step}
\Omega\big(j_{0}(\bar{g}_{\mathbf{n}},\bar{f}_{\mathbf{n}}),(\varphi,\pi)\big)=\omega_{n}\bar{T}_{n}(t_{0})\varphi_{\mathbf{n}}+\omega^{-1}_{n}\dot{\bar{T}}_{n}(t_{0})\pi_{\mathbf{n}},
\end{equation}
where we have used that $(-\Delta)^{\pm1/2}X_{\mathbf{n}}={\omega^{\pm1}_n}X_{\mathbf{n}}$, as well as the orthonormality of {the eigenfunctions} $X_{\mathbf{n}}$ with respect to the $L^{2}$-product on $\Sigma$. {Hence}, for $T_{n}(t_0)=1/\sqrt{2\omega_{n}}$ and $\dot{T}_{n}(t_{0})=-i\sqrt{\omega_{n}/2}$, the annihilation-like variables $a_{\mathbf{n}}$ {reproduce in fact} the massless free annihilation-like variables $a^{0}_{\mathbf{n}}$. 

{By constructing the $j_0$-Fock representation, we get the annihilation and creation operators $\hat{a}^{0}(\overline{\gamma^{+}})$ and $\hat{a}^{0\,\dagger}(\gamma^{+})$ 
defined by $j_0$ [see {Eq.} (\ref{anni-crea-L2})], where $\gamma^{+}=(\gamma-ij_{0}\gamma)/2$ and $\gamma\in \Gamma$. Then, introducing a Fourier decomposition, we obtain $\hat{a}^{0}_{\mathbf{n}}$ and $\hat{a}^{0\, \dagger}_{\mathbf{n}}$, that are nothing but the result of promoting the observables (\ref{S3T3-annhi-var-gen-T}) and their complex conjugates [with $j_{t_0}=j_0$, $T_{n}(t_0)=1/\sqrt{2\omega_{n}}$, and $\dot{T}_{n}(t_{0})=-i\sqrt{\omega_{n}/2}$] to quantum operators. Explicitly, $\hat{a}^{0}_{\mathbf{n}}$ and $\hat{a}^{0\, \dagger}_{\mathbf{n}}$ are given by
\begin{equation}
\label{anni-crea-j0-X}
\hat{a}^{0}_{\mathbf{n}}=\frac{1}{\sqrt{2\omega_{n}}}(\omega_{n}\hat{\varphi}_{\mathbf{n}}+i\hat{\pi}_{\mathbf{n}}), \quad 
\hat{a}^{0\,\dagger}_{\mathbf{n}}=\frac{1}{\sqrt{2\omega_{n}}}(\omega_{n}\hat{\varphi}^{\dagger}_{\mathbf{n}}-i\hat{\pi}^{\dagger}_{\mathbf{n}}),
\end{equation}
where the action of $\hat{\varphi}_{\mathbf{n}}$ and $\hat{\pi}_{\mathbf{n}}$ {on} the Hilbert space is obtained from the Fourier decomposition of Eqs. (\ref{the-gaussian-measure}) and (\ref{the-conf-mom-rep}), for the complex structure {characterized by} $\mathfrak{a}=\mathfrak{c}=0$, $\mathfrak{b}=[-h\Delta]^{-1/2}$, and $\mathfrak{d}=-[-h\Delta]^{1/2}$.} 

{The} time evolution of $\hat{a}^{0}_{\mathbf{n}}$ {is} dictated by a Bogoliubov transformation of the form (\ref{evol-annihi-op}), {namely}
\begin{equation}
\label{evol-annihi-var-gene}
\hat{a}_{\mathbf{m}}(t)=\sum_{\mathbf{{n}}}\left(A_{\mathbf{m{n}}}(t,t_0)\,\hat{a}^{0}_{\mathbf{{n}}}+B_{\mathbf{m{n}}}(t,t_0)\, \hat{a}^{0\, \dagger}_{\mathbf{m}}\right),
\end{equation}
with Bogoliubov coefficients 
\begin{equation}
\label{bogo-decomp-x}
A_{\mathbf{m{n}}}(t,t_0)=\alpha_{{n}}(t,t_{0})\delta_{{\mathbf{m{n}}}},\quad B_{\mathbf{m{n}}}(t,t_0)=\beta_{{n}}(t,t_{0})\,(X_{\mathbf{m}},\bar{X}_{\mathbf{{n}}}),
\end{equation}
where $(X_{\mathbf{m}},\bar{X}_{\mathbf{{n}}})$ corresponds to $(X_{m},\bar{X}_{{n}})=\delta_{-m{n}}$ for the circle case, $(X_{\vec{m}},\bar{X}_{\vec{{n}}})=\delta_{-\vec{m}\vec{{n}}}$ for the three-torus, {and} $(X_{n\ell m},\bar{X}_{n'\ell'm'})=(-1)^{m}\delta_{nn'}\delta_{\ell \ell'}\delta_{-mm'}$ for the three-sphere. {The} coefficients $\alpha_{{n}}(t,t_{0})$ and $\beta_{{n}}(t,t_{0})$ are given by 
\begin{equation}
\label{alph-bet-coeff-example}
\alpha_{{n}}(t,t_{0})=i[\bar{T}_{{n}}(t_{0})\dot{T}_{{n}}(t)-T_{{n}}(t)\dot{\bar{T}}_{{n}}(t_0)],\quad \beta_{{n}}(t,t_{0})=i[\bar{T}_{{n}}(t_{0})\dot{\bar{T}}_{{n}}(t)-\bar{T}_{{n}}(t)\dot{\bar{T}}_{{n}}(t_0)].
\end{equation}
A straightforward calculation shows that $\vert\alpha_{{n}}(t,t_{0})\vert^{2}-\vert\beta_{{n}}(t,t_{0})\vert^{2}=1$. 

{Notice that, instead of considering {a} Fourier decomposition with respect to the set of complex functions $\{X_{\mathbf{n}}\}$, one {can decide to} perform the expansion of the configuration and momentum of the field in terms of explicitly real functions. {In that case}, the Fourier coefficients {become} real as well, and no reality conditions need be  imposed. Then, the corresponding} Bogoliubov coefficients {turn out to be} of the form 
\begin{equation}
\label{Bogo-coeff-expan-real-var}
A_{\mathbf{m{n}}}(t,t_0)=\alpha_{{n}}(t,t_{0})\delta_{{\mathbf{m{n}}}},\quad B_{\mathbf{m{n}}}(t,t_0)=\beta_{{n}}(t,t_{0})\delta_{{\mathbf{m{n}}}}.
\end{equation}
For instance, in the $S^1$ case, the {non-zero modes of the} system can be described in terms {of} real canonically conjugate variables $(q_{{n}},\tilde{q}_{{n}},p_{{n}},\tilde{p}_{{n}})$ {related to} the complex variables $(\varphi_{{n}},\pi_{{n}})$ by 
$q_{{n}}=\sqrt{2}{\rm{Re}}(\varphi_{{n}})$,  $\tilde{q}_{{n}}=\sqrt{2}{\rm{Im}}(\varphi_{{n}})$, $p_{{n}}=\sqrt{2}{\rm{Re}}(\pi_{{n}})$, and 
$\tilde{p}_{{n}}={\sqrt{2}}{\rm{Im}}(\pi_{{n}})$, restricting now $n$ to be a positive integer, $n\in \mathbb{N}^+$. Since $\bar{\eta}_{{n}}=\eta_{-{n}}$, for $\eta=\varphi,\pi$, the operators associated with $\sqrt{2}{\rm{Re}}(\eta_{{n}})$ and $\sqrt{2}{\rm{Im}}(\eta_{{n}})$ are respectively given by the self-adjoint operators $(\hat{\eta}_{{n}}+\hat{\eta}_{-{n}})/\sqrt{2}$ and $(\hat{\eta}_{{n}}-\hat{\eta}_{-{n}})/(i\sqrt{2})$. We use this {canonical transformation} to recast the Schr\"{o}dinger representation, with fundamental operators $\hat{\varphi}_{{n}}$ and $\hat{\pi}_{{n}}$, in terms of the self-adjoint operators $\hat{q}_{{n}}$, $\hat{p}_{{n}}$, $\hat{\tilde{q}}_{{n}}$, and $\hat{\tilde{p}}_{{n}}$. The $j_0$-annihilation operators are given by
\begin{equation}
\label{anni-op-r-example}
\hat{b}_{{n}}=\frac{1}{\sqrt{2{n}} }({n}\hat{q}_{{n}}+i\hat{p}_{{n}}), \quad 
\hat{\tilde{b}}_{{n}}= \frac{1}{\sqrt{2{n}}}({n}\hat{\tilde{q}}_{{n}}+i\hat{\tilde{p}}_{{n}}),
\end{equation}
and the creation operators are {provided} by their adjoints, $\hat{b}^{\dagger}_{{n}}$ and $\hat{\tilde{b}}^{\dagger}_{{n}}$.

{According} to {Eqs.} (\ref{evol-annihi-var-gene}) and (\ref{bogo-decomp-x}), {the} time evolution of $\hat{a}^{0}_{m}=(\omega_{m}\hat{\varphi}_{m}+i\hat{\pi}_{m})/\sqrt{2\omega_{m}}$ {(for all $m\in\mathbb{Z}$ and with $\omega_m= \vert m \vert$)} is given by the Bogoliubov transformation 
\begin{equation}
\label{ann-evol-flat-CS}
\hat{a}_{m}(t)=\alpha_{m}(t,t_{0})\hat{a}^{0}_{m}+\beta_{m}(t,t_{0})\hat{a}^{0\,\dagger}_{-m}.
\end{equation}
The expression for $\hat{a}^{\dagger}_{m}(t)$ is obtained by taking the adjoint of {Eq.} (\ref{ann-evol-flat-CS}). It is not difficult to see that $\hat{a}^{0}_{m}$ and $\hat{a}^{0\, \dagger}_{-m}$ are related {to} the {annihilation and creation operators} $(\hat{b}_{m},\hat{\tilde{b}}_{m})$ and $(\hat{b}^{\dagger}_{m},\hat{\tilde{b}}^{\dagger}_{m})$ by
\begin{equation}
\label{anni-crea-rel-exampl}
\hat{a}^{0}_{m}=\frac{1}{\sqrt{2}}(\hat{b}_{m}+ i\hat{\tilde{b}}_{m}),\quad \hat{a}^{0\, \dagger}_{-m}=\frac{1}{\sqrt{2}}(\hat{b}^{\dagger}_{m}+ i\hat{\tilde{b}}^{\dagger}_{m}),
\end{equation}
{for all positive integers $m$ (for negative $m$, $\hat{a}^{0}_{m}$ and $\hat{a}^{0\, \dagger}_{-m}$ can be found from the adjoint of the above relations).} 
{The} time evolution of $\hat{b}_{m}$ and $\hat{\tilde{b}}_{m}$ {can be determined} by substituting {Eq.} (\ref{anni-crea-rel-exampl}) into {Eq.} (\ref{ann-evol-flat-CS}),
\begin{equation}
\label{ann-evol-flat-CS-exampl}
\hat{b}_{m}(t)=\alpha_{m}(t,t_{0})\hat{b}_{m}+\beta_{m}(t,t_{0})\hat{b}^{\dagger}_{m},\quad
\hat{\tilde{b}}_{m}(t)=\alpha_{m}(t,t_{0})\hat{\tilde{b}}_{m}+\beta_{m}(t,t_{0})\hat{\tilde{b}}^{\dagger}_{m}.
\end{equation}
Thus, in contrast with the expression (\ref{ann-evol-flat-CS}), where the modes $m$ and $-m$ are coupled, the evolution of the annihilation operators $\hat{b}_{m}$ and $\hat{\tilde{b}}_{m}$ is fully decoupled from the rest. Although the Bogoliubov coefficients $A_{m{n}}(t,t_0)$ are the same ones as for $\hat{a}_{m}(t)$, the coefficients of the antilinear part are now given by  
\begin{equation}
\label{BC-flat-CS-real-qp}
B_{m{n}}(t,t_0)=\beta_{{n}}(t,t_{0})\delta_{m{n}}.
\end{equation}

The $j_0$-Fock representation is, by construction, invariant under the isometries of the spatial manifold $\Sigma$ ($S^{1}$, $S^{3}$, or $T^{3}$, depending on the case). This property, however, turns out {not to be} enough to guarantee the uniqueness of the {representation.} Indeed, there are infinitely many complex structures which do not belong to the equivalence class of $j_0$ {but are symmetry} invariant. Thus, one has to look for extra requirements in order to select a unique preferred {Fock} representation. {A natural requirement is to demand that the classical symplectic transformations associated with the time evolution are properly quantized as unitary operators ({note} that it is pointless to ask for time invariance, since time-translation symmetry is broken by the non-stationarity of the system).} {So, we restrict our attention to} invariant Fock representations {that} admit, in addition, a unitary implementation of the dynamics. 

{In} summary, we require that (1) the vacuum state be invariant under the {(spatial) isometries of the} manifold $\Sigma$, and that (2) the dynamics dictated by the field equation (\ref{eq-mot-scaled-field-ii}) be unitarily {implementable}. {Remarkably}, the $j_0$-Fock representation is the unique (up to unitary equivalence) symmetry invariant representation of the CCRs where a unitary implementation of {the} time evolution is available (i.e. {it} is the unique Fock representation satisfying the criteria of invariance and {of} unitarity). Furthermore, {no canonical transformations (except for trivial ones) can} lead to a field description from which an invariant Fock representation admitting a unitary implementation of the dynamics could be defined; i.e. the $\psi$-description is unique, up to trivial canonical transformations. The removal of the ambiguities in the quantization of scalar fields with time dependent mass is {discussed in  Ref.} \cite{CMSV} for the case of the circle topology, {in Refs.} \cite{CMV-PRD81,CMOV-JCAP10,CMOV-PRD83} for the case of the three-sphere topology, and {in Refs.} \cite{CCMMV-T3,MenaMarugan:2013tba,CCMMV-T3-ii} for the case of the three-torus topology. In all of these cases, it is sufficient (but not necessary) that the function $s(t)$ possesses a second derivative which is integrable in every compact subinterval of the time domain. 

The rest of this work {is} an overview of these uniqueness results {obtained}  within the context of cosmology; the arena in which the studies were motivated and developed. {We} will present a compilation of the uniqueness {results} attained for the quantization of Gowdy models, and {of} (test) scalar fields propagating in {FLRW} spacetimes, de Sitter spacetimes, and anisotropic Bianchi I universes.

\section{Uniqueness of the description for quantum Gowdy cosmologies}
\label{sec:Gowdy}

{Symmetry reduced models in general relativity  have {received} great attention,  as a suitable arena where {one can} study issues that {may} play a central role in a future} quantum theory of gravity. On the one hand, this allows us to discuss with specific examples conceptual and technical {problems} that arise when {one tries} to conciliate gravity and quantum mechanics. On the other hand, these reduced models are usually of physical relevance in cosmology or in astrophysical situations. The so-called midisuperspace models \cite{Torre-MidiSupMod,FernandoBarbero:2010qy}, coming from reductions that keep an infinite number of degrees of freedom, are {especially relevant from the technical {point of view}, since they capture at least some of} the field complexity of general relativity. Among this kind of models, the simplest model with applications in cosmology is the family of Gowdy spacetimes \cite{Gowdy} with linear polarization and {with} the spatial topology of a three-torus, $T^3$. This is the model on which  {we will focus our discussion in this section, in order to illustrate the results obtained in recent years about the uniqueness of the quantization of fields in cosmological scenarios}. The removal of {quantization} ambiguities for the rest {of} Gowdy spacetimes, namely the $S^{1}\times S^{2}$ and $S^{3}$ models, {can be} addressed in a very similar {manner.} After gauge fixing, the Gowdy $T^{3}$ model is classically equivalent to $2+1$ gravity coupled to an axially symmetric scalar field \cite{Pierri}. So, by quantizing this field in the fictitious $(2+1)$ background, one obtains a quantum description of the Gowdy cosmology. It is precisely in this way that a quantization for the polarized Gowdy model was introduced {in Ref.} \cite{Pierri}. However, the {proposed} {quantization suffered from a serious drawback: the classical dynamics was not implementable as  unitary transformations} \cite{CCH-GT3}.  By a convenient scaling of the basic field, rendering the fictitious spacetime as a static background, an alternate quantization {that solves the commented} problem was {constructed in Refs.} \cite{CCM-PRD73-i,CCM-PRD73-ii}, providing {in this way a} consistent quantum description of an inhomogeneous cosmological model. It has been shown that the attained quantization is, in fact, the unique Fock representation of the CCRs {which is} invariant under the gauge group that remains {in} the model after gauge fixing, and {such that it admits} a unitary implementation of {the} time evolution \cite{CCMV-gt3-uniq,CMV-PRD75}. Remarkably, {these} {criteria} of invariance and {of} unitarity {proved successful not only to handle} the issue of {the} uniqueness of the representation of the CCRs,  but {in addition} singled out {in a unique way} the field parametrization {that must be adopted for the consistent description of the model} \cite{CMV-PRD75}. Let us briefly discuss {this Gowdy} model, its quantization, and the {mentioned} uniqueness {result.}

After a {partial} gauge fixing, which removes all but a homogeneous constraint, the line element of the linearly polarized Gowdy $T^3$ cosmological spacetimes can be written as \cite{CM}
\begin{equation}
ds^{2}=e^{{\tilde{\gamma}}-\phi/\sqrt{p}}\left(-dt^{2}+d\theta^{2}\right)+e^{-\phi/\sqrt{p}}t^{2}p^{2}d\sigma^{2}+e^{\phi/\sqrt{p}}d\delta^{2}.
\end{equation}
Here, $(\partial/\partial \sigma)^{a}$ and $(\partial/\partial \delta)^{a}$ are the two Killing vector fields of the model, $p$ {denotes} a strictly positive homogeneous constant of motion {that is present in the system}, and the function $\phi$ depends on the time coordinate $t >0$ and the angle $\theta\in S^{1}$. Except for its zero mode, containing a degree of freedom $Q$ that is conjugate to $P{=}\ln p$, the field ${\tilde{\gamma}}$ {is totally} determined {by} $p$, $\phi$, and its canonical momentum $P_{\phi}$ \cite{CM}. The phase space $\tilde{\Gamma}$ of the midisuperspace model is coordinatized by the canonical pairs $(Q,P)$ and $(\phi,P_{\phi})$. {As we have said}, there is still a global constraint on the system, $C_{0}=\oint d\theta P_{\phi}\phi' /\sqrt{2\pi}$, {that generates} translations in $S^1$, so that physical states are restricted to lie in a submanifold of $\tilde{\Gamma}$. {Here, the prime denotes the derivative with respect to $\theta$}. {The} time evolution is dictated by the (explicitly time {dependent}) reduced Hamiltonian $H=\oint d\theta [P^{2}_{\phi}+t^{2}\phi^{\prime\,2}]/(2t)$. The independence of the Hamiltonian on the ``point particle'' degrees of freedom $(Q,P)$ implies that these are
 constants of motion {(in consonance with our previous comment about the constancy of $p$)}. Thus, a non-trivial evolution may only take place in the field sector $\Gamma=\{(\phi,P_{\phi})\}$. Since the homogeneous degrees of freedom $(Q,P)$ are non-dynamical and can be separately quantized by using standard methods of quantum mechanics, we will obviate them {in the following} and concentrate our discussion on the field sector.

The reduced Hamiltonian gives the field equations $P_{\phi}=t\dot{\phi}$ and $\dot{P}_{\phi}=t\phi''$. So, the dynamics of $\phi$ is governed by
\begin{equation}
\label{phi-eq-gt3}
\ddot{\phi}+\frac{1}{t}\dot{\phi}-\phi''=0.
\end{equation}
Hence, the field sector of the model can be viewed as that of an axisymmetric, massless, free scalar field $\phi$ propagating in a $(2+1)$-dimensional flat background $ds^{2}_{0}=-dt^{2}+d\theta^{2}+t^{2}d\sigma^{2}$. The smooth real solutions to {Eq.} (\ref{phi-eq-gt3}) have the form $\phi(t,\theta)=\sum_{n\in \mathbb{Z}}[b_{n}f_{n}(t)\exp(in\theta)+{\rm{c.c.}}]$, where $b_{n}$ are (complex) constants and
\begin{equation}
f_{n}(t)=\frac{H_{0}(\vert n \vert t)}{\sqrt{8}}\quad n\neq 0,\quad  f_{0}(t)=\frac{1-i\ln t}{\sqrt{4\pi}}.
\end{equation}
{In this formula,} $H_0$ {is} the zeroth-order Hankel functions of the second kind \cite{Abramowitz}. Neglecting the zero mode, {the} time evolution from initial time $t_0$ to the final time $t$ is dictated by a Bogoliubov transformation $b_{k}(t)=\tilde{\alpha}_{k}(t,t_{0})b_{k}+\tilde{\beta}_{k}(t,t_{0})\bar{b}_{-k}$, {the} antilinear part {of which} is given by
\begin{equation}
\tilde{\beta}_{k}(t,t_{0})=\frac{i\pi\vert k\vert}{4}\left[t_{0}\bar{H}_{1}(\vert k \vert t_{0})\bar{H}_{0}(\vert k \vert t)-t\bar{H}_{0}(\vert k \vert t_{0})\bar{H}_{1}(\vert k \vert t)\right],
\end{equation}
where $H_1$ is the first-order Hankel function of the second kind \cite{Abramowitz}. Since the sequence $\{\tilde{\beta}_{k}(t,t_{0})\}$ fails to be square summable {for all $t$} {and} $t_{0}$ \cite{CCH-GT3}, {the time evolution is not implementable} as a unitary transformation on the kinematical Fock space constructed from the complex structure $\tilde{J}$, defined by the {families of} positive and negative frequency {modes} $\tilde{u}^{+}_{n}(t,\theta)=f_{n}(t)\exp(in\theta)$ and $\tilde{u}^{-}_{n}(t,\theta)=\overline{u^{+}_{n}(t,\theta)}$. Moreover, the failure of a unitary implementation of {the} time evolution {persists} on the physical Hilbert space of quantum states \cite{Torre-GT3}, defined by the kernel of  $\hat{C}_0$, the quantum counterpart of the remaining constraint $C_{0}$.

Note that, by scaling the field by $\sqrt{t}$, one gets scaled solutions $u^{+}_{n}(t,\theta)=\sqrt{t}f_{n}(t)\exp(in\theta)$. From the asymptotic behavior of {the} Hankel {function} $H_{0}(\vert n \vert t)$ in the regime of large wave numbers $\vert n \vert$, it follows that $u^{+}_{n}(t,\theta)$ {behaves} in the ultraviolet limit as the standard modes of a free scalar field in a two-dimensional flat background ({equivalent} to a three-dimensional formulation with axial {symmetry), namely} $\exp(-i\pi/4)u^{+}_{n}(t,\theta)\approx \exp(-i\vert n\vert t+in\theta)/\sqrt{4\pi\vert n \vert t}$. This, together with the freedom available to redefine the classical phase space through time dependent canonical transformations, motivates {the consideration of} the canonical transformation{\footnote{The change $\psi=\sqrt{t}\phi$ was discussed for the first time in Ref. \cite{Berger}, but just within the study of the WKB regime.}}
\begin{equation}
\label{ct-gowdy-t3}
\psi=\sqrt{t}\phi,\quad P_{\psi}=\frac{1}{\sqrt{t}}\left(P_{\phi}+\frac{\phi}{2}\right),
\end{equation}
{in order} to arrive at a unitary theory. The {contribution} to $P_{\psi}$ {that is linear in $\phi$} is chosen so that the new Hamiltonian does not contain products of the field with its momentum \cite{CCM-PRD73-i}: $H_{\psi}=\oint d\theta[P^{2}_{\psi}+\psi^{\prime\,2}+\psi^{2}/(4t^{2})]/2$. Note that $H_{\psi}$ corresponds to the Hamiltonian of an axially symmetric massless scalar field, subject to a time varying potential $V(\psi)=\psi^{2}/8t^{2}$, propagating in a fictitious, $(2+1)$-dimensional static background $d\bar{s}^{2}_{0}=-dt^{2}+d\theta^{2}+d\sigma^{2}$. By introducing the complex structure $j_0$ [{see Eq.} (\ref{CS-scaled-field-ii})] {on} phase space $\Gamma=\{(\psi,P_{\psi})\}$, it can be shown that the resulting $j_0$-Fock representation admits a unitary implementation of the dynamics  \cite{CCM-PRD73-i,CCM-PRD73-ii}. Specifically, as we have seen in Sec. \ref{subsec:KG-time-mass}, the annihilation and creation-like variables defined by $j_0$ at an arbitrary (but fixed) initial reference time $t_0$ {are}
\begin{equation}
\label{jo-annihi-crea-var}
a_{n}=\frac{1}{\sqrt{2\vert n\vert}}(\vert n \vert \psi_{n}+iP^{n}_{\psi})\quad {\mbox{and}} \quad\bar{a}_{-n}=\frac{1}{\sqrt{2\vert n\vert}}(\vert n \vert \psi_{n}-iP^{n}_{\psi}),
\end{equation}
{where $\psi_{n}$ and $P^{n}_{\psi}$ are the Fourier coefficients of the field $\psi$ and its momentum, respectively, i.e. $\psi=\sum_{n\in \mathbb{Z}}\psi_{n}\exp(in\theta)/\sqrt{2\pi}$ and $P_{\psi}=\sum_{n\in \mathbb{Z} }P^{n}_{\psi}\exp(in\theta)/\sqrt{2\pi}$, that satisfy the canonical relations $\{\psi_{n},P_{\psi}^{-m}\}=\delta_{n}^{m}$.}
{The variables \eqref{jo-annihi-crea-var}} evolve in time according to $a_{k}(t)=\alpha_{k}(t,t_{0})a_{k}+\beta_{k}(t,t_{0})\bar{a}_{-k}$. Since the time dependent mass function is $s(t)=1/(4t^{2})$ here, the Bogoliubov coefficients {turn out to be} given by \cite{CCM-PRD73-i,CCM-PRD73-ii} 
\begin{equation}
\alpha_{n}(t,t_{0})=c(x_{n})\bar{c}(x^{0}_{n})-d(x_{n})\bar{d}(x^{0}_{n}),\quad
\beta_{n}(t,t_{0})=d(x_{n})c(x^{0}_{n})-d(x^{0}_{n})c(x_{n}),
\end{equation}
with $x_{n}{=}\vert n \vert t$, $x^{0}_{n}{=}\vert n \vert t_{0}$, and 
\begin{equation}
d(x)=\sqrt{\frac{\pi x}{8}}\left[\left(1+\frac{i}{2x}\right)\bar{H}_{0}(x)-i\bar{H}_{1}(x)\right],\quad c(x)=\sqrt{\frac{\pi x}{2}}H_{0}(x)-d^{*}(x).
\end{equation}

It is not difficult to see that $\vert c(x)\vert^{2}-\vert d(x)\vert^{2}=1$. Note {also} that $\beta_{n}=\beta_{-n}$, so that we can {consider just} the sequence $\{\beta_{n}(t,t_{0})\}$ with $n\in \mathbb{N}^{+}$. From {the} asymptotic expansions of the Hankel functions {for large arguments} \cite{Abramowitz}, {one gets} \cite{CCM-PRD73-i} $\vert d(x_{n})\vert^{2}=1/(4x_{n})^{4}+o(1/x_{n}^{5})$. {Then we see that}, given any fixed $T>0$, the sequence $\{d(\vert n \vert T)\}$ is square summable. The square summability of $\{d(x_{n})\}$ and $\{d(x^{0}_{n})\}$, together with the relationship $\vert c\vert^{2}=1+\vert d \vert^{2}$, imply that $\{\beta_{n}(t,t_{0})\}$ is square summable for all positive $t_{0}$ and $t$ \cite{CCM-PRD73-i,CCM-PRD73-ii}. Hence, {the} time evolution turns out to be unitarily implementable on the kinematical Hilbert space ${\cal{F}}_{0}$ of the $j_0$-Fock representation. Moreover, a direct calculation shows that the evolution leaves invariant the {constraint that remains} on the system,
\begin{equation}
\label{q-const}
\hat{C}_{0}=\sum_{n=1}^{\infty}n(\hat{a}^{\dagger}_{n}\hat{a}_{n}-\hat{a}^{\dagger}_{-n}\hat{a}_{-n}),
\end{equation}
{that implements quantum mechanically} the condition that the total ($\theta$-)momentum of the field $\psi$ vanish. This invariance  ensures that the dynamics is unitarily implementable not just on ${\cal{F}}_{0}$, but also on the physical Hilbert space ${\cal{F}}_{\rm{phys}}$, defined as the kernel of the constraint (\ref{q-const}). 

{Although} we have specified a Fock representation {that satisfies} the requirements of invariance and of unitary {implementability} of the dynamics, namely the $j_0$-Fock representation, {it might exist} another invariant complex structure $j$ {that admits a unitary dynamics but, however, is not equivalent to $j_0$.} Remarkably, {this cannot be} the case, {as} it is shown in {Ref.} \cite{CCMV-gt3-uniq}. {Let us emphasize this result: any other} compatible invariant complex structure $j$ {that allows} for a unitary implementation of {the} time evolution turns out to be in the equivalence class of $j_0$. Indeed, a thorough analysis \cite{CCMV-gt3-uniq} establishes that every compatible invariant complex structure $j$ is related to $j_0$ by a symplectic transformation $K_{j}$ (i.e. $j=K_{j}j_{0}K^{-1}_{j}$) that is block diagonal, with $4\times 4$ blocks of the form
\begin{equation}
\label{j-jo-rel}
(K_{j})_{n}=\begin{pmatrix}
 ({\cal{K}}_{j})_{n} & {\mathbf{0}} \\
 {\mathbf{0}} & ({\cal{K}}_{j})_{n} \end{pmatrix}
,\quad  ({\cal{K}}_{j})_{n}=\begin{pmatrix}
 \kappa_{n} & \lambda_{n} \\
\bar{\lambda}_{n} &  \bar{\kappa}_{n} \end{pmatrix},
\end{equation}
where $\vert\kappa_{n}\vert^{2}-\vert\lambda_{n}\vert^{2}=1$. {Recall then} that, given a symplectic transformation $S$ and two complex structures, $j$ and $j_0$, related by another symplectic transformation $K_j$, {namely} $j=K_{j}j_{0}K^{-1}_{j}$, the antilinear part $(S+jSj)/2$ is Hilbert-Schmidt with respect to the inner product $\langle \, \cdot\, , \, \cdot \, \rangle_{j}$ [{see Eq.} (\ref{herm-inner-prod-svp})] if and only if the $j_0$-antilinear part of $K_{j}^{-1}SK_{j}$ is Hilbert-Schmidt with respect to  $\langle \, \cdot\, , \, \cdot \, \rangle_{j_{0}}$ (see, for instance, {Ref.} \cite{CCMV-gt3-uniq}). By applying this result, {with the relation between} complex structures {provided} by the symplectic transformation (\ref{j-jo-rel})] and {by} the symplectic transformation {that corresponds to} time evolution,
\begin{equation}
\label{evol-wrt-j}
(U)_{n}(t,t_{0})=\begin{pmatrix}
 {\cal{U}}_{n}(t,t_{0}) & {\mathbf{0}} \\
 {\mathbf{0}} & {\cal{U}}_{n}(t,t_{0})\end{pmatrix}
,\quad  {\cal{U}}_{n}(t,t_{0})=\begin{pmatrix}
 \alpha_{n}(t,t_{0}) & \beta_{n}(t,t_{0}) \\
\bar{\beta}_{n}(t,t_{0}) &  \bar{\alpha}_{n}(t,t_{0}) \end{pmatrix},
\end{equation} 
one arrives {at the conclusion} that the existence of a unitary implementation of the dynamics with respect to $j$ {amounts to the unitary implementation} of $U^{(j)}(t,t_{0})=K_{j}^{-1}U(t,t_{0})K_{j}$ with respect to $j_0$ for all possible values of $t_0$ and $t$. From {Eqs.} (\ref{j-jo-rel}) and (\ref{evol-wrt-j}), it is straightforward to see that the antilinear part of $U^{(j)}(t,t_{0})$ {is}
\begin{equation}
\beta^{(j)}_{n}(t,t_{0})=\bar{\kappa}^{2}_{n}\beta_{n}(t,t_{0})-\lambda^{2}_{n}\bar{\beta}_{n}(t,t_{0})+2i \bar{\kappa}_{n}\lambda_{n}{\rm{Im}}[\alpha_{n}(t,t_{0})].
\end{equation}
{A rigorous} analysis on the behavior of $\beta^{(j)}_{n}(t,t_{0})$ in the asymptotic regime demonstrates that the sequence $\{\beta^{(j)}_{n}(t,t_{0})\}$ is square summable (i.e., the $S^{1}$-invariant $j$-Fock quantization admits a unitary implementation of {the} time evolution) if and only if the sequence $\{\vert \lambda_{n}\vert^{2}\}$ is summable  \cite{CCMV-gt3-uniq}. Since the summability {of this sequence} is the condition for unitary equivalence of the  Fock representations determined by $j_0$ and $j$, we then conclude that (modulo unitary equivalence) there is {just} a unique {compatible,} invariant complex structure that permits the unitary implementation of the dynamics. {That is}, the $j_0$-Fock representation is  the unique (up to unitary mappings) $S^{1}$-invariant representation {which admits} a unitary implementation of the {dynamical transformations}. By employing the algebraic state defined by {this} $j_0$-Fock representation, {one can specify} a $S^{1}$-invariant, unitary functional representation {of the model} \cite{CCMV-GT3-SR}. 

{In the previous discussion, the statement of uniqueness was} circumscribed to the canonical description of phase space in terms of the fundamental field variables $(\psi,P_{\psi})$. However, 
{one is certainly allowed to consider {other} different variables, related e.g. by means of linear canonical transformations}. In fact, it is precisely this freedom {what} we have used to reformulate the system in terms of the $(\psi,P_{\psi})$-{variables} [see Eq. (\ref{ct-gowdy-t3})]. {Since  time dependent canonical transformations modify the dynamics, there is still then the possibility that invariant, unitary Fock representations {can} exist for a different set of basic, canonically conjugate field variables, {valid as well for} parameterizing the phase space. That is to say, it could} happen that the requirements of invariance and {of} unitarity will not suffice to remove the ambiguity in the choice of basic field variables (at least of a certain type, e.g. linear with respect to the original ones), forcing us to seek for additional ``judicious'' extra criteria in order to select a preferred set of fundamental variables. Fortunately, this is not the case. Indeed, the unique field description 
{for} which a $S^{1}$-invariant, unitary Fock representation can be specified is {precisely} the {$\psi$-description \cite{CMV-PRD75}}. 

{An analysis similar to the one that we have presented above has also been} performed to achieve a unique quantum description {of} the linearly-polarized Gowdy $S^{1}\times S^{2}$ and $S^{3}$ cosmological models \cite{BVV,CMV-GowS1S2S3}. {For that purpose, the result of uniqueness has been} extended to axisymmetric fields with a time dependent mass {equal to} $(1+\csc^{2}t)/4$ on $S^2${, case which describes the field sector} (after a {suitable scaling}) of the Gowdy models with the spatial topology of a three-handle and a three-sphere  \cite{BVV0,BVV,CMV-GowS1S2S3}. The uniqueness {proven} for the Gowdy models {has also been} generalized to scalar fields with arbitrary mass terms{\footnote{To be precise, it is required that the time dependent function $s(t)$ possess a first derivative {that} is integrable in every compact subinterval of the time domain.}} {on} $S^1$ (and naturally continued to axisymmetric fields on the two-sphere) \cite{CMSV}. Moreover, the {criteria} of invariance and {of} unitarity {have been} successfully extended {as well so as} to remove the ambiguities in the quantization of scalar fields with time dependent mass propagating in static backgrounds with the spatial topology of either a three-sphere \cite{CMV-PRD81,CMOV-JCAP10,CMOV-PRD83} or a three-torus \cite{CCMMV-T3,CCMMV-T3-ii}, thus providing a unique preferred Fock representation for test KG fields in {FLRW} and de Sitter spacetimes, as we will see in the following two sections.

\section{Scalar fields in FLRW spacetimes: invariance, unitarity, and uniqueness}
\label{sec:FLRW}

For a free scalar field $\phi$ with mass $m$ propagating in an expanding {FLRW} universe, the number density of created particles diverges, so that the Bogoliubov transformation dictating {the} time evolution turns out to be {non}-unitary {on} the Hilbert space of the quantum theory (see, for instance, {Refs.} \cite{non-impl-i,non-impl-i1,non-impl-ii,non-impl-ii1,non-impl-ii2}). However, for spatially compact {FLRW} spacetimes, {with} slices $\Sigma=S^3$ (closed {FLRW} universes) or $\Sigma=T^3$ (flat {FLRW} universes), a Fock representation with the properties of (i) invariance under the isometries of the spatial manifold $\Sigma$, {and} (ii)  a unitary implementation of the dynamics, can be specified in the scaled field description, $\psi=a\phi$. The Fock quantization is, in fact, the unique (up to unitary mappings) representation of the CCRs {that satisfies our criteria} of invariance and {of} unitarity. Moreover, the (scaled) field description $\psi$ is the unique {one} (up to trivial canonical transformations) {for} which an invariant Fock representation with unitary dynamics can be specified. In this section, we will 
overview these results about the uniqueness of the quantization of (test) KG fields in closed and (spatially compact) flat {FLRW} spacetimes. 

\subsection{Closed FLRW spacetimes}
\label{sec:Closed-FLRW}

Let us first consider the case of closed universes. As we have mentioned, after rescaling the field with the conformal factor we get a KG-field $\psi$, subject to a time varying potential $V(\psi)=s(t)\psi^{2}/2$, propagating in a globally hyperbolic, static $(3+1)$-{dimensional} background $(M\approx \mathbb{I}\times S^{3} ,g_{{\alpha\beta}})$, where $g_{{\alpha\beta}}$ is given by {Eq.} (\ref{static-metric}) with $t\in \mathbb{I}=\mathbb{R}^{+}$ and 
\begin{equation}
\label{S3-metric}
h_{ab}{dx^a dx^b}={d\chi^2+\sin^{2}(\chi)d\theta^2+\sin^{2}(\chi)\sin^{2}(\theta)d\sigma^2.}
\end{equation}
Here, $\sigma\in S^{1}$ and $\chi,\theta\in (0,\pi)$. The field $\psi$ satisfies the linear wave equation (\ref{eq-mot-scaled-field-ii}), with $\Delta$ being the LB operator on $S^{3}$ [here, we {will} assume that $s(t)$ is a sufficiently regular function, {with the specific conditions on it} given below]. Since the metric is $SO(4)$-invariant, so is the {LB operator}, and we thus have that the group of rotations $SO(4)$ is a group of symmetries of the field dynamics. {Hence, we will} look for a complex structure {that determines a} representation where both the dynamics and the group of $SO(4)$ symmetries {can be} unitarily implemented. {For this purpose}, we consider the complex structure (\ref{CS-scaled-field-ii}) with $h=\sin^{2}(\theta)\sin^{4}(\chi)$ (and $\Delta$ the LB operator on $S^{3}$). Since this complex structure provides a $SO(4)$-invariant representation, we just need to check whether or not {the time evolution is unitarily implementable. Indeed, the} dynamics turns out to {admit a unitary implementation} in the massless free field representation defined by $j_0$, {as we will now show}  ({for a detailed} proof, see {Ref.} \cite{CMV-PRD81}). 

Following our discussion in Sec. \ref{subsec:KG-time-mass}, let us adopt a description of the field in terms of harmonics. As it is known, the (hyper)spherical harmonics $X_{\mathbf{n}}=Q_{n\ell m}$ of order $(n, \ell, m)$ on $S^{3}$ are eigenfunctions of $\Delta$, with eigenvalues $-n(n+2)$. {They form} an orthonormal basis for the expansion of scalar functions on the three-sphere  (see, for instance, {Refs.} \cite{GS-HypHarm,Halli-Haw-PRD31,Jantzen-JMP,Lifshitz-Khala,Lifshitz-Khala1}). Here, $n\in\mathbb{N}$, $0\leq\ell\leq n$, and $-\ell\leq m\leq \ell$. The harmonics $Q_{n\ell m}$, normalized with respect to the $L^{2}$-product on $S^{3}$, read
\begin{equation}
\label{Hyp-SH}
Q_{n\ell m}(\chi,\theta,\sigma)=2^{\ell}(\ell!)\sqrt{\frac{2(n-\ell)!(n+1)}{\pi(n+\ell+1)!}}\sin^{\ell}(\chi)C^{(\ell+1)}_{n-\ell}[\cos(\chi)]Y_{\ell m}(\theta,\sigma),
\end{equation}
where $Y_{\ell m}$ {are} the spherical harmonics on $S^{2}$ and $C^{(\ell+1)}_{n-\ell}[\cos(\chi)]$ are the Gegenbauer polynomials \cite{Abramowitz,Gradshteyn}. The scalar harmonics $Q_{n\ell m}$ span an irreducible $(n+1)^{2}$-dimensional representation of $SO(4)$ for each fixed $n$. The behavior of the spherical harmonics under complex conjugation, $\bar{Y}_{\ell m}=(-1)^{m}Y_{\ell -m}$, is inherited by the scalar harmonics on $S^{3}$, $\bar{Q}_{n\ell m}=(-1)^{m}Q_{n\ell -m}$. In terms of the real basis {of} scalar harmonics, the field $\psi$ {can be} written as  
\begin{equation}
\label{s3frw-decomp}
\psi(t,x)=\sum_{n,\ell} q_{n\ell 0}Q_{n\ell 0}+\sqrt{2}\sum_{n,\ell, m>0} q_{n\ell m}{\rm{Re}}[Q_{n\ell m}] + \sqrt{2}\sum_{n,\ell, m>0} q_{n\ell -m}{\rm{Im}}[Q_{n\ell m}],
\end{equation}
where the coefficients $q_{n\ell m}$ are real functions of time only, {because} $\psi$ is a real field. From the field equation (\ref{eq-mot-scaled-field-ii}), {using} complex conjugation and {the} orthogonality properties of $Q_{n\ell m}$, it follows that all modes $q_{n\ell m}$ with the same $n$ satisfy the same equation of motion, {namely}
\begin{equation}
\ddot{q}_{n\ell m}+(\omega^{2}_{n}+s)q_{n\ell m}=0;\quad \omega^{2}_{n}{=}n(n+2).
\end{equation} 
{Thus}, the modes $q_{n\ell m}$ obey completely decoupled equations of motion, {that} depend only on $n$. Clearly, the configuration space of the theory is in one-to-one correspondence with the space of all real coefficients $\{q_{n\ell m}\}$. {We will denote it as} ${\cal{Q}}=\oplus_{n}{\cal{Q}}_{n}$, where ${\cal{Q}}_{n}$ is the $(n+1)^{2}$-dimensional linear space spanned by the configuration modes $q_{n\ell m}$ with the same label $n$. The {variables} canonically conjugate to the configurations $q_{n\ell m}$ are the momenta $p_{n\ell m}=\dot{q}_{n \ell m}$. From the basic PB $\{\varphi(x),\pi(x')\}=\delta(x-x')$ and the orthogonality of the scalar harmonics, one can see that
\begin{equation}
\{q_{n\ell m},p_{n'\ell' m'}\}=\delta_{nn'}\delta_{\ell \ell'}\delta_{mm'},\quad \{q_{n\ell m},q_{n'\ell' m'}\}=0,\quad \{p_{n\ell m},p_{n'\ell' m'}\}=0.
\end{equation}
{So, in} the canonically conjugate variables $q_{n\ell m}$ and $p_{n\ell m}$, the phase space $\Gamma$ can be decomposed as the direct sum $\Gamma=\oplus_{n}\Gamma_{n}$, with $\Gamma_{n}={\cal{Q}}_{n}\oplus {\cal{P}}_{n}$, where ${\cal{P}}_{n}$ is the linear space of dimension $(n+1)^{2}$ spanned by the momentum modes $p_{n\ell m}$. Both the configuration and momentum spaces ${\cal{Q}}_{n}$ and ${\cal{P}}_{n}$ carry an irreducible representation of $SO(4)$ of dimension $(n+1)^{2}$ which, in fact, is the same for the two spaces.

Let us introduce {now} the complex structure (\ref{CS-scaled-field-ii}). For all modes {with $n\neq 0$}, the real variables of configuration and momentum, $q_{n\ell m}$ and $p_{n\ell m}$, are related {to the annihilation and creation-like variables defined by $j_0$ as follows:}
\begin{equation}
\label{ac-vs-cm}
a_{n\ell m}=\frac{1}{\sqrt{2\omega_{n}}}(\omega_{n}q_{n\ell m}+ip_{n\ell m}),\quad \bar{a}_{n\ell m}=\frac{1}{\sqrt{2\omega_{n}}}(\omega_{n}q_{n\ell m}-ip_{n\ell m}).
\end{equation}
{For} the sake of simplicity in the presentation, we will drop the $n=0$ mode {in the rest of our discussion.} As a single decoupled mode from the rest of degrees of freedom, it can be quantized [at least for non-negative functions $s(t)$] by using the Schr\"{o}dinger representation of ordinary quantum mechanics {on the} Hilbert space $L^{2}(\mathbb{R},dq_{000})$.

We now construct the $j_0$-Fock representation of the CCRs. {As  discussed in Sec. \ref{sec:quant-syst}, the relevant Hilbert space is ${\mathscr{H}}_{j_0}$; that is, the space of square integrable complex functions with respect to the Gaussian measure (\ref{the-gaussian-measure}), with $\frak{b}^{-1}=(-h\Delta)^{1/2}$, on the infinite dimensional} linear space ${\cal{Q}}$. The configuration and momentum are represented as in {Eq.} (\ref{the-conf-mom-rep}), with $\frak{c}=0$. By performing the expansion in modes, one gets that the measure and the representation of the fundamental operators are 
\begin{equation}
d\varrho=\prod_{n,\ell ,m}\sqrt{\frac{\omega_{n}}{\pi}}\exp(-\omega_{n}q^{2}_{n\ell m})dq_{n\ell m},
\end{equation}
\begin{equation}
\label{z1}
\hat{q}_{n\ell m}\Psi=q_{n\ell m}\Psi,\quad \hat{p}_{n\ell m}=-i\frac{\partial}{\partial q_{n\ell m}}\Psi+i\omega_{n}q_{n\ell m}\Psi,
\end{equation}
with $\Psi\in {\mathscr{H}}_{j_0}$. The annihilation and creation operators are given in terms of the self-adjoint operators $\hat{q}_{n\ell m}$ and $\hat{p}_{n\ell m}$ by $\hat{a}_{n\ell m}=(\omega_{n}\hat{q}_{n\ell m}+i\hat{p}_{n\ell m})/\sqrt{2\omega_{n}}$ and {the adjoint of this definition for} $\hat{a}^{\dagger}_{n\ell m}$. The vacuum state $\Psi_{0}$ is the state annihilated by all {the operators} $\hat{a}_{n\ell m}$, namely the state {satisfying the condition} $\omega_{n}\hat{q}_{n\ell m}\Psi_{0}=-i\hat{p}_{n\ell m}\Psi_0$, which in turn implies that $\partial \Psi_{0}/\partial q_{n\ell m}=0$ (i.e., up to a constant phase, $\Psi_{0}$ is the unit constant function, since $d\rho$ is a probabilistic measure and $\Psi_0$ is normalized). Fock states are generated by repeatedly applying $\hat{a}^{\dagger}_{n\ell m}$ on $\Psi_0$. The {Hilbert space ${\mathscr{H}}_{j_0}$, together with the action (\ref{z1}) of {the} operators $\hat{q}_{n\ell m}$ and $\hat{p}_{n\ell m}$, or equivalently} with the {action of the} set of annihilation and creation operators $\{\hat{a}_{n\ell m},\hat{a}^{\dagger}_{n\ell m}\}$, constitute the $j_0$-Fock representation. 

In view of the decoupling between degrees of freedom, and since the dynamical equations are independent of $\ell$ and $m$, one can check that {the} time evolution of the annihilation and creation operators {is given} by a Bogoliubov transformation of the form{\footnote{Indeed, one can easily verify that the Bogoliubov coefficients have the form $A_{n\ell mn'\ell'm'}=\alpha_{n'}\delta_{nn'}\delta_{\ell \ell'}\delta_{mm'}$ and $B_{n\ell mn'\ell'm'}=\beta_{n'}\delta_{nn'}\delta_{\ell \ell'}\delta_{mm'}$ [{see Eq.} (\ref{Bogo-coeff-expan-real-var})].}}
\begin{equation}
\label{FLRW-S3-Bogo-transf}
\hat{a}_{n\ell m}(t)=\alpha_{n}(t,t_0)\hat{a}_{n\ell m}+\beta_{n}(t,t_0)\hat{a}^{\dagger}_{n\ell m},\quad \hat{a}^{\dagger}_{n\ell m}(t)=\bar{\alpha}_{n}(t,t_0)\hat{a}^{\dagger}_{n\ell m}+\bar{\beta}_{n}(t,t_0)\hat{a}_{n\ell m}.
\end{equation}
{The} Bogoliubov coefficients $\alpha_{n}$ and $\beta_{n}$ are {determined by Eq.} (\ref{alph-bet-coeff-example}),{where $T_{n}$ is related to the solutions of} the differential equation
\begin{equation}
\ddot{q}_{n}+(\omega^{2}_{n}+s)q_{n}=0
\end{equation}
{by $q_{n}(t)=A_{n}T_{n}(t)+\bar{A}_{n}\bar{T}_{n}(t)$.} Since $T_{n}(t_0)=1/\sqrt{2\omega_{n}}$ and $\dot{T}_{n}(t_{0})=-i\sqrt{\omega_{n}/2}$, we have that $q_{n}(t_{0})=(A_{n}+\bar{A}_{n})/\sqrt{2\omega_{n}}$ {and} $\dot{q}_{n}(t_{0})=i\sqrt{\omega_{n}/2}(\bar{A}_{n}-A_{n})$. 

{On the other hand,} writing $T_{n}(t)=\exp[\omega_{n}\Theta(t)]/\sqrt{2\omega_{n}}$, we get from {Eq.} (\ref{eq-of-mot-Tn}) that $\Theta_n$ {must} obey the equation 
\begin{equation}
\label{exp-fact-func}
\omega_{n}\ddot{\Theta}_{n}+\omega^{2}_{n}\dot{\Theta}^{2}_{n}+\omega_{n}^{2}+s=0.
\end{equation}
{The} initial conditions now read $\Theta_{n}(t_{0})=0$ and $\dot{\Theta}_{n}(t_{0})=-i$. The Bogoliubov coefficients $\alpha_n$ and $\beta_n$ are obtained by substituting $T_{n}(t)=\exp[\omega_{n}\Theta(t)]/\sqrt{2\omega_{n}}$ into {Eq.} (\ref{alph-bet-coeff-example}). {In} particular, we get that
\begin{equation}
\beta_{n}(t,t_{0})=\frac{1}{2}e^{\omega_{n}\bar{\Theta}_{n}(t)}\left[1+i\dot{\bar{\Theta}}_{n}(t)\right].
\end{equation}
{The} time evolution will be implemented as a unitary {transformation on} the Hilbert space {if and only if $\{\vert\beta_{n}\vert^{2}\}$ is summable, i.e.}
 \begin{equation}
 \label{u-impl-S3}
\sum_{n=1}^{\infty}\sum_{\ell =0}^{n}\sum_{m=-\ell}^{\ell}\vert \beta_{n}(t,t_0)\vert^{2}=\frac{1}{4}\sum_{n=1}^{\infty}g_{n}e^{2\omega_{n}{\rm{Re}}[\bar{\Theta}_{n}(t)]}\vert1+i\dot{\bar{\Theta}}_{n}(t)\vert^{2}<\infty,
\end{equation}
where $g_{n}=(n+1)^{2}$ is the degeneracy factor counting the number of degrees of freedom with the same dynamics. Introducing the function $R_{n}(t){=}i\omega_{n}[1-i\dot{\Theta}_{n}(t)]$ and using that $\Theta_{n}(t_{0})=0$, condition (\ref{u-impl-S3}) {can be} rewritten as follows:
\begin{equation}
\label{cond-sqs-S3}
\sum_{n=1}^{\infty}\left(\frac{g_{n}}{\omega^{2}_{n}}\right)e^{2\int_{t_0}^{t}{\rm{Re}}[R_{n}]}\vert R_{n}(t)\vert^{2}<\infty.
\end{equation}
From {Eq.} (\ref{exp-fact-func}) and the {mentioned initial condition on} $\dot{\Theta}_{n}$, it follows that the functions $R_n$ satisfy the first order differential equations 
\begin{equation}
\label{Riccati}
\dot{R}_{n}-2i\omega_{n}R_{n}+R^{2}_{n}+s=0,
\end{equation}
with initial {condition} $R_{n}(t_{0})=0$. Let us consider the functions 
\begin{equation}
\label{asympt-sol}
\bar{R}_{n}(t)=-e^{2i\omega_{n}t}\int_{t_0}^{t} e^{-2i\omega_{n}\tau} s(\tau)\,d\tau.
\end{equation}
By assuming that the derivative of $s(t)$ exists and is integrable in every closed interval $[t_{0},t]$, one can integrate by parts {Eq.} (\ref{asympt-sol}) and check that there is a function $C(t)$, independent of $n$, such that $\vert \bar{R}_{n}(t)\vert\leq C(t)/\omega_{n}$. Thus,  in the asymptotic regime (i.e., {for asymptotically} large $n$), $\bar{R}_{n}^{2}$ is negligible compared with $\omega_{n}\bar{R}_n$. Besides, since $\bar{R}_{n}(t)$ is {a solution to the equation} $\dot{R}_{n}-2i\omega_{n}R_{n}+s=0$ with $\bar{R}_{n}(t_{0})=0$, we then conclude that the functions $\bar{R}_{n}(t)$ can be taken (up to higher order corrections) as asymptotic solutions {to Eq. (\ref{Riccati}).} Hence, apart from subdominant terms, condition (\ref{cond-sqs-S3}) {amounts to} requiring that the sequence $\{C(t)/\omega_{n}\}$ be square summable, {something that} certainly holds. So, {the} time evolution {admits a unitary implementation} in the $j_{0}$-Fock representation.

So far, we have seen that there is a representation that satisfies the requirements of symmetry invariance [$SO(4)$-invariance] and of unitary implementability of the dynamics, namely the $j_0$-Fock representation. The next question to answer  is whether these two properties are enough to remove completely the inherent ambiguity in the representation of the CCRs or not.  Are there distinct (i.e. not unitarily equivalent) representations with the same  properties of invariance and unitarity? Remarkably, the answer is in the negative. Though there are infinitely many {inequivalent} $SO(4)$-invariant Fock representations, the requirement of a unitary {implementability} of the field dynamics singles out a unique {family} of {unitarily equivalent} representations {(specifically, equivalent to the $j_{0}$-Fock representation}). Indeed, as it is shown in {Ref.} \cite{CMV-PRD81}, the annihilation operators $\hat{a}^{(j)}_{n\ell m}$ defined by an arbitrary $SO(4)$-invariant complex structure $j$ are related {to} the $j_0$-annihilation and creation operators, $\hat{a}_{n\ell m}$ and $\hat{a}^{\dagger}_{n\ell m}$, by a Bogoliubov transformation $\hat{a}^{(j)}_{n\ell m}=\kappa_{n}\hat{a}_{n\ell m}+\lambda_{n}\hat{a}^{\dagger}_{n\ell m}$, where $\kappa_{n}$ and $\lambda_{n}$ are time independent complex coefficients satisfying $\vert \kappa_{n}\vert^{2}-\vert \lambda_{n}\vert^{2}=1$. The antilinear part of the time evolved {operator} $\hat{a}^{(j)}_{n\ell m}$ is given by 
\begin{equation}
\label{antilinear-ann-inv}
\beta^{(j)}_{n}(t,t_{0})=\kappa^{2}_{n}\beta_{n}(t.t_{0})-\lambda^{2}_{n}\bar{\beta}_{n}(t,t_{0})+2i\kappa_{n}\lambda_{n}{\rm{Im}}[\bar{\alpha}_{n}(t,t_{0})],
\end{equation}
where, as we have already seen, $\sqrt{g_{n}}\beta_{n}(t,t_{0})$ is square summable. {Then,} performing an analysis along the lines of {Ref.} \cite{CMV-PRD81}, one {can show that $\sqrt{g_{n}}\beta^{(j)}_{n}(t,t_{0})$ is square summable if and only if the sequence $\{\sqrt{g_{n}}\lambda_{n}\}$ is square summable, assuming} that the second derivative of the mass function $s(t)$ exists and is integrable in every compact subinterval of $\mathbb{I}$ \cite{CMV-PRD81}. Hence, the Fock representation constructed from the $SO(4)$-invariant complex structure $j$ will {allow to} implement the dynamics as a unitary mapping if and only if {$j$ is equivalent to $j_0$}. Thus, up to unitary transformations, the $j_0$-Fock representation is unique: {the} {criteria} of symmetry invariance and {of unitary implementability of the} time evolution {select} a unique preferred representation of the CCRs.

Two remarks about the $j_0$-Fock quantization are in order. 

(1) {Let us} recall that given a free scalar field $\phi$ propagating in a globally hyperbolic spacetime, a Hadamard representation of the CCRs can be specified \cite{Wald-book} by looking for a vacuum state {with a} two-point function $\langle \phi(x)\phi(y)+\phi(y)\phi(x)\rangle$ {that} has a short-distance behavior of the Hadamard type \cite{Hadamard}. Although the Hadamard criterion does not suffice to pick out a unique preferred quantization in general, it has been shown that, for the case of free scalar fields in spacetimes with compact Cauchy slices, all Hadamard vacua belong to the same class of unitarily equivalent states \cite{Wald-book}. Since this result applies to a free scalar field propagating in a closed {FLRW} spacetime, we have at our disposal two different criteria (the Hadamard approach, on the one hand, and the discussed {requirements} of invariance and unitarity, on the other hand) in order to select a unique preferred quantization of the scalar field. {One} may be wonder whether the unitary and the Hadamard quantizations are in conflict or not. The answer, as it is shown in {Ref.} \cite{CMOV-PRD86}, is that no conflict arises between the {two} approaches. {In fact, since} Hadamard states are unitarily equivalent to adiabatic vacuum states \cite{Junker1,Junker2}, one can {proceed to} translate the form of adiabatic states from the original $\phi$-description to the scaled $\psi$-description $(\psi=a\phi)$ {and then realize} the equivalence of the resulting quantization with the $j_0$-Fock representation \cite{CMOV-PRD86}. That is, when the Hadamard quantization is reformulated in the scaled field description, the resulting representation of the CCRs is related {to} the $j_0$-Fock representation by means of a unitary transformation. So, in the framework of the scaled $\psi$-description, the Hadamard and the unitary quantizations allow for {equivalent} physical predictions. It is in this sense that one can assure that there is no tension between the invariant, unitary $j_0$-Fock representation and the Hadamard quantization. 

(2) As we have mentioned in Sec. \ref{subsec:canon-quant}, the choice of fundamental classical variables involves an inherent ambiguity in the quantization of both mechanical and field systems. On account of this ambiguity, it is natural to ask whether a quantization with $SO(4)$ invariance and unitary dynamics can be achieved for a distinct pair of fundamental canonically conjugate variables, say $(\zeta,P_{\zeta})$, related {to} $(\varphi,\pi)$ by a time dependent canonical transformation, compatible with the symmetries of the field equations and with all linear structures on phase space, namely
\begin{equation}
\label{t-dep-CT}
\zeta=F(t)\varphi,\quad P_{\zeta}=\frac{\pi}{F(t)}+G(t)\sqrt{h}\varphi.
\end{equation}
{Here,} $F$ and $G$ are restricted to be {smooth real} functions of time, with $F(t)$ different from zero everywhere.  Without loss of generality [since the initial values $F(t_0)$ and $G(t_0)$ define an irrelevant time independent linear canonical transformation{\footnote{A time independent, linear canonical transformation does not modify {the} spatial symmetries, nor the dynamics. The quantum representation for {the} transformed and the original fields is actually the same (see, for instance, {Ref.} \cite{CMV-PRD75}).}}] one can set $F(t_0)=1$ and $G(t_0)=0$. {In this way, relationships (\ref{t-dep-CT}) carry all the time dependence of the possible change of variables}. By analyzing the new dynamics obtained with the transformation (\ref{t-dep-CT}), it has been demonstrated in {Refs.} \cite{CMOV-JCAP10,CMOV-PRD83} that {no transformation of} {this type} (apart for the identity) can lead to a classical evolution that admits a unitary implementation with respect to any of the Fock representations defined by a $SO(4)$ invariant complex structure. Hence, the {criteria of invariance and {of} unitarity {fix} not only the representation of the CCRs (up to unitary transformations) but, remarkably, the choice of field description as well. In this sense,  the ambiguities in the quantization process are fully removed.}

\subsection{Flat FLRW spacetimes}
\label{sec:Flat-FLRW}

Let us now consider a {real} scalar field $\phi$ propagating in a flat {FLRW} background with the spatial topology of a three-torus. By scaling the field with the conformal factor, the system {can be described as} a KG-field $\psi$, subject to a time dependent potential $V(\psi)=s(t)\psi^{2}/2$, propagating in a fictitious $(3+1)$-{dimensional} static spacetime $(M ,g_{{\alpha\beta}})$, with $M\approx \mathbb{I}\times T^{3}$ and $g_{{\alpha\beta}}$ given by {Eq.} (\ref{static-metric}), with $t\in \mathbb{I}=\mathbb{R}^{+}$ and {an induced spatial metric} $h_{ab}$ {equal to} the standard metric of the three-{torus}. The canonical phase space {can be described as in the discussion below} {Eq.} (\ref{static-metric}) {of} Sec. \ref{subsec:KG-time-mass}. The field dynamics, {that} is governed by {Eq.} (\ref{eq-mot-scaled-field-ii}) with $\Delta$ being the LB operator on $T^{3}$, is invariant under the group of isometries of the three-torus {formed by} rigid rotations in each of the periodic spatial directions {that diagonalize} the {spatial metric}, $R_{\theta_{i}}:x_{i}\to x_{i}+\theta_{i}$, with $\theta_{i}\in S^{1}$, and $i=1,2,3$. {We will denote the} composition $R_{\theta_1}\circ R_{\theta_2}\circ R_{\theta_3}$ by $R_{\vec{\theta}}$, with $\vec{\theta}=(\theta_{1},\theta_{2},\theta_{3})$.

The configuration and momentum can be decomposed as in {Eq.} (\ref{decomp-FLRW-S3-T3}), {taking as the} basis {$\{X_{\mathbf{n}}(x)\}$} of the space of square integrable functions on $T^{3}$ {the} eigenfunctions {of the LB operator} $\exp(i\vec{n}\cdot\vec{x})/(2\pi)^{3/2}$, where $\vec{n}=(n_{1},n_{2},n_{3})$, $n_{i}\in \mathbb{Z}$ ($i=1,2,3$), {and the corresponding eigenvalues are equal to} $-\omega^{2}_{n}=-\vec{n}\cdot\vec{n}$. Since the field is {real}, we have that the {associated} complex Fourier coefficients $\varphi_{\vec{n}}$ and $\pi_{\vec{n}}$ {satisfy} $\bar{\varphi}_{\vec{n}}=\varphi_{-\vec{n}}$ and $\bar{\pi}_{\vec{n}}=\pi_{-\vec{n}}$. To avoid {having to deal with these} reality conditions, we will expand the configuration and momentum in the {alternative basis of real eigenfunctions} $\{\cos(\vec{n}\cdot \vec{x}),\sin(\vec{n}\cdot \vec{x})\}$. Furthermore, since the {ultraviolet obstructions to the} unitary implementation of {the} time evolution {do} not depend on the removal of a finite number of degrees of freedom {from the system}, we will ignore the zero mode $\vec{n}=(0,0,0)$ from now on. {The exclusion of this mode does} not alter the field properties of the system and it can be quantized separately. In terms of Fourier modes corresponding to sines and cosines, the configuration and momentum are given by \cite{CCMMV-T3,MenaMarugan:2013tba,CCMMV-T3-ii} 
\begin{equation}
\label{conf-exp-T3}
\varphi(t,\vec{x})=\frac{1}{\pi^{3/2}}{\sum_{\vec{n}}}\left[q_{\vec{n}}(t)\cos(\vec{n}\cdot \vec{x})+\tilde{q}_{\vec{n}}(t)\sin(\vec{n}\cdot \vec{x})\right],
\end{equation}
\begin{equation}
\label{mom-exp-T3}
\pi(t,\vec{x})=\frac{\sqrt{h}}{\pi^{3/2}}{\sum_{\vec{n}}}\left[p_{\vec{n}}(t)\cos(\vec{n}\cdot \vec{x})+\tilde{p}_{\vec{n}}(t)\sin(\vec{n}\cdot \vec{x})\right].
\end{equation}
{Only triples} $\vec{n}$ of integers {in which the} first non-zero component is positive are contained in the sum. {All} different {triples} satisfying this restriction are to be summed over ({once each of them}). Since $\pi=\sqrt{h}\dot{\varphi}$, we have that {$p_{\vec{n}}=\dot{q}_{\vec{n}}$ and $\tilde{p}_{\vec{n}}=\dot{\tilde{q}}_{\vec{n}}$.} From the basic PB $\{\varphi(x),\pi(x')\}=\delta(x-x')$ and {Eqs.} (\ref{conf-exp-T3}) {and} (\ref{mom-exp-T3}), one can check that the only non-vanishing PB are {$\{q_{\vec{n}},p_{\vec{n}'}\}=\{\tilde{q}_{\vec{n}},\tilde{p}_{\vec{n}'}\}=\delta_{\vec{n}\vec{n}'}$.}

The equations of motion for the field modes coincide in each eigenspace of the LB operator, with eigenvalue {$-\omega^{2}_{n}$} [see Eq. (\ref{eq-of-mot-Tn})], 
\begin{equation}
\label{eq-mot-conf}
\ddot{q}_{\vec{n}}=-(\omega^{2}_{n}+s)q_{\vec{n}},\qquad \ddot{\tilde{q}}_{\vec{n}}=-(\omega^{2}_{n}+s)\tilde{q}_{\vec{n}}.
\end{equation}

It is worth noticing that, in contrast with the situation encountered for closed universes, where the number of independent eigenfunctions with the same eigenvalue is simple to derive in an exact form, in flat universes the degeneracy $g_n$ of each eigenspace of the LB operator presents a {complicated} dependence on the label $n$ because of {the existence of} accidental degeneracy. {Indeed}, apart from the {triples} related by permutations of the components, or by a flip of sign in one of the components, one can find {triples} which lead to the same eigenvalue. {Nonetheless, even though} the exact dependence of the degeneracy with $n$ cannot be given explicitly, an inspection of the asymptotic behavior of $g_n$ allows {us to conclude} that the sequence formed by $g_{n}/\omega^{4}_{n}$ is, in fact, summable \cite{CCMMV-T3-ii} (see also {Ref.} \cite{CCMMV-T3}). The argument is as follows. Recall that the {triples} are restricted to have {a positive integer as their} first non-vanishing component. However, since there exist two modes for each value of $\vec{n}$, namely the cosine and sine modes, we can assign these two modes to the pair of vectors $(\vec{n},-\vec{n})$. So, in spite of the {existing} restriction, we can make correspond modes to all vectors with integer components, {with the zero excluded}. Let $D_{N}$ be the number of modes {for which the eigenvalue function} $\omega_{n}$ is in the interval $(N,N+1]$, with $N$ a natural number. Geometrically, $D_N$ is nothing but the number of vertices of the cubic lattice with step equal to one that are contained between the sphere of radius $N$ and the sphere of radius $N+1$ (including {the surface of this latter sphere}). Therefore, $D_N$ increases with $N$ like $N^2$ and, consequently, the sum $\sum_{N}(D_{N}/N^{4})$ is finite. Since $1/\omega_{n}$ is strictly decreasing with $n$, we have that $\sum_{n}(g_{n}/\omega^{4}_{n})\leq \sum_{N}(D_{N}/N^{4})$, {inequality from which} the result follows.

As in our previous discussions, let us {call} $t_0\in \mathbb{I}$ the initial reference time. {At $t=t_0$ we} introduce the complex structure $j_0$ {defined in Eq.}  (\ref{CS-scaled-field-ii}). {Since this} complex structure {is totally} determined by the spatial metric, {it} is invariant under the three-torus isometries. The $j_0$ annihilation-like variables are
\begin{equation}
a_{\vec{n}}=\frac{1}{\sqrt{2\omega_{n}}}(\omega_{n}q_{\vec{n}}+ip_{\vec{n}}),\quad \tilde{a}_{\vec{n}}=\frac{1}{\sqrt{2\omega_{n}}}(\omega_{n}\tilde{q}_{\vec{n}}+i\tilde{p}_{\vec{n}}).
\end{equation}
The corresponding creation-like variables are given by the complex conjugates $\bar{a}_{\vec{n}}$ and $\bar{\tilde{a}}_{\vec{n}}$. The annihilation and creation-like variables provide a complete set of coordinates on phase space. The action {of} $j_0$ is diagonal in these variables; namely, $j_{0}(a_{\vec{n}})=ia_{\vec{n}}$, $j_{0}(\bar{a}_{\vec{n}})=-i\bar{a}_{\vec{n}}$, and likewise for $\tilde{a}_{\vec{n}}$ and $\bar{\tilde{a}}_{\vec{n}}$. {Given that} the different Fourier modes decouple in the dynamics and, in addition, the equations of motion for {them} depend only on the LB {operator}, we get that the Bogoliubov coefficients of the time evolution transformation from $t_0$ to $t$ have the form (\ref{Bogo-coeff-expan-real-var}), {namely} 
\begin{equation}
\label{FLRW-T3-Bogo-transf}
a_{\vec{n}}(t)=\alpha_{n}(t,t_0)a_{\vec{n}}+\beta_{n}(t,t_0)\bar{a}_{\vec{n}},\quad \bar{a}_{\vec{n}}(t)=\bar{\alpha}_{n}(t,t_0)\bar{a}_{\vec{n}}+\bar{\beta}_{n}(t,t_0)a_{\vec{n}},
\end{equation}
and similarly for $\tilde{a}_{\vec{n}}$ and $\bar{\tilde{a}}_{\vec{n}}$. The Bogoliubov coefficients $\alpha_{n}$ and $\beta_{n}$ are given as in {Eq.} (\ref{alph-bet-coeff-example}), with $T$ being replaced {with} $q$, when time evolution is that of the pair $(a_{\vec{n}},\bar{a}_{\vec{n}})$, and {with} $\tilde{q}$ when the considered evolved pair is $(\tilde{a}_{\vec{n}},\bar{\tilde{a}}_{\vec{n}})$. 

Taking into account the degeneracy $g_n$ of the eigenspaces of the LB operator, {we have that} the time evolution is unitary implementable in the Fock representation defined by $j_0$ if and only if the sequence $\{g_{n}\vert\beta_{n}(t,t_{0})\vert^{2}\}$ is summable for all possible values of $t\in \mathbb{I}$. From the analysis performed in Sec. \ref{sec:Closed-FLRW}, it follows that, in the ultraviolet regime {and} provided that the function $s(t)$ has an integrable first derivative in every closed time subinterval, the antilinear coefficients $\beta_{n}(t,t_{0})$ behave as $\beta_{n}(t,t_{0})=O(\omega^{-2}_{n})$, where the symbol $O$ indicates the asymptotic order. {Therefore, the unitarity of the dynamics} depends upon the asymptotic behavior of the sequence $\{g_{n}/\omega^{4}_{n}\}$. Since {this sequence} is certainly summable, as we {have} seen above, we conclude that {the} time evolution is unitarily implementable with respect to the complex structure $j_0$. 

{In this way}, we have at hand a Fock representation which is invariant under the isometries of the three-torus {consisting of} the transformations $R_{\vec{\theta}}$, and {which allows for} a unitary implementation of the dynamics. In order to know whether or not this representation is unique, {we need} to examine how many classes of {unitarily} equivalent representations {have elements in} the whole family of {invariant and unitary} Fock representations. Remarkably, the answer is that {this} family consists of {representations in} one and only one equivalence {class}. The sketch of the proof is as follows (see {Refs.}\cite{CCMMV-T3,CCMMV-T3-ii} for details).

The first step is to characterize the {compatible} complex structures that are invariant under the group of transformations $R_{\vec{\theta}}$. A careful analysis shows that, for every invariant complex structure $j$ (compatible with the symplectic structure), the annihilation and creation-like variables defined by $j$ are related {to the corresponding variables for} $j_0$ by a Bogoliubov transformation which only {mixes} modes with the same labels $\vec{n}$ (and the mixing {depends only on this label}) \cite{CCMMV-T3,CCMMV-T3-ii}. Specifically, 
\begin{equation}
\begin{pmatrix}
a^{(j)}_{\vec{n}} \\
\bar{a}^{(j)}_{\vec{n}}\end{pmatrix}=
\begin{pmatrix}
  \kappa_{\vec{n}} & \lambda_{\vec{n}}\\
\bar{\lambda}_{\vec{n}}  & \bar{\kappa}_{\vec{n}}
 \end{pmatrix}\begin{pmatrix}
a_{\vec{n}} \\
\bar{a}_{\vec{n}}\end{pmatrix}, \quad \begin{pmatrix}
\tilde{a}^{(j)}_{\vec{n}} \\
\bar{\tilde{a}}^{(j)}_{\vec{n}}\end{pmatrix}=
\begin{pmatrix}
  \kappa_{\vec{n}} & \lambda_{\vec{n}}\\
\bar{\lambda}_{\vec{n}}  & \bar{\kappa}_{\vec{n}}
 \end{pmatrix}\begin{pmatrix}
\tilde{a}_{\vec{n}} \\
\bar{\tilde{a}}_{\vec{n}}\end{pmatrix},\end{equation}
where {the} coefficients  $\kappa_{\vec{n}}$ and $\lambda_{\vec{n}}$ satisfy that  $\vert\kappa_{\vec{n}}\vert^{2}- \vert\lambda_{\vec{n}}\vert^{2}=1$ for all $\vec{n}$. Thus, the antilinear part of the time evolved annihilation-like variables $a^{(j)}_{\vec{n}}$ and $\tilde{a}^{(j)}_{\vec{n}}$ is given by
\begin{equation}
\label{antilinear-ann-t3}
\beta^{(j)}_{\vec{n}}(t,t_{0})=\kappa^{2}_{\vec{n}}\beta_{n}(t,t_{0})-\lambda^{2}_{\vec{n}}\bar{\beta}_{n}(t,t_{0})+2i\kappa_{\vec{n}}\lambda_{\vec{n}}{\rm{Im}}[\bar{\alpha}_{n}(t,t_{0})].
\end{equation}

The next step in the proof is to suppose that $\beta^{(j)}_{\vec{n}}(t,t_{0})$ in {Eq.} (\ref{antilinear-ann-t3}) is square summable (i.e. we suppose that the $j$-Fock representation {allows}, indeed, {for} a unitary implementation of the dynamics). It is a simple matter to see that the square summability of $\beta^{(j)}_{\vec{n}}(t,t_{0})$ {and that of $\beta_{n}(t,t_{0})$, that we have already proven,} imply that $\{(\lambda_{\vec{n}}/\kappa_{\vec{n}}){\rm{Im}}[\bar{\alpha}_{n}(t,t_{0})]\}$ is square summable. By using the ultraviolet behavior of alpha, $\alpha_{n}(t,t_{0})=\exp[-i\omega_{n}(t,t_{0})]+O(\omega^{-1}_{n})$, and assuming that the mass function $s(t)$ possesses a second derivative which is integrable in every closed subinterval of $\mathbb{I}$, an average in time together with a suitable application of Luzin's theorem \cite{Kolmo-Fom} shows {then} that $\{\vert \lambda_{\vec{n}}/\kappa_{\vec{n}}\vert^{2}\}$ is summable \cite{CCMMV-T3,CCMMV-T3-ii}. {Employing} that  $\vert\kappa_{\vec{n}}\vert^{2}- \vert\lambda_{\vec{n}}\vert^{2}=1$, one can see that $\{\lambda_{\vec{n}}\}$ {must be} square summable as well. {But this summability is precisely the necessary and sufficient condition for the representations} given by $j$ and $j_0$ {to be} unitarily equivalent. {Therefore}, the family of invariant {and dynamically} unitary Fock representations {contains representations in only} one equivalence class, {namely that of $j_0$.}

The requirements of invariance under spatial symmetries and of {unitary implementability of the} time evolution {select}, apart from a unique family of unitarily equivalent Fock representations, a unique preferred canonical pair of field variables. Indeed, {consider} the most general form of a time dependent, linear canonical transformation scaling the field,
\begin{equation}
\xi=F(t)\varphi,\quad P_{\xi}=\frac{\pi}{F(t)}+G(t)\sqrt{h}\varphi,
\end{equation}
where $F(t)$ and $G(t)$ are assumed to be twice differentiable real functions, {with} initial values {set} (without loss of generality) {to} $F(t_{0})=1$ and $G(t_{0})=0$, and {such that} $F(t)$ is non-vanishing. {Then,} one can show \cite{CCMMV-T3,CCMMV-T3-ii}  that a unitary evolution with respect to a complex structure {that is $R_{\vec{\theta}}$ -invariant} is only possible when $F(t)$ and $G(t)$ are the unit and {the} zero constant functions, respectively. That is to say, no time dependent scaling or redefinition of the field momentum is allowed. In total, we have that the ambiguities in the Fock quantization coming from the scaling of the field and from the choice of {representation} are fully removed. 

\section{Uniqueness for scalar fields in de Sitter spacetime}
\label{sec:deSitter}

Let us now consider the propagation of a minimally coupled, massless, and real scalar field $\phi$ in de Sitter spacetime, the maximally symmetric spacetime of positive constant curvature. In conformal time, the metric can be written in the form
\begin{equation}
{g_{\alpha\beta}dx^{\alpha}dx^{\beta}=a^{2}(t)\left[-dt^2+h_{ab}dx^adx^b\right],}
\end{equation}
where the conformal factor {is} $a^{2}(t)=12R^{-1}\sin^{-2}(t)$, with $R$ {denoting the constant spacetime} curvature. {Besides,} $h_{ab}$ is the standard metric of the three-sphere, given in {Eq.} (\ref{S3-metric}). Scaling the field {with} $a(t)$, one gets that the dynamics of the {new} field $\psi=a\phi$ is dictated by {Eq.} (\ref{eq-mot-scaled-field-ii}), {namely $\ddot{\psi}-\Delta\psi+s(t)\psi=0.$} Here, $s(t)=1-Ra^{2}/6$ and $\Delta$ is the LB operator on $S^{3}$, {the eigenfunctions of which} are the harmonics $Q_{n\ell m}$ given in {Eq.} (\ref{Hyp-SH}), $\Delta Q_{n\ell m}=-\omega^{2}_{n} Q_{n\ell m}$ with $\omega^{2}_{n}=n(n+2)$. Owing {to} the complete analogy of the system with that of a (re-scaled) KG field propagating in a closed {FLRW} spacetime, discussed in Sec. \ref{sec:Closed-FLRW}, {for which} a unique preferred Fock representation was specified for quite generic mass terms $s(t)$, we {could already} claim that the {criteria of unitary dynamics and {of} invariance of the vacuum under the {spatial} symmetries single} out a {unique} quantization for the scalar field in de Sitter spacetime. For the sake of completeness, however, we {will} briefly discuss the quantization in a slightly different way, {by} considering complex modes and (see below) a frequency {that differs} from $\omega_{n}$ in the {introduction of} annihilation and creation-like variables.

Since the harmonics provide a complete and orthonormal set for the expansion of functions on the three-sphere, we can write the field $\psi$ as
\begin{equation}
\label{sf-decomp-dS}
{\psi(t,x)=\sum_{n,\ell, m}\varphi_{n \ell m}(t) Q_{n \ell m}(x).}
\end{equation}
{Since} $\psi$ is a real field {and} $\bar{Q}_{n\ell m}=(-1)^{m}Q_{n\ell -m}$, we have that the time dependent coefficients  $\varphi_{n \ell m}$ must satisfy the reality conditions $\bar{\varphi}_{n\ell m}=(-1)^{m}\varphi_{n\ell -m}$. Introducing the decomposition (\ref{sf-decomp-dS}) into the field {equation}, we get that the time depending complex coefficients $\varphi_{n\ell m}$ obey the equation of motion (\ref{eq-of-mot-Tn}) which, after substituting $a(t)=12R^{-1}\sin^{-2}(t)$ {and $\omega^{2}_{n}=n(n+2)$}, reads
\begin{equation}
\label{dS-modes}
\ddot{\varphi}_{\mathbf{n}}+\left[\tilde{\omega}^{2}_{n}-2\sin^{-2}t\right]\varphi_{\mathbf{n}}=0,
\end{equation}
where $\mathbf{n}$ {collectively denotes the} {triple of indices $(n,\ell,m)$  and 
$\tilde{\omega}^{2}_{n}{=}(\omega^{2}_{n}+1)=(n+1)^{2}$.} The general solution to {Eq.} (\ref{dS-modes}) {is} \cite{CdBMV-deSitter} 
\begin{equation}
\label{sol-modes-dS}
\varphi_{\mathbf{n}}(t)=A_{\mathbf{n}}\sqrt{\sin t} P^{\mu}_{\nu}(-\cos t)+B_{\mathbf{n}}\sqrt{\sin t} Q^{\mu}_{\nu}(-\cos t),
\end{equation}
where $P^{\mu}_{\nu}$ and $Q^{\mu}_{\nu}$ are the associated Legendre functions \cite{Abramowitz,Gradshteyn} with $\nu=n+1/2$ and $\mu=3/2$, whereas $A_{\mathbf{n}}$ and $B_{\mathbf{n}}$ are arbitrary complex constants.  {Given that the field momentum is} $\pi=\sqrt{h}\dot{\psi}$, we have that its Fourier coefficients $\pi_{\mathbf{n}}=\int \pi\bar{Q}_{\mathbf{n}}d^{3}x$, {that satisfy} the reality conditions $\bar{\pi}_{n\ell m}=(-1)^{m}\pi_{n\ell -m}$, are related {to} $\varphi_{\mathbf{n}}$ by $\pi_{\mathbf{n}}=\dot{\varphi}_{\mathbf{n}}$. From {this relationship and Eq.} (\ref{sol-modes-dS}), one gets that {the} time evolution from an arbitrary initial reference time $t_0$ to a final time $t$ {takes the form} \cite{CdBMV-deSitter}
\begin{equation}
\label{qp-evol-dSitter-i}
\begin{pmatrix}
\varphi_{\mathbf{n}}(t) \\
\pi_{\mathbf{n}}(t) \end{pmatrix}
={\cal{T}}_{n}(t,t_{0}) \begin{pmatrix}
\varphi_{\mathbf{n}}(t_{0}) \\
 \pi_{\mathbf{n}}(t_{0})\end{pmatrix},
 \quad
{\cal{T}}_{n}(t,t_{0})=W_{n}(t)W^{-1}_{n}(t_0),
\end{equation}
with
\begin{equation}
\label{qp-evol-dSitter-ii}
W_{n}(t)=\begin{pmatrix} R^{\mu}_{\nu}(-\cos t)& S^{\mu}_{\nu}(-\cos t)\\
 \dot{R}^{\mu}_{\nu}(-\cos t)&  \dot{S}^{\mu}_{\nu}(-\cos t)  \end{pmatrix},
\end{equation}
where $R^{\mu}_{\nu}(-\cos t)$, $S^{\mu}_{\nu}(-\cos t)$, and their time {derivatives are}
\begin{eqnarray}
R^{\mu}_{\nu}(-\cos t)&{=}&\sqrt{\sin t} P^{\mu}_{\nu}(-\cos t), \quad S^{\mu}_{\nu}(-\cos t)=\sqrt{\sin t} Q^{\mu}_{\nu}(-\cos t), \nonumber \\
\dot{R}^{\mu}_{\nu}(-\cos t)&=&\frac{1}{\sqrt{\sin{\eta}}}\left[(\nu+1/2)\cos t P^{\mu}_{\nu}(-\cos t)+(\nu+\mu)P^{\mu}_{\nu-1}(-\cos t)\right],\nonumber \\
\dot{S}^{\mu}_{\nu}(-\cos t)&=&\frac{1}{\sqrt{\sin{\eta}}}\left[(\nu+1/2)\cos t Q^{\mu}_{\nu}(-\cos t)+(\nu+\mu)Q^{\mu}_{\nu-1}(-\cos t)\right]. \label{RS-and-deriv}
\end{eqnarray}

Let us now introduce the annihilation and creation-like variables
\begin{equation}
\label{ann-crea-var-dS}
a_{\mathbf{n}}=\frac{1}{\sqrt{2\tilde{\omega}_{n}}}(\tilde{\omega}_{n}\varphi_{\mathbf{n}}+i\pi_{\mathbf{n}}),\quad \bar{a}_{\mathbf{n}}=\frac{1}{\sqrt{2\tilde{\omega}_{n}}}(\tilde{\omega}_{n}\bar{\varphi}_{\mathbf{n}}-i\bar{\pi}_{\mathbf{n}}).
\end{equation}
The {relation between the variables (\ref{ann-crea-var-dS}) and the corresponding ones associated {to} the frequency $\omega_{n}$, i.e.} $b_{\mathbf{n}}=(\omega_{n}\varphi_{\mathbf{n}}+i\pi_{\mathbf{n}})/{\sqrt{2\omega_{n}}}$ and $\bar{b}_{\mathbf{n}}=(\omega_{n}\bar{\varphi}_{\mathbf{n}}-i\bar{\pi}_{\mathbf{n}})/{\sqrt{2\omega_{n}}}$ ({where we exclude} the zero mode), is given by {a} Bogoliubov transformation $b_{\mathbf{n}}=\tilde{\alpha}_{n}a_{\mathbf{n}}+\tilde{\beta}_{n}\bar{a}_{\mathbf{n}}$ {characterized by} $\tilde{\alpha}_{n}=(\tilde{\omega}_{n}+\omega_{n})/2\sqrt{\tilde{\omega}_{n}\omega_{n}}$ and $\tilde{\beta}_{n}=(\tilde{\omega}_{n}-\omega_{n})/2\sqrt{\tilde{\omega}_{n}\omega_{n}}$. A direct calculation shows that the square of the antilinear part, which is given by $\vert \tilde{\beta}_{n}\vert^{2}=1/(2y_{n})+(y_{n}/2)-1$, where $y_{n}{=}(1-\tilde{\omega}^{-2}_{n})^{1/2}$, behaves as $O(\tilde{\omega}_{n}^{-4})$ in the asymptotic regime, so that $\{(n+1)^{2}\vert\tilde{\beta}_{n}\vert^{2}\}$ is summable [recall that the degeneracy factor {is $g_n=(n+1)^2$}]. Hence, the complex structures defined {by the choice of} annihilation and creation-like variables $(a_{\mathbf{n}},\bar{a}_{\mathbf{n}})$ and $(b_{\mathbf{n}},\bar{b}_{\mathbf{n}})$ give rise to unitary equivalent Fock representations. Since the variables $(b_{\mathbf{n}},\bar{b}_{\mathbf{n}})$ are nothing but those associated with the complex structure $j_0$, they define an $O(4)$-invariant Fock representation. {On the other hand, since the variables $(a_{\mathbf{n}},\bar{a}_{\mathbf{n}})$} are defined by a complex structure 
$j_{\tilde{0}}$ (equivalent to $j_0$) which {also depends} on the LB operator only, we conclude that the $j_{\tilde{0}}$-Fock representation is $O(4)$-invariant as well.

From {Eqs.} (\ref{qp-evol-dSitter-i}) and (\ref{ann-crea-var-dS}), and using the reality conditions for the Fourier coefficients $\varphi_{\mathbf{n}}$ and $\pi_{\mathbf{n}}$, {we get} that the annihilation and creation-like variables evolve according to
\begin{equation}
\label{ac-evol-dSitter-i}
\begin{pmatrix}
a_{n\ell m}(t) \\
(-1)^{m}\bar{a}_{n\ell -m}(t) \end{pmatrix}
={\mathbf{U}}_{n}(t,t_{0}) \begin{pmatrix}
a_{n\ell m}(t)  \\(-1)^{m}\bar{a}_{n\ell -m}(t)
\end{pmatrix},
\end{equation}
where
\begin{equation}
\label{ac-evol-dSitter-ii}
{\mathbf{U}}_{n}(t,t_{0})={\cal{M}}_{n} {\cal{T}}_{n}(t,t_{0}) {\cal{M}}^{-1}_{n}{=}\begin{pmatrix} \alpha_{n}(t,t_{0}) & \beta_{n}(t,t_{0})\\
\bar{\beta}_{n}(t,t_{0}) & \bar{\alpha}_{n}(t,t_{0})\end{pmatrix};\quad {\cal{M}}_{n}=\frac{1}{\sqrt{2\tilde{\omega}_{n}}}\begin{pmatrix} \tilde{\omega}_{n} & i\\
\tilde{\omega}_{n}  & -i\end{pmatrix}.
\end{equation}
The {coefficients $\alpha_{n}$ and  $\beta_{n}$} can be explicitly obtained from the matrices $W_n$, relations (\ref{RS-and-deriv}), and the matrices ${\cal{T}}_{n}$ and ${\cal{M}}_{n}$. Taking into account the asymptotic behavior of the functions {$P_{\nu}$ and $Q_{\nu}$} at large values of the degree $\nu=n+1/2$ (see for instance {Ref.} \cite{Gradshteyn}), a lengthy but direct calculation shows that $\beta_{n}(t,t_{0})$ is of order $O(n^{-2})$ in the ultraviolet regime \cite{CdBMV-deSitter}. Hence, we have that $\sqrt{g_{n}}\beta_{n}(t,t_{0})$ is of order $O(n^{-1})$, and consequently
\begin{equation}
\sum_{n =0}^{\infty}\sum_{\ell=0}^{n}\sum_{m=-\ell}^{\ell}\vert \beta_{n}(t,t_{0})\vert ^2=\sum_{n =0}^{\infty}(n+1)^{2}\vert \beta_{n}(t,t_{0})\vert ^2<\infty,
\end{equation}
for all values of $t_0$ and $t$. Thus, the dynamics is unitarily {implementable} in the 
$j_{\tilde{0}}$-Fock representation. {This conclusion corrects the claims} of {Ref.} \cite{VV-dS}, where it is {argued} that one cannot attain quantum unitarity of the evolution for the massless field in de Sitter spacetime, independent of the field redefinition $\phi\to f(t)\phi$. {We} have seen that {it is perfectly possible to find suitable canonical variables and construct for them a well-defined Fock quantization such that the dynamics admits a} unitary implementation.

It it worth {pointing out that our} unitarity result holds as well for any massive free field. {The} Fock representation defined by the choice (\ref{ann-crea-var-dS}) of annihilation and creation-like variables provides a quantum description where {the} time evolution of {a} free massive field admits a unitary implementation. Indeed, it can be verified from the asymptotic behavior of $P^{\mu}_{\nu}$ and $Q^{\mu}_{\nu}$ that, for any constant value of the parameter $\mu$, including complex numbers and the massive case $\mu=(9/4-12m^{2}/R)^{1/2}$, the corresponding beta coefficient in the Bogoliubov transformation of {the} time evolution satisfies that $\beta_{n}(t,t_{0})=O(n^{-2})$ when $n\to \infty$ for all $t$ and $t_0$ \cite{CdBMV-deSitter}. Hence, given a scalar field with $m\geq 0$, {there} exists at least one Fock representation where {the} time evolution is {implementable} as a unitary operator. Moreover, in view of the uniqueness of the $j_0$-Fock representation for scalar fields with time dependent mass on $S^{3}$ [{see} Sec. \ref{sec:Closed-FLRW}], and that $j_{\tilde{0}}$ {is equivalent to $j_{0}$}, {one can conclude} {that the $j_{\tilde{0}}$-Fock representation} {provides} {the unique (up to unitary equivalence) Fock representation of the CCRs {that satisfies} the criteria} of invariance and {of} unitarity for KG fields (massive or not) in de Sitter spacetime. Furthermore, the unique field description from which an invariant {Fock representation with unitary dynamics} can be specified is the $\psi$-description.

{Finally, let us discuss briefly the connection between this quantization and} {the choice of} (translated) Hadamard states (in the $\psi$-description). For massive free fields in de Sitter spacetime, the standard Fock representation is accomplished by using mode solutions of the form (\ref{sol-modes-dS}), with $\mu=(9/4-12m^{2}/R)^{1/2}$ and 
\begin{equation}
A_{\mathbf{n}}=\sqrt{\frac{\pi}{4}\frac{\Gamma(n-\mu+3/2)}{\Gamma(n+\mu+3/2)}}e^{i\pi \mu/2},\qquad B_{\mathbf{n}}=-\frac{2i}{\pi}A_{\mathbf{n}},
\end{equation}
{that} provide a unique invariant solution under the full $O(1,4)$ symmetry group, satisfying the Hadamard condition \cite{Allen,Allen-Folacci}. Explicitly, 
{this solution is defined by the modes}
\begin{equation}
\label{BD-sol}
\chi_{\mathbf{n}}(t)=A_{\mathbf{n}}\left[R^{\mu}_{\nu}(-\cos t)-\frac{2i}{\pi}S^{\mu}_{\nu}(-\cos t)\right].
\end{equation}
The vacuum of the corresponding Fock representation is known as the Bunch-Davies vacuum \cite{Bunch-Davies}. {The} question naturally arises of whether or not the $j_{\tilde{0}}$-Fock quantization is unitarily equivalent to the representation based on the Bunch-Davies vacuum, {that} follows from the requirement of full de Sitter invariance and the Hadamard condition in the massive field case. The answer, as it is shown in {Ref.} \cite{CdBMV-deSitter}, is that these two quantizations turn out to be unitarily equivalent. So, no {tension} arises between the
 $j_{\tilde{0}}$-Fock representation (specified by {imposing the requirements} of invariance and {of} unitarity {of the dynamics}) and the standard Fock representation defined by the Bunch-Davies vacuum. 

For the massless case, the {definition of the} Bunch-Davies vacuum breaks down and there is no de Sitter-invariant Hadamard vacuum \cite{Allen}. However, since {this is due to the dynamics of the zero mode only \cite{Allen,Allen-Folacci}, we have that proper solutions for the zero mode (or a proper independent quantization of that single mode)} together with the ${\mathbf{n}}\neq 0$ solutions (\ref{BD-sol}) {still} provide a complete set of well-defined solutions, and hence a well-defined quantization. Thus, as it is shown in {Ref.} \cite{Allen-Folacci}, one can arrive at a one-parameter family of solutions for the zero {mode} {such that, together with the solutions (\ref{BD-sol}) for ${\mathbf{n}}\neq 0$, one obtains} $O(4)$-invariant Hadamard {vacua}. A direct calculation  then shows that the $j_{\tilde{0}}$-Fock representation for the massless KG field  is, in fact, unitarily equivalent to the representation defined by {these} $O(4)$-invariant Hadamard vacua \cite{CdBMV-deSitter} (let us remark that the particular quantization used for the zero mode is irrelevant, {except perhaps in what concerns whether it satisfies or not} the Stone-von Neumann conditions, {because} unitary equivalence depends on the behavior of states in the ultraviolet regime). {In this way}, the complex structures $j_{H}$ characterizing the $O(4)$-invariant Hadamard vacua belong to the equivalence class of the complex structure
 $j_{\tilde{0}}$, which in turn is {the} equivalence class of {$j_{0}$}.

\section{Uniqueness for scalar fields in Bianchi I universes}
\label{sec:BI-univ}

The {criteria of symmetry invariance and {of a} unitary dynamics {have} been successfully {imposed} not only in homogeneous and isotropic backgrounds, but also in anisotropic spacetimes with shear, which are then not conformally symmetric. Specifically, the criteria {have been} employed to remove the ambiguities in the quantization of scalar fields propagating in Bianchi I spacetimes \cite{Unique-Bianchi-I}. This section} {gives an overview of the result of uniqueness for these anisotropic scenarios.} 

Let us consider a free real scalar field $\phi$ with mass $m$ propagating in a Bianchi I spacetime with spatial sections of three-torus topology. In {a diagonal system of} coordinates, and {with} $t\in {\mathbb{I}}$, where $\mathbb{I}$ is a connected interval of the real line, the metric of {the} Bianchi I universes can be written in {the form}
\begin{equation}
\label{Bianchi-I-metric}
{g_{\alpha\beta}dx^{\alpha}dx^{\beta}=-N^{2}(t)dt^2+\sum_{i=1}^{3}a^{2}_{i}(t)(dx^{i})^2,}
\end{equation}
where $N(t)$ is the lapse function and $a_{i}(t)$ are the scale factors {of the three diagonal directions}, {with $i=1,2,3$}. The LB operator {$\Delta$} is essentially self-adjoint on the space of square integrable functions with respect to the measure {$\sqrt{h}d^{3}x=\prod_{i} a_i dx^i$} and, owing to the compactness of the spatial sections, it has a discrete spectrum. A complete set of eigenfunctions {of} the LB operator is provided by the set of plane waves {$\exp(i\vec{k}\cdot \vec{x})$}  in $T^{3}$, where $\vec{k}=(k_{1},k_{2},k_{3})\in \mathbb{Z}^{3}$. {The corresponding eigenvalue is}
\begin{equation}
\label{LB-ev}
\Delta_{k}=-\sum_{i=1}^{3}\left(\frac{k_{i}}{a_{i}}\right)^{2}.
\end{equation}
Note that there are different wave vectors $\vec{k}$ with the same eigenvalue $\Delta_k$ (i.e. the spectrum is {degenerate}). The eigenspaces of $\Delta$ provide irreducible representations of the group formed by the composition of rigid rotations in the three principal spatial directions of the tori, which are the Killing symmetries of the Bianchi I universes. 

Since the field $\phi$ is {real}, we {will consider} its Fourier expansion in terms of the real basis of cosine and sine functions {obtained from the real and imaginary parts of the} plane waves. Thus, the configuration and momenta of the field at time $t$ are (ignoring {again} the zero mode {in our discussion}) 
\begin{equation}
\label{conf-B1}
\varphi(t,\vec{x})=\frac{1}{\sqrt{4\pi^{3}}}\sum_{\vec{k}\in {\mathbb{L}}}\left[q^{(1)}_{\vec{k}}(t)\cos(\vec{k}\cdot\vec{x})+q^{(2)}_{\vec{k}}(t)\sin(\vec{k}\cdot\vec{x})\right],
\end{equation}
\begin{equation}
\label{mom-B1}
\pi(t,\vec{x})=\frac{1}{\sqrt{4\pi^{3}}}\sum_{\vec{k}\in {\mathbb{L}}}\left[p^{(1)}_{\vec{k}}(t)\cos(\vec{k}\cdot\vec{x})+p^{(2)}_{\vec{k}}(t)\sin(\vec{k}\cdot\vec{x})\right],
\end{equation}
where $\mathbb{L}$ stands for the lattice
\begin{equation}
{\mathbb{L}}{=}\{\vec{k}\vert k_{1}>0\}\cup \{\vec{k}\vert k_{1}=0,k_{2}>0\}\cup \{\vec{k}\vert k_{1}=0=k_{2},k_{3}>0\},
\end{equation}
introduced to avoid duplication of the real modes in the Fourier expansion. From the basic brackets at equal time $\{\varphi(\vec{x}),\pi(\vec{x}')\}=\delta(\vec{x}-\vec{x}')$ and {Eqs.} (\ref{conf-B1}) {and} (\ref{mom-B1}), it follows that the only {non-zero PB} are
\begin{equation}
\{q^{(l)}_{\vec{k}},p^{(l')}_{\vec{k}'}\}=\delta^{ll'}\delta_{\vec{k}\vec{k}'},
\end{equation}
where $l,l'=1,2$. {To simplify} the notation, in what follows we will denote any of the canonical pairs $(q^{(1)}_{\vec{k}},p^{(1)}_{\vec{k}})$ and $(q^{(2)}_{\vec{k}},p^{(2)}_{\vec{k}})$ {by} $(q_{\vec{k}},p_{\vec{k}})$, {unless otherwise stated}. 

{In order to achieve} {a quantum description {that satisfies} the requirements of invariance under {the} spatial symmetries and {of a} unitary dynamics}, and inspired by the case of free scalar fields propagating in {FLRW} spacetimes, {we introduce} a time dependent transformation of the field canonical pairs, regarded as variables that change with the considered section of constant time $t$. However, in contrast to the isotropic case, where the transformation is just a time dependent scaling of the configuration field variable, we now consider a transformation {that}, on account of the lack of isotropy, is also mode dependent ({for} a discussion on time and mode dependent canonical transformations see {Ref.} \cite{CFdBMM}). Specifically, we {introduce} the  canonical change \cite{Unique-Bianchi-I}
\begin{equation}
\label{mode-dep-transf}
\tilde{q}_{\vec{k}}=\sqrt{b(\tau,\hat{k})}\,q_{\vec{k}},\quad \tilde{p}_{\vec{k}}=\frac{1}{2\sqrt{b(\tau,\hat{k})}}\left[2p_{\vec{k}}+q_{\vec{k}}\,\frac{d}{d\tau}\ln b(\tau,\hat{k})\right].
\end{equation}
Here, $\hat{k}$ stands for the unit vector $\vec{k}/k$, $\tau$ is the harmonic time, defined by $N(t)dt=a^{3}(\tau)d\tau$ with $a^{3}(\tau){=}a_{1}(\tau)a_{2}(\tau)a_{3}(\tau)$, and 
\begin{equation}
b(\tau,\hat{k})=a^{3}\sqrt{\sum_{i=1}^{3}\left(\frac{\hat{k}_{i}}{a_{i}}\right)^{2}},
\end{equation}
where $\hat{k}_{i}$ denotes the components of the unit vector $\hat{k}$. {In the following, we will} denote the function $b(\tau,\hat{k})$ just by $b_{\hat{k}}$, in order to {shorten the} notation. Notice that, in the limit of isotropy ($a=a_{1}=a_{2}=a_{3}$), $b_{\hat{k}}$ becomes {just} $b_{\hat{k}}=a^{2}$, and we recover the scaling employed in isotropic scenarios ({see} Sec. \ref{sec:FLRW}). The transformation (\ref{mode-dep-transf}) respects the symmetries of the LB operator $\Delta$ (i.e. the spatial Killing symmetries). {By} setting $N=a^3$, one can see that the dynamics of the new canonical pairs $(\tilde{q}^{(l)}_{\vec{k}},\tilde{p}^{(l)}_{\vec{k}})$ is governed by the {respective} Hamiltonians \cite{Unique-Bianchi-I}
\begin{equation}
\label{Hamil-BI}
H^{(l)}=\frac{1}{2}\sum_{\vec{k}\in\mathbb{L}}b_{\hat{k}}\left[\left(\tilde{p}^{(l)}_{\vec{k}}\right)^{2}+[k^{2}+s_{\hat{k}}(\tau)]\left(\tilde{q}^{(l)}_{\vec{k}}\right)^{2}\right],\quad l=1,2,
\end{equation}
with
\begin{equation}
\label{time-mode-dep-mass}
s_{\hat{k}}(\tau)=m^{2}\left(\frac{a^{3}}{b_{\hat{k}}}\right)^{2}+\frac{3}{4}\left(\frac{\dot{b}_{\hat{k}}}{b^{2}_{\hat{k}}}\right)^{2}-\frac{1}{2}\frac{\ddot{b}_{\hat{k}}}{b^{3}_{\hat{k}}}.
\end{equation}
{Here, the} dot stands for the derivative with respect to the harmonic time $\tau$. Note that, up to the factor $b_{\hat{k}}$, $H^{(l)}$ {is} a sum of Hamiltonians of harmonic oscillator type, one for each mode, {with} masses {that} depend on time as well as on $\hat{k}$. From {Eq.} (\ref{Hamil-BI}) one gets that  {the Hamiltonian} equations of motion for the {modes are} $\dot{\tilde{q}}_{\vec{k}}=b_{\hat{k}}\tilde{p}_{\vec{k}}$ and $\dot{\tilde{p}}_{\vec{k}}=-b_{\hat{k}}(k^2+s_{\hat{k}})\tilde{q}_{\vec{k}}$. By writing their real solutions in the form
\begin{equation}
\label{generic-sol-BI}
\tilde{q}_{\vec{k}}(\tau)=Q_{\vec{k}}e^{i\Theta^{q}_{\vec{k}}(\tau)}+\bar{Q}_{\vec{k}}e^{-i\bar{\Theta}^{q}_{\vec{k}}(\tau)}, \quad
\tilde{p}_{\vec{k}}(\tau)=k\left(P_{\vec{k}}e^{i\Theta^{p}_{\vec{k}}(\tau)}+\bar{P}_{\vec{k}}e^{-i\bar{\Theta}^{p}_{\vec{k}}(\tau)}\right),
\end{equation}
where $Q_{\vec{k}}$ and $P_{\vec{k}}$ are complex constants, and $\Theta^{{\varepsilon}}_{\vec{k}}(\tau)$ are complex functions (${\varepsilon}=q,p$), it can be shown \cite{Unique-Bianchi-I} that {the} solutions with initial conditions $\Theta^{{\varepsilon}}_{\vec{k}}(\tau_{0})=0$ and $\dot{\Theta}^{{\varepsilon}}_{\vec{k}}(\tau_{0})=kb_{\hat{k}}(\tau_{0})$ have an asymptotic (ultraviolet) behavior given by
\begin{equation}
\label{theta-asympt-BI}
\Theta^{{\varepsilon}}_{\vec{k}}(\tau)=k\eta_{\vec{k}}(\tau)+O(k^{-1}),\quad \eta_{\vec{k}}(\tau)=\int_{\tau_{0}}^{\tau}d\tau' b_{\hat{k}}(\tau'),
\end{equation}
provided that the function $s_{\hat{k}}(\tau)$ possesses a first derivative which is integrable in every closed interval of the time domain $\mathbb{I}$. From {Eq.} (\ref{generic-sol-BI}) {and} the above initial conditions, it is not difficult to see that
\begin{equation}
\label{qp-evol-BI}
\begin{pmatrix}
\tilde{q}_{\vec{k}} \\
\tilde{p}_{\vec{k}} \end{pmatrix}_{\tau}
={\cal{T}}_{\vec{k}}(\tau,\tau_{0}) \begin{pmatrix}
\tilde{q}_{\vec{k}} \\
\tilde{p}_{\vec{k}} \end{pmatrix}_{\tau_{0}}
; \quad
{\cal{T}}_{\vec{k}}(\tau,\tau_{0})=\begin{pmatrix} {\rm{Re}}\left[e^{i\Theta^{q}_{\vec{k}}(\tau)}\right] & \frac{1}{k}{\rm{Im}}\left[e^{i\Theta^{q}_{\vec{k}}(\tau)}\right]\\\frac{-[k^{2}+s_{\hat{k}}(\tau_{0})]}{k}{\rm{Im}}\left[e^{i\Theta^{p}_{\vec{k}}(\tau)}\right] &  {\rm{Re}}\left[e^{i\Theta^{p}_{\vec{k}}(\tau)}\right]  \end{pmatrix}.
\end{equation}

{Symmetry} invariant complex structures {on} phase space ({and hence} symmetry invariant Fock representations) for the Bianchi I universes are totally characterized by {definitions of annihilation and creation-like variables} of the {linear form} \cite{Unique-Bianchi-I}
\begin{equation}
\label{inv-rep}
\begin{pmatrix}
a^{(l)}_{\vec{k}} \\
\bar{a}^{(l)}_{\vec{k}} \end{pmatrix}
={\cal{M}}_{\vec{k}} \begin{pmatrix}
\tilde{q}^{(l)}_{\vec{k}} \\
\tilde{p}^{(l)}_{\vec{k}} \end{pmatrix}
, \quad
{\cal{M}}_{\vec{k}}{=}\begin{pmatrix} f_{\vec{k}} & g_{\vec{k}} \\ \bar{f}_{\vec{k}} & \bar{g}_{\vec{k}}.  \end{pmatrix},
\end{equation} 
{with}
\begin{equation}
\label{restric-fg-BI}
f_{\vec{k}}\bar{g}_{\vec{k}}-g_{\vec{k}}\bar{f}_{\vec{k}}=-i.
\end{equation}
{This latter restriction} ensures {the} standard {PB} relationship $\{a^{(l)}_{\vec{k}} ,\bar{a}^{(l')}_{\vec{k}'} \}=-i\delta^{ll'}\delta_{\vec{k}\vec{k}'}$.

Let us now consider families of representations {that are} connected by a unitary implementable dynamics. Given the Cauchy initial section $\Sigma_{\tau_{0}}$, we specify an invariant Fock representation for the {KG} system by adopting as annihilation and creation-like variables the set formed by $a^{(l)}_{\vec{k}}(\tau_{0})$ and $\bar{a}^{(l)}_{\vec{k}}(\tau_{0})$, regarded as coefficients in an expansion of the field in an appropriate basis of solutions. Recall that the initial variables are related {to} their evolved ones, $a^{(l)}_{\vec{k}}(\tau)$ and $\bar{a}^{(l)}_{\vec{k}}(\tau)$, by a Bogoliubov transformation. The linearity of this transformation and of the space of solutions imply that $a^{(l)}_{\vec{k}}(\tau)$ and $\bar{a}^{(l)}_{\vec{k}}(\tau)$ can be used as a new set of annihilation and creation-like coefficients on the initial time section $\Sigma_{\tau_{0}}$. This new set defines a distinct Fock representation of the system that, by construction, is obtained from the previous one by dynamical evolution. The different representations specified in this way will be equivalent if and only if the introduced dynamics is implementable as a unitary operator on the Fock space associated {to} any of them. {Let us present} {an explicit construction of these families
of representations.}

From {Eqs.} (\ref{qp-evol-BI}) and (\ref{inv-rep}), it follows that
\begin{equation}
\label{ani-crea-evol-BI-i}
\begin{pmatrix}
a_{\vec{k}} (\tau)\\
\bar{a}_{\vec{k}}(\tau) \end{pmatrix} 
={\mathbf{U}}_{\vec{k}}(\tau,\tau_{0}) 
\begin{pmatrix}
a_{\vec{k}} (\tau_{0})\\
\bar{a}_{\vec{k}}(\tau_{0}) \end{pmatrix}, 
\end{equation} 
where 
\begin{equation}
\label{ani-crea-evol-BI-ii}
{\mathbf{U}}_{\vec{k}}(\tau,\tau_{0})={\cal{M}}_{\vec{k}}(\tau){\cal{T}}_{\vec{k}}(\tau,\tau_{0}){\cal{M}}^{-1}_{\vec{k}}(\tau_{0}) 
{=}\begin{pmatrix} \alpha_{\vec{k}}(\tau,\tau_{0}) & \beta_{\vec{k}}(\tau,\tau_{0}) \\ \bar{\beta}_{\vec{k}}(\tau,\tau_{0}) & \bar{\alpha}_{\vec{k}}(\tau,\tau_{0}).  \end{pmatrix}.
\end{equation}
We have ignored the superscript $l$ in {Eq.} (\ref{ani-crea-evol-BI-i}) since ${\cal{M}}_{\vec{k}}$ {and }the evolution transformation ${\cal{T}}_{\vec{k}}$ {are independent of it}. The Bogoliubov coefficients alpha and beta of {the} time evolution can be directly read off from {Eq.} (\ref{ani-crea-evol-BI-ii}). In particular, we {have}
\begin{eqnarray}
\label{antilenear-BI}
i\beta_{\vec{k}}(\tau,\tau_{0})&=&f_{\vec{k}}(\tau)g_{\vec{k}}(\tau_{0}){\rm{Re}}\left[e^{i\Theta^{q}_{\vec{k}}(\tau)}\right]-g_{\vec{k}}(\tau)f_{\vec{k}}(\tau_{0}){\rm{Re}}\left[e^{i\Theta^{p}_{\vec{k}}(\tau)}\right] \nonumber \\
\, &-& \frac{1}{k}f_{\vec{k}}(\tau)f_{\vec{k}}(\tau_{0}){\rm{Im}}\left[e^{i\Theta^{q}_{\vec{k}}(\tau)}\right]
-\frac{k^{2}+s_{\hat{k}}(\tau_{0})}{k}g_{\vec{k}}(\tau)g_{\vec{k}}(\tau_{0}){\rm{Im}}\left[e^{i\Theta^{p}_{\vec{k}}(\tau)}\right].
\end{eqnarray} 
We now impose the condition of unitary implementability of {the} time evolution, namely $\sum_{\vec{k}\in\mathbb{L}}\vert \beta_{\vec{k}}(\tau,\tau_{0})\vert^{2}<\infty$ for all times $\tau$. From {Eqs.} (\ref{theta-asympt-BI}) and (\ref{antilenear-BI}), a rigorous analysis shows \cite{Unique-Bianchi-I} that the necessary and sufficient conditions to attain a square summable {sequence} $\{\beta_{\vec{k}}(\tau,\tau_{0})\}$ in a consistent {and non-trivial} way [that is, {satisfying} the restriction (\ref{restric-fg-BI}) {and avoiding} a trivialization of the dynamics] are
\begin{enumerate}
\item{{There exists a subset ${\mathbb{L}}'$ of ${\mathbb{L}}$, differing from the latter set} in a finite number of elements at most, {such that}
\begin{equation}
\label{cond-i-BI}
f_{\vec{k}}(\tau)=\sqrt{\frac{k}{2}}e^{iF_{\vec{k}}(\tau)}+k\theta^{f}_{\vec{k}}(\tau),\quad g_{\vec{k}}(\tau)=\frac{i}{\sqrt{2k}}e^{iF_{\vec{k}}(\tau)}+\theta^{g}_{\vec{k}}(\tau),
\end{equation}
{for all $\vec{k}\in {\mathbb{L}}'$. Here,} $F_{\vec{k}}$ is {an undetermined} phase, whereas $k\theta^{f}_{\vec{k}}(\tau)$ and $\theta^{g}_{\vec{k}}(\tau)$ are subdominant terms in the asymptotic limit $k\to \infty$.}
\item{The subdominant terms $k\theta^{f}_{\vec{k}}(\tau)$ and $\theta^{g}_{\vec{k}}(\tau)$ must satisfy {for all $\tau$} the square summability condition
\begin{equation}
\label{cond-ii-BI}
\sum_{\vec{k}\in {\mathbb{L}}'}k\vert \theta^{f}_{\vec{k}}(\tau)+i\theta^{g}_{\vec{k}}(\tau)\vert^{2}<\infty.
\end{equation}}
\item{To {satisfy} restriction (\ref{restric-fg-BI}), the subdominant terms {must} be related by
\begin{equation}
\label{cond-iii-BI}
{\rm{Re}}\left[i\left(e^{-iF_{\vec{k}}}+\sqrt{2k}\bar{\theta}^{f}_{\vec{k}}\right)\theta^{g}_{\vec{k}}\right]={\rm{Re}}\left[\theta^{f}_{\vec{k}}e^{-iF_{\vec{k}}}\right].
\end{equation}}
\end{enumerate}
In short, the elements of a time dependent family of invariant Fock representations {that satisfy} the above three (necessary and sufficient) conditions will be related by a dynamical evolution that can be implemented as a unitary endomorphism in Fock space. Among all possible Fock representations {that verify} conditions (\ref{cond-i-BI}), (\ref{cond-ii-BI}), and (\ref{cond-iii-BI}), {we find the analogue of the $j_0$-Fock} representation, which is given by
\begin{equation}
\label{jo-like-rep}
\mathring{f}_{\vec{k}}=\sqrt{\frac{k}{2}},\quad \mathring{g}_{\vec{k}}=\frac{i}{\sqrt{2k}}.
\end{equation}
From {Eqs.} (\ref{theta-asympt-BI}), (\ref{antilenear-BI}), and (\ref{jo-like-rep}), one can check {that} $\vert \beta_{\vec{k}}\vert=\vert s_{\hat{k}}(\tau_{0})O(k^{-2})\vert$, so that $\{\beta_{\vec{k}}\}$ is a square summable {sequence}. Remarkably,  {the representations of any of the possible quantizations with a unitary dynamics are all equivalent among them} \cite{Unique-Bianchi-I} (i.e. there is just one family of unitary equivalent invariant Fock representations {allowing the dynamics to be unitarily implementable}). In particular, all {of them} are unitarily equivalent to the {analogue of the $j_0$-Fock} representation, at any value of the time parameter. More explicitly, let us consider any Fock representation belonging to a family {that is} connected by a {unitarily} implementable dynamics, with annihilation and creation-like variables $(a_{\vec{k}},\bar{a}_{\vec{k}})$. From {Eq.} (\ref{inv-rep}), it is {straightforward} to see that these annihilation and creation-like variables at generic time $\tau$ are related {to those of the (analogue of the)  $j_0$-Fock representation}, $(\mathring{a}_{\vec{k}},\bar{\mathring{a}}_{\vec{k}})$, by means of the Bogoliubov transformation
\begin{equation}
\label{rel-ac-var-BI}
\begin{pmatrix}
a_{\vec{k}} \\
\bar{a}_{\vec{k}} \end{pmatrix}_{\tau}
={\cal{K}}_{\vec{k}} (\tau)\begin{pmatrix}
\mathring{a}_{\vec{k}} \\
\bar{\mathring{a}}_{\vec{k}} \end{pmatrix}_{\tau}
; \quad
{\cal{K}}_{\vec{k}}={\cal{M}}_{\vec{k}}\mathring{{\cal{M}}}^{-1}_{\vec{k}}{=}\begin{pmatrix} \kappa_{\vec{k}} & \lambda_{\vec{k}} \\ \bar{\lambda}_{\vec{k}} & \bar{\kappa}_{\vec{k}}.  \end{pmatrix}.
\end{equation} 
{The} Bogoliubov coefficients kappa and lambda {are}
\begin{equation}
\label{kappa-lambda-BI}
\kappa_{\vec{k}} (\tau)=\frac{1}{\sqrt{2k}}\left[f_{\vec{k}}(\tau) -ikg_{\vec{k}}(\tau)\right],\quad
\lambda_{\vec{k}} (\tau)= \frac{1}{\sqrt{2k}}\left[f_{\vec{k}}(\tau) +ikg_{\vec{k}}(\tau)\right].
\end{equation}
Taking into account the asymptotic behavior of $f_{\vec{k}}$ and $g_{\vec{k}}$, given by {Eq.} (\ref{cond-i-BI}), we get from {Eq.} (\ref{kappa-lambda-BI}) that
\begin{equation}
\label{lambda-sqs}
\sum_{\vec{k}\in \mathbb{L}'}\vert \lambda_{\vec{k}} (\tau) \vert^{2}=\frac{1}{2}\sum_{\vec{k}\in \mathbb{L}'}k\vert \theta^{f}_{\vec{k}}(\tau)+i\theta^{g}_{\vec{k}}(\tau)\vert^{2}.
\end{equation}
The finiteness of this sum is {just} the necessary and sufficient condition (\ref{cond-ii-BI}) for {the} unitary implementability of the dynamics, unitarity that {indeed} we are assuming for the quantization determined by $(a_{\vec{k}}(\tau),\bar{a}_{\vec{k}}(\tau))$. Since {this finiteness implies that the sum of $\vert \lambda_{\vec{k}} (\tau) \vert^{2}$ over all $\vec{k}\in \mathbb{L}$ also converges,} we conclude that the considered quantizations are unitarily equivalent. {Hence,} the invariant Fock quantization defined by $(\mathring{a}_{\vec{k}},\bar{\mathring{a}}_{\vec{k}})$ is, {up to unitary transformations, the unique one that allows for a unitary implementation of the dynamics.}

\section{Conclusions}
\label{sec:diss}

In this work, we have presented an overview of the uniqueness results attained in recent years for the {Fock quantization of Gowdy cosmological models, {and of (test) real KG} fields minimally coupled to {FLRW}, de Sitter, and Bianchi I spacetimes, accomplished by imposing the requirements of (i) invariance}  under the isometries of the spatial sections and (ii) unitary {implementability of the Heisenberg} dynamics. For the cases of Gowdy models and (test) KG fields in {FLRW} and de Sitter backgrounds, the uniqueness {of the quantum representation {follows} from the removal of the ambiguities in the quantization of scalar fields with time dependent mass in spatially compact ultrastatic spacetimes, {as} a consequence of the fact that Gowdy  models and KG fields in conformally ultrastatic spacetimes can be mapped, after a suitable transformation, to a system of the aforementioned type. Let us emphasize that this transformation is itself {completely} determined by the requirements of invariance and unitarity. So, no ambiguities are left in the process. The proposed criteria {single} out a preferred description of the system, by means of a preferred choice of {the set of} variables {that are to} be quantized, and fixes a unique {family of equivalent Fock representations}.} For the case of test KG fields in Bianchi I universes, the performed analysis is slightly different because of the lack of isotropy. {This requires a non-local} time dependent canonical transformation {(more specifically,} a time dependent canonical transformation such that the change of the configuration and momentum variables is mode dependent) {that defines} a new set of canonical variables {supporting} a family of symmetry invariant Fock representations {that allow for} a unitary implementation of the dynamics. Remarkably, {this family is contained just in} one equivalence class of unitarily equivalent Fock representations, with the {analogue of the $j_0$-Fock} quantization {in it}. {Although} the ambiguity in the representation of the CCRs {is} fully removed {in this manner}, it is worth mentioning that the ambiguity concerning {the} parametrization of {the} phase space {for} Bianchi I universes remains to be elucidated.

For the sake of clarity and completeness of the presentation, we have discussed in some detail the classical and quantum theories of scalar fields in globally hyperbolic spacetimes. In particular, we have {commented the two kinds} of ambiguities {that arise} in the canonical quantization program, namely (i) the choice of fundamental observables {(at least among a family of candidates related by linear transformations)} and (ii) the selection of a Hilbert space representation of the CCRs. {Special attention has been given to the introduction of complex structures {on} phase space, and to} the role that they play in the quantum theory (both in the construction of a Hilbert space representation {and in the ambiguity for the parameterization of the CCRs}). The Bogoliubov transformations {that encode the} classical time evolution and the unitary implementability of the dynamics {have also been discussed}. For non-stationay systems, {or more} specifically for scalar fields with {a} time dependent mass in ultrastatic spacetimes $\mathbb{I}\times \Sigma$, we {have} introduced the complex structure {$j_0$, that} is the simplest complex structure {which gives} rise to an invariant Fock quantization ($j_0$ commutes with the isometries of the spatial manifold $\Sigma$). We sketched the $j_0$-Fock representation for {KG} fields with time dependent mass in ultrastatic backgrounds with the spatial topology of a circle, a three-sphere, and a three-torus. {That} is, we explicitly displayed the $j_0$-Fock representation {for} cases that describe the local degrees of freedom of {linearly polarized} Gowdy models,{\footnote{We obviated the case of (axisymmetric) KG fields with time dependent mass on $S^2$, {that describes} the Gowdy $S^{1}\times S^{2}$ and $S^{3}$ cosmological models, {because this can be done by simply replacing in Sec. \ref{subsec:KG-time-mass} the harmonics $X_{\mathbf{n}}=Q_{n\ell m}$ on $\Sigma=S^{3}$ with} the spherical harmonics $X_{\mathbf{n}}=Y_{\ell m}$ on $\Sigma=S^{2}$.}} and {KG} fields in {FLRW} and de Sitter spacetimes. 

Although we have exploited the uniqueness of the $j_0$-Fock representation in very specific ultrastatic backgrounds (namely, $\Sigma=T^{3}$ and $\Sigma=S^{n}$, with $n=1,2,3$), it is worth emphasizing that the {discussed} uniqueness result, attained by imposing the {criteria} of symmetry invariance and {of a} unitary dynamics, {is not restricted}  to these spatial topologies, but {has been extended} to ultrastatic spacetimes with arbitrary compact Riemannian sections $\Sigma$ of dimension $d\leq 3$ \cite{CMOV-cqg,CMOV-PRD86}. In particular, the $j_0$-Fock representation for a KG field with mass $m$ in $(1+1)$ Minkowski spacetime $M\cong S^{1}\times \mathbb{R}$, is the unique Fock representation satisfying the {requirements} of invariance and unitarity. This is of course not in conflict with the standard Fock representation specified by imposing Poincar\'{e} invariance. Indeed, the complex structure $j_0$ and the standard complex structure $j_{m}(\varphi,\pi)=(-(-\Delta+m^{2})^{-1/2}\pi,(-\Delta+m^{2})^{-1/2}\varphi)$ belong to the same equivalence class.{\footnote{{These} complex structures are related {via} a Bogoliubov transformation {with} $\beta_{k}=(\omega_{k}-k)/(2\sqrt{k\omega_{k}})$, where $\omega_{k}=\sqrt{k^{2}+m^{2}}$. {In} the ultraviolet regime $\beta^{2}_{k}\approx m^{4}/(16k^{4})+O(m^{6}/k^{6})$, {behavior from which our statement follows.}} Apart from this simple example, it is worth {remarking that, more generally,} no conflict arises between the Hadamard approach (reformulated in terms of the scaled field {of the $\psi$-description}) and the {criteria} of invariance and {of dynamical} unitarity, at least for the cases of scalar fields in closed {FLRW} \cite{CMOV-PRD86} and de Sitter spacetimes {\cite{CdBMV-deSitter}.} 

As we mentioned in the Introduction, the {criteria} of symmetry invariance and {of unitary implementability} of the quantum evolution {have} been successfully extended to {select} a unique preferred Fock representation for fermion fields in cosmological scenarios \cite{Uniq-for-DF,Uniq-for-DF1,Uniq-for-DF3,Uniq-for-DF4,Uniq-for-DF2,EMP,EMP2,EMP3}. {This and} the uniqueness results here reviewed have been fruitfully exploited e.g. within the Hybrid Quantization Approach \cite{hybrid-app} in order to deal with (both scalar and fermionic) perturbations in quantum cosmology (see, for instance, Refs. \cite{HQ-various1,HQ-various2,HQ-various3,HQ-various31,HQ-various4,HQ-various5,HQ-various6,HQ-various7}). 

{Finally, notwithstanding the repeatedly verified effectiveness and  robustness of {these} criteria of invariance and {of} unitarity, there are still many interesting questions, applications, and extensions to be addressed.} {Some of them are the following.}

\begin{itemize}
\item{{\emph{{Generalizations to other dimensions}}}: As we pointed out, the uniqueness of the $j_0$-Fock representation extends to spacetimes with arbitrary compact Riemannian sections $\Sigma$ of dimension $d\leq 3$. The proof of this result is based on the behavior of {the} time evolution in the ultraviolet regime. {Using this behavior, the satisfaction of the condition that guarantees the unitary implementability} of the dynamics depends critically on the dimension of $\Sigma$ and, though {the condition} is fulfilled for $d\leq 3$ \cite{CMOV-cqg}, {in general it is} not satisfied in dimensions greater or equal {than} four. In such cases, an open issue is whether one can still find a {different} Fock representation {that} leads to a unitary evolution and analyze whether its equivalence class is {singled out} uniquely by our {criteria of invariance and of unitarity.}}
\item{{\emph{Other backgrounds}}: Other interesting backgrounds where the {criteria of invariance and of unitarity} can be tested are shear free anisotropic spacetimes, like Bianchi III cosmologies. }
\item{{\emph{Other fields}}: Even though the discussion has been focused primarily on the uniqueness of scalar and fermionic fields, there seems to be no obstacles (neither conceptual nor technical) to extend the analysis to other kind of fields, for example Maxwell fields, applying to them the {proposed criteria to pick out} a unique preferred Fock quantization.}
\end{itemize}

\acknowledgments
The authors acknowledge L. Castell\'o Gomar, A. Corichi, B. Elizaga Navascu\'es, M. Fern\'andez-M\'endez, L. Fonseca, M. Mart\'{\i}n-Benito, D. Mart\'{\i}n-de Blas, J. Olmedo, S. Prado,  P. Tarr\'{\i}o, and T. Thiemann for scientific collaborations and conversations. This work was supported by Project. No. MINECO FIS2017-86497-C2-2-P from Spain, Project DGAPA-UNAM IN113618 from Mexico, and COST Action CA16104 GWverse, supported by COST (European Cooperation in Science and Technology).

\end{document}